\documentclass[12pt,twoside,openright]{report}

\usepackage{hyperref}

\usepackage{amsmath,amsfonts,amssymb,xspace,mathrsfs}
\usepackage[dvips]{color}

\usepackage{fancyheadings}
\usepackage{HDR}
\usepackage{emlines}
\usepackage{epsfig,xspace}

\def\gr{\color[named]{Green} }

\def\red{\color[named]{Red} }

\def\blu{\color[named]{Blue} }

\newcommand{\be}{\begin{equation}}
\newcommand{\ee}{\end{equation}}
\newcommand{\beq}{\begin{eqnarray}}
\newcommand{\eeq}{\end{eqnarray}}
\newcommand{\bea}[2]{\be\label{#2}\begin{array}{#1}}
\newcommand{\eea}{\end{array}\ee}

\def\lfig#1#2#3#4#5{
 \begin{figure}
 \refstepcounter{figure}
 \label{#4}
 \addtocounter{figure}{-1}
 \epsfxsize=#3
 \centerline{\epsfbox{#2}}
 \vspace{#5}
 {\bf \caption{{\rm #1}}}
 \end{figure}
}

\def\figlabel#1{\xdef#1{\the\figno}}
\def\encadremath#1{\vbox{\hrule\hbox{\vrule\kern8pt\vbox{\kern8pt
\hbox{$\displaystyle #1$}\kern8pt}
\kern8pt\vrule}\hrule}}

\chardef\tempcat=\the\catcode`\@ \catcode`\@=11
\def\cyracc{\def\u##1{\if \i##1\accent"24 i%
    \else \accent"24 ##1\fi }}
\newfam\cyrfam

\newcommand{\Nint}{\mathbb{N}}

\def\Im{\,{\rm Im}\, }
\def\Re{\,{\rm Re}\, }

\def\rangl{\right\rangle }
\def\langl{\left\langle }
\def\({\left(}
\def\){\right)}
\def\[{\left[}
\def\]{\right]}
\def\<{\left\langle}
\def\>{\right\rangle}
\def\p{\partial}

\def\11{1\!\! 1}

\def\hf{{1\over 2}}
\newcommand{\half}{\frac{1}{2}}

\newcommand{\expe}[1]{{\bf E}\!\left( #1\right)}

\def\frc#1#2{{\textstyle{#1\over#2}}}
\newcommand{\nn}{\nonumber}

\DeclareMathOperator{\Li}{Li}
\DeclareMathOperator{\Td}{Td}
\DeclareMathOperator{\ch}{ch}
\DeclareMathOperator{\rk}{rk}

\renewcommand{\d}{\mathrm{d}}

\newcommand{\IR}{\mathbb{R}}
\newcommand{\IC}{\mathbb{C}}
\newcommand{\IZ}{\mathbb{Z}}

\newcommand{\abb}{{\bf a}}
\newcommand{\bbb}{{\bf b}}
\newcommand{\cbb}{{\bf c}}
\newcommand{\dbb}{{\bf d}}
\newcommand{\sk}{\mathcal{SK}}

\def\eps{\varepsilon}

\def\vrh{\varrho}
\def\vph{\varphi}

\def\varpi{\mathrm{t}}

   \def\CX {{\cal X}}
   \def\CY {{\cal Y}}
   
\newcommand{\de}{\mathrm{d}}

\newcommand{\I}{\mathrm{i}}
\newcommand{\cA}{\mathcal{A}}
\newcommand{\cB}{\mathcal{B}}
\newcommand{\cC}{\mathcal{C}}
\newcommand{\cD}{\mathcal{D}}

\newcommand{\cF}{\mathcal{F}}
\newcommand{\cG}{\mathcal{G}}
\newcommand{\cH}{\mathcal{H}}
\newcommand{\cJ}{\mathcal{J}}
\newcommand{\cK}{\mathcal{K}}
\newcommand{\cL}{\mathcal{L}}
\newcommand{\cM}{\mathcal{M}}
\newcommand{\cN}{\mathcal{N}}
\newcommand{\cO}{\mathcal{O}}

\newcommand{\cQ}{\mathcal{Q}}

\newcommand{\cT}{\mathcal{T}}
\newcommand{\cU}{\mathcal{U}}
\newcommand{\cV}{\mathcal{V}}
\newcommand{\cW}{\mathcal{W}}
\newcommand{\cX}{\mathcal{X}}

\newcommand{\cZ}{\mathcal{Z}}
\newcommand{\lb}{\mathscr{L}}

\def\bX{\bar X}
\def\bY{ \bar Y }
\def\bZ{ \bar Z  }

\def\bF{\bar F}
\def\bH{\overline{H}}

\def\bi{\bar {\i}}
\def\bj{\bar {\j}}

\def\ba{\bar a}

\def\bc{\bar c}

\def\bw{\bar w}

\def\bz{\bar z}
\def\bu{\bar u}
\def\bv{\bar v}

\def\hA{\hat A}
\def\hB{\hat B}

\def\hH{\hat H}

\def\hR{\hat R}
\def\hphi{\hat \phi}

\def\hg{\hat g}
\newcommand{\hU}{\hat{\mathcal{U}}}
\newcommand{\hCX}{\mathcal{X}}
\def\hcF{\hat \cF}
\def\hgam{\hat \gamma}

\def\tA{{\tilde A}}

\def\tD{{\tilde D}}

\newcommand{\tsigma}{\tilde\sigma}

\newcommand{\tzeta}{{\tilde\zeta}}
\newcommand{\txi}{\tilde\xi}

\newcommand{\CP}{\IC P^1}

\def\tleta{\tilde\eta}

\def\rf{r^{\flat}}

\def\ui#1{^{[#1]}}

\def\nui#1{\nu_{[#1]}}
\def\xii#1{\xi_{[#1]}}
\def\mui#1{\mu^{[#1]}}
\def\etai#1{\eta_{[#1]}}
\def\txii#1{{\tilde\xi}^{[#1]}}
\def\ai#1{{\alpha}^{[#1]}}
\def\tai#1{\tilde{\alpha}^{[#1]}}

\def\Hij#1{H^{[#1]}}
\def\Hp{H_{\scriptscriptstyle{\smash{(1)}}}}
\def\Hpij#1{\Hij{#1}_{\scriptscriptstyle{\smash{(1)}}}}

\def\Hnpij#1{\Hij{#1}_{\scriptscriptstyle{\smash{(0)}}}}

\def\hHij#1{\hat H^{[#1]}}

\def\Gi#1{G^{[#1]}}

\def\cij#1{c}
\def\ci#1{c}
\def\ctxi{c_0}

\def\Gg{\cG_{\gamma}}

\def\Thg#1{\Theta_{\gamma_{#1}}}
\def\hng#1{n_{\gamma_{#1}}}

\def\Igg#1{\cJ_{\gamma_{#1}}}

\def\ellg#1{\ell_{#1}}

\def\Thkli#1{\Theta_{\gamma_{#1}}}

\def\Thkl{\Theta_{\gamma}}

\def\gnkl#1{\Omega(#1)}
\def\bgnkl#1{\overline{\Omega}(#1)}
\def\bOm#1#2{\overline{\Omega}_{#1}(#2)}
\def\hnkl{\Omega(\gamma)}

\def\Xikl{\Xi_{\gamma}}
\def\ellg#1{\ell_{#1}}

\def\Xiijg#1#2{\Xi_{\gamma_{#2}}^{[#1]}}
\def\Xigi#1{\Xi_{\gamma_{#1}}}

\newcommand{\kahler}{{K\"ahler}\xspace}
\newcommand{\hk}{{hyperk\"ahler}\xspace}
\newcommand{\qk}{{quaternion-K\"ahler}\xspace}

\def\ZNS#1{\cZ^{(#1)}_{\rm NS5}}
\def\hZNS#1{\hat\cZ^{(#1)}_{\rm NS5}}
\def\Ztop{\cZ_{\rm top}}

\def\MH{\cM_{\rm HM}}
\def\MV{\cM_{\rm VM}}
\def\KC{\cK_C}
\def\KK{\cK_K}
\def\QC{\cQ_C}
\def\QK{\cQ_K}
\def\Mgt{\cM^{\rm gt}_{4d}}
\def\Mthree{\cM^{\rm gt}_{3d}}

\def\hkm{\mathfrak{K}}
\def\qkm{\cQ}
\def\hkt{\cZ_{\hkm}}
\def\qkt{\cZ_{\qkm}}

\def\CY{\mathfrak{Y}}
\def\CYm{\mathfrak{\hat Y}}

\def\Cns{\cC_{\rm NS5}}

\def\gfinv{n^{(0)}}

\def\signs{\sigma_{\text{NS5}}}
\def\ThD{\Theta_{\text{D}}}
\def\Thns{\Theta}
\def\thD{\theta_{\text{D}}}
\def\thns{\theta}
\def\phiD{\phi^{\text{D}}}
\def\phins{\phi}

\def\tzp{\varpi_+^{c,d}}
\def\tzm{\varpi_-^{c,d}}
\def\tzpm{\varpi_\pm^{c,d}}

\def\tzpcd#1{\varpi_+^{#1}}
\def\tzmcd#1{\varpi_-^{#1}}
\def\tzpmcd#1{\varpi_\pm^{#1}}

\def\gsl{g}
\def\Fcl{F^{\rm cl}}
\def\Fclp{{F'}^{{\rm cl}}}
\def\bFcl{\bF^{\rm cl}}

\def\sigp{{\theta}}
\def\phip{{\theta'}}

\def\pTheta{\Theta'}

\def\cff{\beta}
\def\cfff{\gamma}

\def\Prz{{\rm Prz}}
\def\dPrz{{\rm dPrz}}
\def\todaQ{T}

\begin{document}
\renewcommand{\baselinestretch}{1.0} \normalsize

\thispagestyle{empty}

\

\vskip 3cm

\begin{center}
{\Huge \bfseries
Twistor Approach}
\\
\
\\
{\Huge \bfseries to String Compactifications:}
\\ \
\\
{\Huge \bfseries  a Review}

\vskip 1cm

{\Large Sergei Alexandrov}

\vskip 0.5cm

%

{\it Universit\'e Montpellier 2, Laboratoire Charles Coulomb \\
UMR 5221, F-34095, Montpellier, France}
\end{center}

\vskip 0.3cm

\centerline{\bf Abstract}

\begin{quote}
We review a progress in obtaining the complete non-perturbative effective action
of type II string theory compactified on a Calabi-Yau manifold. This problem is equivalent
to understanding quantum corrections to the metric on the hypermultiplet moduli space.
We show how all these corrections, which include D-brane and NS5-brane instantons,
are incorporated in the framework of the twistor approach, which provides a powerful
mathematical description of hyperk\"ahler and quaternion-K\"ahler manifolds.
We also present new insights on S-duality, quantum mirror symmetry,
connections to integrable models and topological strings.

%
\end{quote}

\renewcommand{\baselinestretch}{1.0} \normalsize
\tableofcontents
\newpage\thispagestyle{empty}

\clearpage

\begin{chapternon}{Introduction}

String theory is the most promising candidate for the theory of quantum gravity. At Planck scale
it provides a consistent and unified description of all interactions.
However, to connect this high energy description to the low energy world accessible to current observations
and to make falsifiable predictions is an extremely difficult problem.
One of the main complications is that the theory is formulated initially in ten-dimensional spacetime
and should be compactified down to four dimensions.
The resulting theory, of course, crucially depends on the internal manifold one compactifies on.
On one hand, this opens plenty of possibilities and makes string theory very rich.
On the other hand, this raises the question of how to choose the right compactification.

Furthermore, even if we could answer this question, one would still have to find
the corresponding effective theory in four dimensions. Although in the classical regime it can be obtained by the simple
Kaluza-Klein reduction, generically, even its low energy approximation is affected
by quantum corrections. In string theory they can be of two types: either weighted by the parameter $\alpha'$
controlling string tension and counting loops on the string world sheet,
or dependent on the string coupling constant $g_s$ counting string loops in the target space.
Moreover, besides the usual loop corrections, the low energy effective action receives non-perturbative contributions.
They represent the most non-trivial part of the problem. In particular, this is due to the absence
of a string instanton calculus which would allow their microscopic calculation, as can be done, for example,
in gauge theories.

As a result, our ability to find the effective theory strongly depends on how complicated the compactification
manifold is chosen to be. For instance, it can be very helpful if the chosen compactification is known
to lead to effective theories satisfying some generic restrictions, such as being invariant under various symmetries or dualities.
One of the very important characteristics of this sort is the amount of supersymmetry preserved after compactification.
The more one has supersymmetry, the more constraints the low energy effective action should satisfy.

In this review we will concentrate on the case with $N=2$ supersymmetry.
This is a somewhat compromised situation between the very rigid and much more simple
case of $N=4$ supersymmetry, on one hand, and the phenomenologically relevant, but extremely complicated
case with $N=1$ supersymmetry or without it at all, on the other hand. Thus, what we are going to consider
is already sufficiently rich to contain a lot of interesting and quite non-trivial physics,
but at the same time it is still amenable to a deep analytical analysis.

Besides, our consideration will be restricted to string theories of type II. As is well known, in this case
in order that the compactified theory has $N=2$ supersymmetry, the compactification manifold should be Calabi-Yau.
Then in the low energy limit one finds $N=2$ supergravity coupled with a set of matter fields.
One of the consequences of $N=2$ supersymmetry is that the low energy effective action is completely determined
by the metric on the moduli space parametrized by four-dimensional scalar fields. Furthermore, the moduli space
is always factorized to two disconnected components corresponding to the moduli spaces of vector and hypermultiplets.
Whereas the first one is classically exact and well understood,
the metric on the hypermultiplet moduli space $\MH$ receives various quantum corrections
and at the full non-perturbative level remains still unknown.
To find its non-perturbative description, and thus the complete low energy effective action
for compactification on arbitrary Calabi-Yau, is our primary goal.

It is useful to know that there is a simplified version of this problem. It appears upon consideration of
a supersymmetric $N=2$ four-dimensional gauge theory compactified on a circle.
The effective theory in three dimensions is also determined by the metric on its moduli space $\Mthree$
corrected by non-perturbative contributions.
As we will see below, the hypermultiplet moduli space and the one appearing in this gauge
theory context are closely related to each other and in some sense the former
can be reduced to the latter in a certain approximation.

Let us briefly compare the two moduli spaces between each other.
First of all, both of them are quaternionic manifolds.
More precisely, due to constraints of $N=2$ supersymmetry,
$\MH$ is {\it quaternion-K\"ahler} (QK) and $\Mthree$ is {\it hyperk\"ahler} (HK).
We will describe in detail these types of geometries in the second chapter.
Here we note that, before even trying to find the non-perturbative geometry of the moduli spaces,
one has to address a pure mathematical problem of how to parametrize such quaternionic manifolds.
It is almost hopeless to find the complete metric, and the non-perturbative description
is feasible only if all information about the metric can be encoded in a certain potential function.
Due to this reason, we first develop the so called {\it twistor approach} to the description
of quaternionic spaces which solves this parametrization problem.
It allows to encode all geometric information on a quaternionic manifold in a set
of holomorphic functions, thereby revealing the well known connection between $N=2$ supersymmetry and holomorphicity.
We postpone the discussion of the twistor approach till chapter \ref{chap_quatern}.
At this point we would like just to mention that it encodes the HK geometry
into a {\it complex symplectic} structure, whereas in the case of QK geometry
it gives rise to a {\it complex contact} structure on its twistor space.
Although the latter is just an odd-dimensional analogue of the former, the difference between them turns out
to be substantial. As a result, QK spaces appear to be more complicated than HK ones
and one may think that the level of this complication is precisely what differs gravity from
gauge theories.

A more concrete understanding of this idea comes when one looks at the non-perturbative
corrections to the classical metrics on $\Mthree$ and $\MH$. All these corrections
are of instantonic type originating either from BPS particles winding around the compactification circle
on the gauge theory side, or from Euclidean D-branes and NS5-branes wrapping non-trivial cycles of
the Calabi-Yau on the string theory side. In a certain sense the effects of D-branes on
$\MH$ can be identified with the effects of the BPS particles on $\Mthree$.
Later we will see how this identification can be made precise.
In particular, the D-instanton corrected contact structure on the twistor space of $\MH$
is reduced to the symplectic structure on $\Mthree$.
However, this does not work anymore upon inclusion of NS5-branes.
There is no a gauge theory analogue for the instantons coming from these branes
and they represent a pure stringy effect. It is responsible for why the string theory story
is so different and so complicated compared to the gauge theory one.

Although the final goal, namely, to find a complete non-perturbative description of the
hypermultiplet moduli space $\MH$, has not been achieved yet, a great progress has been done
in this direction.
This review aims to present these achievements.

$\bullet$
We start with a general overview of Calabi-Yau compactifications of Type II string theories.
We discuss generic restrictions on their moduli spaces, which symmetries they should realize
and describe the hypermultiplet  moduli space at the classical level. Then we present the current status of the problem
and compare it with the gauge theory case.

$\bullet$
In the second chapter we develop an important mathematical machinery
which is necessary to efficiently describe HK and QK geometries. It is based on the so called twistor spaces
associated with the initial quaternionic manifolds. It is this twistor description that was at origin
of the progress presented in the following chapters.

$\bullet$
In the third chapter we show how the classical moduli space can be reformulated using this twistor approach.
Afterwards we include a one-loop correction into this description, which is supposed to exhaust all perturbative
corrections to the moduli space. In this way we explicitly evaluate the full perturbative
metric.

$\bullet$
In the next chapter we turn to the non-perturbative effects where we restrict ourselves
to the instanton corrections due to D-instantons. We show how they are nicely incorporated
at the level of the twistor space and discuss the relation of this construction to its gauge theory cousin
and the so called wall-crossing phenomenon. Then in the same chapter we reveal an intimate relation to integrability
via Thermodynamic Bethe Ansatz (TBA).

$\bullet$
Whereas the construction of the forth chapter was suited for type IIA string theory,
in the fifth chapter we turn to the type IIB formulation. In particular, we discuss how mirror symmetry and
S-duality are realized on $\MH$ and on its twistor space and obtain instanton corrections to the
quantum mirror map.

$\bullet$
The sixth chapter is devoted to the last missing part of the full non-perturbative picture of
the hypermultiplet moduli space --- the effects of NS5-branes.
We discuss their partition function, how they affect the topology of the moduli space and derive
the corresponding instanton corrections in the one-instanton approximation
by applying S-duality to the known D-instantons.
Finally, we uncover an intriguing relation of the NS5-brane corrections to topological strings
and their wave functions.

$\bullet$
Finally, in the last chapter we exemplify some of the previous results on the simplest
case of the four-dimensional hypermultiplet moduli space, which is known by the name of
{\it universal hypermultiplet}. In this case QK spaces allow a detailed description in terms of
solutions of a certain non-linear differential equation.
We establish precise relations between different approaches to encode the QK geometry and
obtain some implications for the instanton corrections.

As can be seen from this overview, most of the results presented here
are derived and formulated using the twistor approach,
what explains also the title of the review. However, it provides only a mathematical framework.
Without explicit microscopic calculations, the only additional physical input which one has at our disposal
is a set of symmetries respected by the moduli space metric.
A remarkable fact is that these symmetries turn out to be sufficient, at least in principle, to determine
the complete non-perturbative geometry of $\MH$ or, more precisely, to parametrize it in terms of
a set of topological characteristics of Calabi-Yau such as Gromov-Witten and Donaldson-Thomas invariants,
which are supposed to be our input data.
Our results show that the idea to use the symmetries of string theory
can be realized by combining it with the twistor approach
and using the fact that any isometry on a QK space is lifted to a {\it holomorphic} action on its twistor space.
This provides a very strong constraint which fixes both the metric and the symmetry transformations
which all get quantum corrections.
In particular, as mentioned above, this strategy allows to find the action
of mirror symmetry at the non-perturbative level.

It is worth also to emphasize that
the non-perturbative description of $\MH$ and $\Mthree$ shows various connections to integrable structures.
Combining these observations with the fact that such complicated system seems to admit an exact description
hints towards integrability of gauge and string theories with $N=2$ supersymmetry.
Although this is already a well established situation in the context of $N=4$ SYM and AdS/CFT \cite{Beisert:2010jr},
here we have a chance to get it with two times less supersymmetries.
While there have been already observed many intriguing connections of $N=2$ theories
to the integrable world \cite{Donagi:1995cf,Alday:2009aq,Nekrasov:2009ui,Cecotti:2010fi},
their complete understanding is still lacking. We hope that this work may shed some light
on this fascinating subject.

\end{chapternon}

\chapter{Moduli spaces in gauge and string theories}
\label{chap_moduli}

\section{Calabi-Yau compactifications of string theory}

As is well known, all five consistent superstring theories are formulated in ten dimensions
(see, for instance, \cite{Green:1987sp,Polchinski:1998rr}).
Therefore, to connect any of them to the four-dimensional world,
one should compactify it on a six-dimensional compact manifold.
Then at low energies in four dimensions one finds an effective field theory which is typically
given by supergravity coupled to various matter fields. In particular, if one restricts to the lowest
order in the string tension $\alpha'$,
all fields come from the massless string sector and there are no terms in
the effective action with more than two derivatives. However, the precise form of the effective action
depends, of course, on the internal compact manifold chosen for compactification, background fluxes, {\it etc}.

Here we are interested in type II superstring theories and in their compactifications
which have $N=2$ supersymmetry in the low energy limit. The presence of 8 unbroken supercharges
imposes certain restrictions on the internal space which are equivalent to the condition
that it is a Calabi-Yau manifold.
In this section we describe the general structure of the effective theory
appearing in Calabi-Yau compactifications with particular
emphasis on its moduli space which completely determines the low energy action.
However, first we need to review some basic facts about Calabi-Yau geometry.

\subsection{Calabi-Yau manifolds and their moduli spaces}

A Calabi-Yau manifold $\CY$ can be defined as a compact complex manifold with a Ricci-flat K\"ahler metric.
It can be shown to have holonomy group contained in $SU(n)$, where $n$ is its complex dimension,
and vanishing first Chern class $c_1(\CY)$.
Here we are interested in the case $n=3$.
It has also been proven that given a K\"ahler manifold
with vanishing first Chern class, in the given K\"ahler class, there is a unique metric which is Ricci-flat
and thus defines a Calabi-Yau manifold.

These properties fix some of the topological characteristics of Calabi-Yau spaces.
For example, the Hodge diamond which provides the dimensions of the Dolbeault cohomology groups $H^{p,q}(\CY)$
takes the following form
\be
\begin{array}{ccccccc}
&&&1&&&
\\
&&0&&0&&
\\
&0&& h^{1,1} &&0&
\\
1&& h^{2,1} && h^{2,1} && 1
\\
&0&& h^{1,1} &&0&
\\
&&0&&0&&
\\
&&&1&&&
\end{array}
\label{Hodgediam}
\ee
It implies several immediate consequences:
\begin{itemize}
\item
Since the Hodge numbers give also dimensions of the homology groups, one observes that
$H_1(\CY)$ and $H_5(\CY)$ are trivial so that any Calabi-Yau does not have non-trivial one-
and five-dimensional cycles.

\item
The Euler characteristic can be easily calculated and is given by
\be
\chi_{\CY}=\sum_{p,q}(-1)^{p+q} h^{p,q}(\CY)=2\(h^{1,1}(\CY)-h^{2,1}(\CY)\).
\ee

\item
The fact $h^{3,0}(\CY)=1$ ensures the existence and uniqueness up to a holomorphic factor
of nowhere vanishing holomorphic (3,0)-form, commonly denoted by $\Omega$.
\end{itemize}

It is clear that $\Omega$ uniquely
determines the complex structure of $\CY$. On the other hand, the information about the K\"ahler structure
is contained in the K\"ahler (1,1)-form $J$.
Since these two structures uniquely determine the Calabi-Yau metric, they parametrize the {\it moduli space}
of Calabi-Yau manifolds.
Thus, the moduli space factorizes into the space of complex structure deformations given locally by
$H^{2,1}(\CY,\IC)$ and the space of K\"ahler class deformations coinciding with
$H^{1,1}(\CY,\IR)$.
In fact, from the string theory point of view it is natural to consider a complexification of
the K\"ahler deformations as well because the metric always appears together with the antisymmetric B-field
from the NS-NS sector which naturally combines with the K\"ahler form as $B+\I J$.
Denoting the resulting moduli spaces of complex structure and complexified K\"ahler deformations by $\KC(\CY)$ and $\KK(\CY)$,
the moduli space which should be analyzed is given by
\be
\cM_{\CY}=\KC(\CY) \times \KK(\CY).
\label{factorMX}
\ee

Both factors in \eqref{factorMX} come with a natural metric which makes them special K\"ahler manifolds.
Such a manifold of complex dimension $n$ has a metric defined by a K\"ahler potential
\be
g_{a\bar b}=\p_a\p_{\bar b}\cK(z,\bz),
\label{metKahl}
\ee
which is completely determined
by a holomorphic, homogeneous of degree 2 function $F(X)$ of $n+1$ variables $X^\Lambda$, $\Lambda=0,1,\dots,n$.
This function is called {\it prepotential} and it defines the K\"ahler potential as follows
\be
\cK=-\log \[2\Im \(\bX^\Lambda F_\Lambda(X)\)\],
\label{Kahlerpot_spec}
\ee
where we denoted $F_\Lambda=\p_{X^\Lambda}F,\ F_{\Lambda\Sigma}=\p_{X^\Lambda}\p_{X^\Sigma}F$, {\it etc}.
The coordinates $z^a$, $a=1,\dots,n$, parametrizing the special K\"ahler manifold are related to the homogeneous
coordinates $X^\Lambda$ as $z^a=X^a/X^0$.

The holomorphic (3,0)-form $\Omega$ and the complexified K\"ahler (1,1)-form provide
natural candidates for the metric, the above defined coordinates and the prepotential.
We describe now these two cases separately, which gives us also an opportunity to
introduce a few useful formulae.

\subsubsection{Complex structure moduli}

The holomorphic (3,0)-form $\Omega$ gives rise to the following K\"ahler potential on $\KC(\CY)$
written as
\be
\cK=-\log \I\int_{\CY} \Omega\wedge \bar\Omega.
\label{potKcomplstr}
\ee
To introduce the corresponding holomorphic coordinates on $\KC(\CY)$,
it is convenient to consider the full de Rham cohomology group
\be
H^3=H^{3,0}\oplus H^{2,1}\oplus H^{1,2}\oplus H^{0,3}
\ee
and its dual $H_3(\CY)$. It is always possible to choose a basis of the latter group
given by three-cycles $\cA^\Lambda$ and $\cB_\Lambda$ such that the only non-vanishing intersection numbers
are $\cA^\Lambda \# \cB_\Sigma=\delta^\Lambda_\Sigma$. They can be thought as A and B-cycles, correspondingly.
Given such a basis, we define
\be
X^\Lambda =\int_{\cA^\Lambda}\Omega,
\qquad
F_\Lambda =\int_{\cB_\Lambda}\Omega.
\label{defXF}
\ee
Of course, these $2(h^{2,1}+1)$ coordinates are not all independent since
the deformations of the complex structure span only a $h^{2,1}$-dimensional space.
As a result, one can always tune the basis of cycles such that all $F_\Lambda$ are functions of $X^\Lambda$.
Moreover, the Riemann bilinear identity
\be
\int_{\CY} \chi \wedge \psi =\sum_\Lambda \(\int_{\cA^\Lambda}\chi \int_{\cB_\Lambda}\psi
-\int_{\cB_\Lambda}\chi\int_{\cA^\Lambda}\psi\)
\label{Riemannident}
\ee
for closed 3-forms $\chi$ and $\psi$,
applied to $\chi=\Omega$, $\psi=\p_\Lambda \Omega$, ensures that
$F_\Lambda$ are derivatives of a homogeneous function which can be obtained as $F=\hf\, X^\Lambda F_\Lambda$.
This function defines the prepotential on the space of complex structure moduli.
Then the K\"ahler potential \eqref{potKcomplstr} coincides with the one following from \eqref{Kahlerpot_spec}.

\subsubsection{K\"ahler moduli}

In this case one should consider instead the even cohomology group
\be
H^{\rm even}=H^{0}\oplus H^{2}\oplus H^{4}\oplus H^{6}.
\label{evencoh}
\ee
Let us choose a basis in this space parametrized by $\omega_I=(1,\omega_i)$ and
$\omega^I=(\omega_{\CY},\omega^i)$ where $\omega_{\CY}$ is the volume form such that
\be
\omega_i\wedge\omega^j=\delta_i^j\omega_{\CY},
\qquad
\omega_i\wedge\omega_j=\kappa_{ijk}\omega^k,
\label{reldifforms}
\ee
and the indices run over $I=(0,i)=0,1,\dots,h^{1,1}(\CY)$.\footnote{From now on, the indices
$\Lambda,\Sigma$ (resp. $a,b$) always run from 0 (resp. 1) till $h^{2,1}(\CY)$,
whereas $I,J$ (resp. $i,j$) go till $h^{1,1}(\CY)$.\label{foot_indices}}
The second equality in \eqref{reldifforms}
defines triple intersection numbers of the Calabi-Yau.
Choosing the dual basis $\gamma^i$ of 2-cycles and a basis $\gamma_i$ of 4-cycles,
they can also be given as
$\kappa_{ijk}=\int_{\CY}\omega_i\wedge \omega_j\wedge\omega_k=\langle \gamma_i,\gamma_j,\gamma_k\rangle$.
Then the natural metric on $\KK(\CY)$ is defined by
\be
g_{i\bj}=\frac{1}{4\cV}\int_{\CY} \omega_i\wedge \star\omega_j=\p_i\p_{\bj}\(-\log 8\cV\),
\label{metKdef}
\ee
where
\be
\cV=\frac16\, \int_{\CY} J\wedge J\wedge J
\ee
is the volume of the Calabi-Yau.

The holomorphic coordinates on $\KK(\CY)$ arise as
\be
v^i\equiv b^i+\I t^i=\int_{\gamma^i}(B+\I J).
\label{coorCK}
\ee
Defining the homogenous coordinates via $X^I=(1,v^i)$, the prepotential on the K\"ahler moduli
space can be written in terms of the intersection numbers
\be
F(X)=-\frac{1}{6}\,\frac{\kappa_{ijk}X^i X^j X^k}{X^0}.
\label{prepcubic}
\ee
Then \eqref{Kahlerpot_spec} indeed leads to the K\"ahler potential from \eqref{metKdef}.
As we will see, the cubic prepotential \eqref{prepcubic} is only an approximation,
valid in the limit of large volume, of the prepotential which appears in string compactifications.

\subsection{Type II supergravities in 10 and 4 dimensions: field content}
\label{subsec-sugra}

To obtain the low energy effective description of compactified type II string theories,
it is convenient to start from the corresponding supergravity theories in 10 dimensions.
There are two versions of ten-dimensional supergravity corresponding to the two versions
of type II superstring --- type IIA and type IIB. The bosonic sector of both of them consists
from NS-NS fields and R-R fields. The former feature the ten-dimensional metric, a two-form
$\hB_2$ and the dilaton $\hphi$,\footnote{In this section all ten-dimensional fields will be labeled
by hat and $p$-forms will carry the low index $p$.} whereas the latter comprise $p$-form potentials $\hA_p$
with $p=1,3$ in type IIA and $p=0,2,4$ in type IIB with additional self-duality constraint
on the field strength of $\hA_4$
\be
\hcF_5 =\star \hcF_5.
\label{selfF5}
\ee

To compactify these theories on a Calabi-Yau $\CY$, one should expand the ten-dimensional
fields in the basis of harmonic forms on $\CY$.
For even $p$-forms, such a basis was already introduced below \eqref{evencoh}. For odd $p$,
one should consider only 3-forms since $H^1(\CY)$ and $H^5(\CY)$ are trivial.
We label the basis 3-forms as $\alpha_\Lambda$ and $\beta^\Lambda$, which are chosen to be
dual to the B and A-cycles, correspondingly. In the following we will be interested only
in the bosonic sector since the fermionic contributions can be restored, at least in principle,
by using supersymmetry. Below we summarize the result of its effective four-dimensional description.

\subsubsection{Type IIA}

The bosonic sector of type IIA supergravity in ten dimensions is described by the following action \cite{Polchinski:1998rr}
\be
\begin{split}
S_{\rm IIA}=&\, \hf\int \biggl[ e^{-2\hphi}\(\hR\star 1 + 4\de \hphi\wedge \star \de\hphi-\hf\,\hH_3\wedge \star \hH_3 \)
\biggr.
\\ & \biggl.
-\bigl(\hcF_2\wedge\star \hcF_2+\hcF_4\wedge\star \hcF_4\bigr)-\hB_2\wedge \de \hA_3\wedge \de \hA_3\biggr],
\end{split}
\ee
where the field strengths for the RR gauge potentials and the antisymmetric B-field are defined as
\be
\hH_3=\de \hB_2,
\qquad
\hcF_2=\de \hA_1,
\qquad
\hcF_4=\de \hA_3-\hA_1\wedge \hH_3
\ee
Upon compactification all fields  are decomposed as shown in the third line of Table \ref{tableIIA}.
In particular, the metric gives rise to the complex scalars $z^a$ and the real scalars $t^i$ because,
as we know from the previous subsection, its deformations, comprising deformations of
the complex structure and of the K\"ahler class, are in one-to-one correspondence with the harmonic
(2,1) and (1,1)-forms, respectively.

\begin{table}[h]
\begin{center}
\begin{tabular}{c|c|c|c|c|c}
Type IIA & \multicolumn{3}{|c|}{NS-NS} & \multicolumn{2}{c}{R-R}
\\[2pt]
\hline
\rule{0pt}{15pt}
10d & $\hg_{XY}$ & $\hB_2$ & $\hphi$ & $\hA_1$ & $\hA_3$
\\[3pt]
\hline
\rule{0pt}{15pt}
4d & $g_{\mu\nu}$ \ $t^i$ \ $z^a$ &
$B_2+b^i\omega_i$ & $\phi$ & $A_1^0$ &
$A_3+A_1^i \wedge \omega_i+\zeta^\Lambda\alpha_\Lambda+\tzeta_\Lambda\beta^\Lambda$
\\[6pt]
\hline\vspace{-0.7cm}
&&&&&
\\
4d dual & ${\gr g_{\mu\nu}}$ \ ${\blu t^i}$ \ ${\red z^a}$ &
$\mathop{\vphantom{A}{\red \sigma}}\limits^{\displaystyle\updownarrow}$ \hspace{0.5cm} ${\blu b^i}$\
& ${\red \phi}$ & ${\gr A_1^0}$ &
$\mathop{\vphantom{A}\rm const}\limits^{\displaystyle\updownarrow}$\hspace{0.2cm}
${\blu A_1^i}$ \hspace{0.9cm} ${\red \zeta^\Lambda}$ \hspace{0.7cm} ${\red \tzeta_\Lambda}$ \ \ \ \
\end{tabular}
\end{center}
\caption{\rm Field content of 10d type IIA supergravity and its Calabi-Yau compactification. The last line
shows the field content after dualization of $p$-forms in four dimensions. The green, blue and
red colors indicate fields from the gravitational, vector and hypermultiplets, respectively.}
\label{tableIIA}
\end{table}

All four-dimensional fields organize in $N=2$ supermultiplets:
\begin{center}
\begin{tabular}{ll}
gravitational multiplet & $(g_{\mu\nu},A_1^0)$
\\
tensor multiplet  & $(B_2,\phi,\zeta^0,\tzeta_0)$
\\
$h^{2,1}$ hypermultiplets & $(z^a,\zeta^a,\tzeta_a)$
\\
$h^{1,1}$ vector multiplets & $(A_1^i,v^i\equiv b^i+\I t^i)$
\end{tabular}
\end{center}
This list however does not include the 3-form $A_3$. The reason is that such a field
carries no degrees of freedom since in four dimensions it can be dualized to a constant.
In the following we will put it to zero,
although in \cite{Louis:2002ny} it has been shown that it can play an important role inducing a gauge charge
for the NS-axion $\sigma$. The latter scalar in turn appears after dualization of the 2-form $B_2$.
As a result of this dualization, the tensor multiplet converts into an additional hypermultiplet and the resulting
field content is shown in the last line of Table \ref{tableIIA} where different colors distinguish different types
of $N=2$ multiplets.

Note that the hypermultiplet dual to the tensor multiplet is always present in spectrum, independently
on the Hodge numbers of the Calabi-Yau. Due to this reason it is called {\it universal hypermultiplet}.
When $h^{2,1}$ vanishes it exhausts all hypermultiplets and the problem of its effective description
represents the simplest case of the problem considered in this work.\label{UHMremark}

\subsubsection{Type IIB}

Now we turn to the type IIB supergravity. The bosonic part of its action reads \cite{Polchinski:1998rr}
\be
\begin{split}
S_{\rm IIB}=&\, \hf\int \biggl[ e^{-2\hphi}\(\hR\star 1 + 4\de \hphi\wedge \star \de\hphi-\hf\,\hH_3\wedge \star \hH_3 \)
\biggr.
\\ & \biggl.
-\(\hcF_1\wedge \star \hcF_1+\hcF_3\wedge\star \hcF_3+\hf\,\hcF_5\wedge\star \hcF_5\)-\hA_4\wedge \de \hH_3\wedge \de \hA_2\biggr],
\end{split}
\label{actIIB10d}
\ee
where
\be
\hH_3=\de \hB_2,
\quad
\hcF_1=\de \hA_0,
\quad
\hcF_3=\de \hA_2-\hA_0\wedge \hH_3,
\quad
\hcF_5=\de \hA_4-\hf\, \hA_2\wedge \hH_3+\hf\, \hB_2 \wedge \de \hA_2,
\ee
and the equations of motion following from the action \eqref{actIIB10d} should be supplemented by
the self-duality condition \eqref{selfF5}.\footnote{It is possible also to write a covariant action
which incorporates the self-duality condition \cite{Dall'Agata:1997ju,Dall'Agata:1998va}.}
As in the type IIA case, we present the result of the expansion of ten-dimensional fields in Table \ref{tableIIB}.

\begin{table}[h]
\begin{center}
\begin{tabular}{c|c|c|c|c|c|c}
Type IIB & \multicolumn{3}{|c|}{NS-NS} & \multicolumn{3}{c}{R-R}
\\[2pt]
\hline
\rule{0pt}{15pt}
10d & $\hg_{XY}$ & $\hB_2$ & $\hphi$ & $\hA_0$ & $\hA_2$ & $\hA_4$
\\[3pt]
\hline
\rule{0pt}{15pt}
4d & $g_{\mu\nu}$ \ $t^i$ \ $z^a$ &
$B_2+b^i\omega_i$ & $\phi$ & $c^0$ & $A_2+c^i\omega_i$ &
$D_2^i\omega_i+\tD_i\omega^i+A_1^\Lambda\alpha_\Lambda+\tA_{1,\Lambda}\beta^\Lambda$
\\[6pt]
\hline\vspace{-0.7cm}
&&&&&
\\
4d dual & ${\gr g_{\mu\nu}}$ \ ${\red t^i}$ \ ${\blu z^a}$ &
$\mathop{\vphantom{A}{\red \psi}}\limits^{\displaystyle\updownarrow}$ \hspace{0.5cm} ${\red b^i}$\
& ${\red \phi}$ & ${\red c^0}$ & $\mathop{\vphantom{A}{\red c_0}}\limits^{\displaystyle\updownarrow}$
\hspace{0.4cm} ${\red c^i}$ &
$\mathop{\vphantom{A}{\red c_i}}\limits^{\displaystyle\updownarrow}$\hspace{1.9cm}
${\gr A_1^0}\quad{\blu A_1^a}$ \hspace{1.4cm}
\end{tabular}
\end{center}
\caption{\rm Field content of 10d type IIB supergravity and its Calabi-Yau compactification. The last line
shows the field content after dualization of $2$-forms in four dimensions. The green, blue and
red colors indicate fields from the gravitational, vector and hypermultiplets, respectively.}
\label{tableIIB}
\end{table}

A new feature here is that the self-duality condition implies that
the scalars $\tD_i$ and the vectors $\tA_{1,\Lambda}$ are dual to
the 2-forms $D_2^i$ and to the vectors $A_1^\Lambda$, respectively. Therefore, they are redundant fields
and do not contribute to the spectrum.
The remaining fields again combine in $N=2$ supermultiplets as follows:
\begin{center}
\begin{tabular}{ll}
gravitational multiplet & $(g_{\mu\nu},A_1^0)$
\\
double-tensor multiplet  & $(B_2,A_2,\phi,c^0)$
\\
$h^{1,1}$ tensor multiplets & $(t^i,b^i,c^i,D_2^i)$
\\
$h^{2,1}$ vector multiplets & $(A_1^a,z^a)$
\end{tabular}
\end{center}
Finally, all 2-forms can be dualized into scalars. This maps the double-tensor and $h^{1,1}$ tensor
multiplets into $h^{1.1}+1$ hypermultiplets. The resulting spectrum is shown
in the last line of Table \ref{tableIIB}.

\bigskip

From this analysis one concludes that  in four dimensions in both cases one obtains
$N=2$ supergravity given by the gravitational multiplet coupled to
two types of matter multiplets:
vector multiplets (VM), whose bosonic sector contains a gauge field and a complex scalar,
and hypermultiplets (HM) each having 4 real scalars.
The number of the multiplets is determined by the Hodge numbers according to the following table:
\begin{center}
\begin{tabular}{c|cc}
& type IIA & type IIB
\\[2pt]
\hline
\rule{0pt}{15pt}
hypermultiplet & $h^{2,1}+1$ & $h^{1,1}+1$
\\
vector multiplets & $h^{1,1}$ & $h^{2,1}$
\end{tabular}
\end{center}

This result is in perfect agreement with the statement of {\it mirror symmetry}
that type IIA string theory compactified on a Calabi-Yau manifold $\CY$ should be
equivalent to type IIB string theory compactified on a mirror Calabi-Yau $\CYm$.
The mirror manifold has swaped Hodge numbers with respect to the original Calabi-Yau
\be
h^{2,1}(\CYm)=h^{1,1}(\CY),
\qquad
h^{1,1}(\CYm)=h^{2,1}(\CY),
\ee
which agrees with the spectrum presented above.

\subsection{Low energy effective action}

The action of $N=2$ supergravity coupled to $n_V$ vector multiplets and $n_H$ hypermultiplets
to the large extent is fixed by supersymmetry. At the two-derivative level, corresponding
to the low energy approximation of string theory, the vector and hypermultiplets
stay decoupled from each other and interact only via gravitational interaction.
Thus, in the Einstein frame the bosonic part of the action splits into three terms
\be
S_{\rm eff}=\hf\int R\star 1 +S_{\rm VM}+S_{\rm HM}.
\ee

The effective action for the vector multiplets is known to have the following general
form \cite{deWit:1984pk,Cecotti:1989qn}\footnote{We use here the notations for indices appropriate for the type IIA case.}
\be
S_{\rm VM}=\int \[ -\hf\Im \cN_{IJ} \cF^I\wedge \star \cF^J+\hf\Re\cN_{IJ}\cF^I\wedge\cF^J
+\cK_{i\bj}(v,\bv)\de v^i \wedge \star\de \bv^{\bj}\],
\label{actvect}
\ee
where the field strength of the graviphoton $\cF^0$ is combined with other gauge fields
and $\cN_{IJ}$ is a complex matrix dependent on the scalars $v^i$ and their complex conjugates.
Moreover, the couplings $\cN_{IJ}$ and $\cK_{i\bj}$ are not arbitrary, but restricted further by $N=2$ supersymmetry.
In particular, $\cK_{i\bj}$ must be the metric on a special K\"ahler manifold parametrized by $v^i$.
In other words, it is determined by a prepotential $F$, a homogeneous holomorphic function,
according to \eqref{Kahlerpot_spec}. Furthermore, the gauge couplings $\cN_{IJ}$ are also determined
by the same prepotential \cite{deWit:1984pk}
\be
\cN_{IJ}(v,\bv) =\bF_{IJ} -
\I\, \frac{ [N \cdot v]_I[N \cdot v]_{J}}{v^K N_{KL}v^{L}},
\ee
where we introduced $N_{IJ}=-2\Im F_{IJ}$ and used $v^I=(1,v^i)$.
Note that $\Im \cN_{IJ}$ is negative definite so that the kinetic term in \eqref{actvect} is well defined.

As to the hypermultiplet effective action, it is given by a non-linear $\sigma$-model
\be
S_{\rm HM}=\int \de^4 x\, g_{\alpha\beta}(q) \p^\mu q^\alpha \p_\mu q^\beta,
\ee
where $q^\alpha$, $\alpha=1,\dots, 4n_H$, denotes the collection of all scalar fields from the hypermultiplet sector.
As in the case of vector multiplets, the target space metric $g_{\alpha\beta}(q)$ is not arbitrary.
The $N=2$ supersymmetry requires that it defines a {\it quaternion-K\"ahler} (QK) manifold \cite{Bagger:1983tt}.
We will study this type of geometry in detail in chapter \ref{chap_quatern}.

Thus, one concludes that the low energy effective action of Calabi-Yau compactifications
of type II superstring theory is completely determined by the metric on the {\it moduli space},
which is parametrized by the scalars of vector and hypermultiplets.
Moreover, since the coupling matrices $\cK_{ij}$ and $g_{\alpha\beta}$, determining the metric,
can depend only on $v^i$ and $q^\alpha$, respectively,
at the two derivative level the complete moduli space is factorized \cite{Bagger:1983tt,deWit:1984px}
\be
\cM_{\rm 4d}=\MV\times \MH,
\label{factmodspace}
\ee
where the first factor is required to be special K\"ahler, whereas the second factor must be
quaternion-K\"ahler. These statements follow just from $N=2$ supersymmetry and therefore should hold
at full quantum level.

The holomorphic prepotential $F$ describing $\MV$ is tree level exact,
i.e. it does not receive any $g_s$-corrections, and can be explicitly computed.
In particular, in type IIB theory, where the vector multiplet moduli space coincides with
the space of complex structure deformation of Calabi-Yau, $\MV=\KC$, the prepotential is also free of
$\alpha'$-corrections. Therefore, in this case the effective action obtained for the vector multiplets
by the usual Kaluza-Klein reduction is exact.
As a result, the prepotential
turns out to be the same as the one defined via \eqref{defXF} by the holomorphic 3-form $\Omega$.

On the other hand, in type IIA theory where $\MV$ appears as the space of complexified K\"ahler class deformations,
$\MV=\KK$, the natural prepotential \eqref{prepcubic} gives only the large volume approximation of
the complete geometry of the moduli space.
The correct prepotential can be derived using
mirror symmetry which requires in particular that the moduli spaces of type IIA and type IIB theories
compactified on mirror Calabi-Yaus coincide. Due to the factorization \eqref{factmodspace},
this implies that
\be
\MV^{\rm A/B}(\CY)=\MV^{\rm B/A}(\CYm),
\qquad
\MH^{\rm A/B}(\CY)=\MH^{\rm B/A}(\CYm).
\label{mirrorsymmoduli}
\ee
As a result, one can find the prepotential in type IIA from the known exact prepotential in the type IIB formulation.
The result is given by the following expression \cite{Candelas:1990rm,Hosono:1993qy}
\be
\label{lve}
F(X)=-\kappa_{ijk}\frac{X^i X^j X^k}{6 X^0}+ \frac12 A_{IJ} X^I X^J
+ \chi_{\CY}\frac{\zeta(3)(X^0)^2}{2(2\pi\I)^3}
-\frac{(X^0)^2}{(2\pi\I)^3}{\sum_{k_i\gamma^i\in H_2^+(\CY)}} \gfinv_{k_i}\,
\Li_3\left( e^{2\pi \I  k_i X^i/X^0}\right).
\ee
It consists of four contributions. The first term is the cubic prepotential \eqref{prepcubic}
which appeared in the analysis of K\"ahler deformations of Calabi-Yau.
The second quadratic term is determined by a real symmetric matrix $A_{IJ}$ and
is usually omitted since it does not affect the K\"ahler potential. However, as we will see
in section \ref{chap_IIB}.\ref{sec_IIBform}, it turns out to be crucial for non-perturbative mirror symmetry.
Therefore, we include it from the very beginning.
The remaining two terms represent $\alpha'$-corrections and thus they are stringy effects
which cannot be obtained by the Kaluza-Klein reduction.
In particular, the third term is the only perturbative $\alpha'$-correction, determined by
the Euler characteristic of the Calabi-Yau, whereas the last term
gives a contribution of worldsheet instantons, which are exponentially suppressed in $\alpha'$.
In this term the sum goes over effective divisors belonging to the second homology group,
the trilogarithm $\Li_3(x)=\sum_{n=1}^\infty x^n/n^3$ encodes multi-covering effects, and
$\gfinv_{k_a}$ are the genus zero Gopakumar-Vafa invariants.
It is useful to note that viewing $\zeta(3)$ as $\Li_3(1)$, one can include the one-loop
correction into the sum by defining $\gfinv_{0}=-\chi_{\CY}/2$.

Anyway, one concludes that the holomorphic prepotential is generically known, being
determined in terms of some topological invariants of the Calabi-Yau threefold,
and therefore the vector multiplet sector is completely under control.

On the other hand, the situation in the hypermultiplet sector
is much more complicated.
The difficulty is twofold. First, $\MH$ is a \qk manifold.
This is a more complicated type of geometry than special K\"ahler.
The main problem with QK manifolds is how to parametrize their metrics.
In the special K\"ahler case this is done through the holomorphic prepotential, but
in the QK case there is no single function playing the role of such prepotential.
We will show below that this problem is solved by using a certain twistor space
construction which provides us with a set of holomorphic functions generalizing
the prepotential of the special K\"ahler geometry.

The second difficulty is that, contrary to $\MV$, $g_{\alpha\beta}(q)$ receives
all possible $g_s$-corrections, both perturbative and non-perturbative \cite{Becker:1995kb}.
The later are especially complicated since the rules of the string instanton calculus
are not established so that the straightforward microscopic calculation cannot be performed.
As a result, we need to use other methods to find these corrections such as various symmetries
and dualities existing in string theory.

As a result, the exact non-perturbative metric on the hypermultiplet moduli space,
and therefore the low energy effective action, remains so far unknown.
This metric will be the central object of our interest and our main purpose will be to describe
the progress which has been achieved in the description of the non-perturbative geometry of $\MH$.

\section{Hypermultiplet moduli space}
\label{sec_HMms}

\subsection{Symmetries and dualities}
\label{subsec_sym}

Before we give explicit expressions for the classical metric on the hypermultiplet moduli space
and its quantum corrections, it is important to understand the general symmetry properties of $\MH$.
In the absence of explicit microscopic calculations, these symmetries are actually the only
tool to access the quantum corrected metric. To help the reader, we summarize in Table \ref{HMfields}
the field content of the hypermultiplet moduli spaces in type IIA and type IIB theories.

\vspace{0.7cm}

\begin{table}[h]
\begin{center}
\begin{tabular}{c|c|c|c|c}
 & \multicolumn{3}{|c|}{NS-NS} & R-R
\\[2pt]
\hline
\rule{0pt}{20pt}
  & ${\mbox{complex structure/K\"ahler}\atop \mbox{deformations of }\displaystyle\CY}$ & dilaton & NS-axion &
${\mbox{periods of $p$-form}\atop \mbox{gauge potentials}}$
\\[6pt]
\hline
\rule{0pt}{15pt}
Type IIA & $z^a$ & $\phi$ & $\sigma$ & $\zeta^\Lambda \quad \tzeta_\Lambda$
\\[6pt]
\rule{0pt}{10pt}
Type IIB & $v^i\equiv b^i+\I t^i$ & $\phi$ & $\psi$ &
$c^0\quad c^i\quad c_i\quad c_0$
\end{tabular}
\end{center}
\caption{\rm Coordinates on $\MH$ and their physical origin.}
\label{HMfields}
\end{table}

\begin{itemize}
\item
{\bf Symplectic invariance}

In the type IIA formulation the four-dimensional fields are defined with respect to a basis of 3-cycles,
$(\cA^\Lambda,\cB_\Lambda)$. However, there is no canonical choice for such a basis.
One can apply to it a symplectic transformation $\cO\in Sp(2h^{2,1}+2,\IZ)$
\be
\label{emag}
\cO:\
\begin{pmatrix}
\cA^\Lambda\\
\cB_\Lambda
\end{pmatrix} \mapsto
\begin{pmatrix} \abb & \bbb \\ \cbb & \dbb \end{pmatrix}
\begin{pmatrix}
\cA^\Lambda\\
\cB_\Lambda
\end{pmatrix},
\ee
where $\abb,\bbb,\cbb$ and $\dbb$ are integer valued $(h^{2,1}+1)\times(h^{2,1}+1)$ matrices obeying
\be
\begin{array}{c}
\abb^{\rm T} \cbb - \cbb^{\rm T} \abb =
\bbb^{\rm T} \dbb - \dbb^{\rm T} \bbb = 0 ,
\\ \vphantom{\mathop{A}\limits^{A^A}}
\abb^{\rm T} \dbb - \cbb^{\rm T} \bbb =  \bf{1} ,
\end{array}
\ee
and the new basis will still satisfy all required properties.
Such transformation of the basis of 3-cycles induces a symplectic transformation of the four-dimensional fields.
In particular $(\zeta^\Lambda,\tzeta_\Lambda)$ form a symplectic vector, $\phi$ and $\sigma$ are invariant,
whereas the transformation of $z^a$ is induced by the symplectic transformation of the vector
$(X^\Lambda,F_\Lambda)$ built from the homogeneous coordinates and the prepotential.
Since physics should not depend on the choice of the basis, the compactified theory,
and the metric on $\MH$ in particular, should be invariant under this
symplectic symmetry.

\item {\bf $SL(2,\IZ)$ duality}

On the other hand, the type IIB string theory is known to be invariant under S-duality \cite{Hull:1994ys},
which is in fact generalized to $SL(2,\IR)$ symmetry of the supergravity action
\eqref{actIIB10d} \cite{Berkovits:1995cb,Berkovits:1998jh}.
After compactification on a Calabi-Yau, this ten-dimensional symmetry
descends to a symmetry of the four-dimensional effective theory.
Whereas in the large volume, small string coupling limit one still has the continuous
$SL(2,\IR)$ symmetry group, one expects that it becomes broken to $SL(2,\IZ)$ by $\alpha'$ and $g_s$ corrections.
This symmetry plays an extremely important role in our approach. However, we postpone
the discussion of its action on the physical fields till chapter \ref{chap_IIB}
because understanding the correct transformation laws requires a thorough analysis of
various subtle issues.

\item {\bf Heisenberg symmetry}

Since the scalars in the last two columns of Table \ref{HMfields} appear as periods or duals of
various gauge fields, the moduli space should inherit a remnant of the gauge symmetries
of the original ten-dimensional theory. As a result, $\MH$ turns out to be invariant under
the so called Peccei-Quinn symmetries which act as certain shifts of the RR-fields and the NS-axion \cite{deWit:1992wf}.
Their action is most easily described in the type IIA formulation where they
form the Heisenberg algebra:
\be
\label{Heisenb}
T_{H,\kappa}:\quad
\bigl(\zeta^\Lambda,\ \tzeta_\Lambda,\ \sigma\bigr)\mapsto
\bigl(\zeta^\Lambda + \eta^\Lambda ,\
\tzeta_\Lambda+ \tleta_\Lambda ,\
\sigma + 2 \kappa - \tleta_\Lambda \zeta^\Lambda + \eta^\Lambda \tzeta_\Lambda
\bigr) ,
\ee
so that
\be
T_{H_2,\kappa_2} T_{H_1,\kappa_1} =
T_{H_1+H_2,\kappa_1+\kappa_2+\frac{1}{2}\langle H_1, H_2\rangle},
\label{grouplaw}
\ee
where $H=(\eta^\Lambda,\tleta_\Lambda)$ and we introduced a symplectic invariant scalar product
\be
\langle H, H' \rangle=\tleta_\Lambda \eta'^\Lambda-\tleta'_\Lambda \eta^\Lambda.
\label{SymplecticPairing}
\ee
At the perturbative level, the parameters $(\eta^\Lambda,\tilde\eta_\Lambda,\kappa)$ can
take any real value. However, as we will discuss below, instanton corrections break
these symmetries to a discrete subgroup with $(\eta^\Lambda,\tilde\eta_\Lambda)\in H^3(\CY,\IZ)$, $\kappa \in \IZ$.

\item {\bf Monodromy invariance}

There is another Peccei-Quinn symmetry which is not contained in the Heisenberg group \eqref{Heisenb}.
Its origin is clear in the type IIB formulation where $\MH$ includes the fields $b^i$ originating from the 2-form $\hB_2$
and therefore inheriting the corresponding gauge symmetry. As usual, the symmetry acts by shifting these fields,
but also affecting some other fields. However, in contrast to the Heisenberg transformations, it is
a continuous symmetry only in the large volume limit, or the large complex structure limit in type IIA.
Upon including any quantum or $\alpha'$ corrections, it is broken to a discrete subgroup
and it is interpreted as monodromy invariance around the large volume point in $\KK$.
The latter is in fact a particular subset of symplectic transformations.
Nevertheless, it is worth to consider it on its own because on the type IIB side
the general symplectic invariance is not explicit.
We will return to this symmetry when we consider the type IIB formulation in more detail in chapter \ref{chap_IIB}.

\item {\bf Mirror symmetry}

Finally, as we already know, mirror symmetry identifies the moduli space in type IIA (type IIB) theory
with the moduli space in type IIB (resp. type IIA) compactified on the mirror Calabi-Yau
\eqref{mirrorsymmoduli}.
This implies that, upon correct identification of the coordinates, the metrics on the moduli space in the two formulations
must be identical. As a result, all symmetries which have been observed in one formulation, should also be present in
the other one. However, being rewritten in the new coordinates, they may become hidden and
are not easy to deal with. Thus, each symmetry is naturally associated with one of the two formulations.
For example, the type IIA formulation is adapted to symplectic invariance, whereas the type IIB formulation
is suited for $SL(2,\IZ)$ duality. On the other hand, mirror symmetry allows to pass from one to another.

\end{itemize}

\subsection{Tree level metric and c-map}
\label{subsec_cmap}

From Table \ref{HMfields} it follows that $\MH$ always contains fields parameterizing either
$\KC$ or $\KK$ depending on whether we are considering type IIA or type IIB formulation, respectively.
Therefore, one can expect that the metric on $\MH$ restricted to these subspaces coincides
with the corresponding special K\"ahler metric determined by a holomorphic prepotential.
In fact, it turns out that at the string tree level the {\it full} metric on $\MH$ is determined
by the prepotential governing the vector multiplets.
Such a map from a special K\"ahler metric to a \qk metric is known as {\it c-map} \cite{Cecotti:1989qn,Ferrara:1989ik}.

The existence of the c-map is not accidental but has a concrete physical explanation.
It relies on compactification to three dimensions and the use of T-duality.
Namely, let us consider a compactification of type IIA and type IIB theories on the {\it same}
Calabi-Yau times a circle $S^1$. The T-duality along the circle relates the two theories so that
\be
{\rm IIA}/\( \CY\times S^1_R\) \simeq {\rm IIB}/\(\CY\times S^1_{1/R}\).
\label{Tdual3d}
\ee
On the other hand, their low energy descriptions can be obtained by compactifying
the four-dimensional effective theories from the previous section on the circle $S^1$.
Upon such compactification, each gauge field gives rise to two massless scalars in three dimensions:
one comes from the component of the four-dimensional gauge field along the circle and the second
appears by dualizing the remaining three-dimensional vector field.
As a result, all vector multiplets in four dimensions descend to hypermultiplets in three dimensions.
Moreover, one additional hypermultiplet arises from the metric.
The corresponding part of the effective action is given by a non-linear $\sigma$-model
with a \qk target space. Thus, in both type IIA and type IIB cases,
the full moduli space in three dimensions is given by a product of
two QK manifolds
\be
\cM_{\rm 3d}=\QC(\CY)\times\QK(\CY)
\label{3dmodspace}
\ee
which are extensions of the complex structure and complexified K\"ahler moduli spaces of the Calabi-Yau, respectively.
One of the factors in \eqref{3dmodspace} corresponds to the HM moduli space $\MH$ of the four-dimensional theory
and the other one comes from the vector and gravitational multiplets in the way just described.
The T-duality \eqref{Tdual3d} simply exchanges the two factors.

This shows that $\MH$ in type IIA (resp. type IIB) theory can be obtained by compactifying the
vector multiplet sector in type IIB (resp. type IIA) compactified on the same Calabi-Yau.
At the classical level one ignores the instanton contributions related to winding modes along the
compactification circle (see below section \ref{sec_compgauge}) and the simple Kaluza-Klein reduction
is sufficient to get the metric on $\MH^{\rm A/B}$ from the metric on $\MV^{\rm B/A}$.
This explains why both of them are described by the same holomorphic prepotential.

To present the result of this c-map procedure, we restrict ourselves to the type IIA formulation.
To get the corresponding result in type IIB theory, it is enough to apply mirror symmetry.
However, we will not touch this second formulation till chapter \ref{chap_IIB}.
In the type IIA case, in the coordinates given in Table \ref{HMfields}, the metric reads
as follows \cite{Cecotti:1989qn,Ferrara:1989ik}
\be
\begin{split}
\de s_{\MH^{\rm tree}}^2=& \, \frac{1}{r^2}\,\de r^2
-\frac{1 }{2r}\, (\Im\cN)^{\Lambda\Sigma}\(\de\tzeta_\Lambda -\cN_{\Lambda\Lambda'}\de\zeta^{\Lambda'}\)
\(\de\tzeta_\Sigma -\bar\cN_{\Sigma\Sigma'}\de\zeta^{\Sigma'}\)
\\ &
+ \frac{1}{16 r^2} \(\de\sigma+\tzeta_\Lambda\de\zeta^\Lambda-\zeta^\Lambda\de\tzeta_\Lambda\)^2
+4\cK_{a\bar b}\de z^a \de \bz^{\bar b}  ,
\end{split}
\label{hypmettree}
\ee
where we introduced a convenient notation for the dilaton $r\equiv e^\phi\sim 1/g_{(4)}^2$, which encodes
also the effective string coupling in four dimensions.

\subsection{Quantum corrections}
\label{subsec_quantcor}

String theory produces two types of corrections to the results obtained by the Kaluza-Klein compactification:
$\alpha'$ and $g_s$-corrections which may be both perturbative and non-perturbative.
$\alpha'$-corrections appear together with the volume of the Calabi-Yau and therefore
they affect only the sector comprising the moduli $t^i$. This is the reason why in the vector multiplet sector
only the holomorphic prepotential of type IIA formulation gets these corrections.
Due to the c-map, in the hypermultiplet sector the situation in reversed: the moduli space metric
is $\alpha'$-exact in type IIA and receives corrections in type IIB.
Due to mirror symmetry, both cases are described by the same metric determined completely by
the prepotential and given in \eqref{hypmettree}.

However, this metric is valid only at tree level of string perturbation theory and further affected
by $g_s$-corrections. The string coupling arises from the dilaton. It is always a part
of the hypermultiplet sector and this is why $\MH$ gets $g_s$-corrections whereas $\MV$ is tree level exact.
Let us make an account of these corrections and their general properties:
\begin{itemize}
\item {\bf Perturbative corrections}

Perturbative $g_s$-corrections come from string loop diagrams.
It is clear that they should represent an expansion in powers of the dilaton, which is the counting parameter
of the string loop expansion.
Besides, they should preserve the continuous Heisenberg symmetry \eqref{Heisenb} and be consistent with the known
results from string loop computations \cite{Antoniadis:1997eg,Antoniadis:2003sw}.

\item {\bf Non-perturbative corrections}

Non-perturbative corrections are always associated with instanton effects.
In our case all relevant instantons have a geometric interpretation and
can be viewed as Euclidean branes wrapping non-trivial cycles of Calabi-Yau
so that they appear as point-like objects from the four-dimensional point of view.
In string theory there are two types of branes which give such contributions:
\begin{itemize}
\item {\bf D-brane instantons}

First, these are D$p$-branes of string theory with $p$ even in type IIA and $p$ odd in type IIB.
They have $p+1$ dimensional world-volumes and therefore are expected to wrap the cycles
of the corresponding dimension.
But since any Calabi-Yau does not have non-trivial 1- and 5-dimensional cycles,
on the type IIA side only D2-branes wrapping non-trivial 3-dimensional cycles of $\CY$
can give non-vanishing contributions to the hypermultiplet metric.
Such D2-branes are labeled by a charge vector $\gamma=(p^\Lambda,q_\Lambda)$
indicating the cycle wrapped by the brane, $q_\Lambda \cA^\Lambda+p^\Lambda \cB_\Lambda$.
Due to this, the branes with vanishing $p^\Lambda$ wrap only A-cycles and can be called A-D2-branes.
It is known that mirror symmetry maps them to D(-1) and D1-branes on the type IIB side.
Similarly, B-D2-branes having non-vanishing $p^\Lambda$ are mapped to D3 and D5 branes.

On general ground it is known that the D-instanton corrections have the following form \cite{Becker:1995kb}
\be
\label{d2quali}
\delta \de s^2\vert_{\text{D2}} \sim e^{ -2\pi|Z_\gamma|/g_s
- 2\pi\I (q_\Lambda \zeta^\Lambda-p^\Lambda\tzeta_\Lambda)} ,
\ee
where ($z^0\equiv 1$)
\be
\label{defZ}
Z_\gamma(z) = q_\Lambda z^\Lambda- p^\Lambda F_\Lambda(z)
\ee
is the central charge function determined by the charge vector. This function gives the ``strength" of the instanton,
whereas the imaginary terms are the so called axionic couplings which show that the RR-fields are
similar to the theta-angle in QCD.
In particular, these couplings are responsible for breaking the continuous Heisenberg symmetry \eqref{Heisenb}
to a discrete subgroup, except the last isometry along $\sigma$.
A similar picture arises on the type IIB side where the role of the $\theta$-angles
is played by the RR-fields $c^0,c^i,c_i$ and $c_0$ and the charge vector $\gamma$ should be replaced
by a charge characterizing a D5-D3-D1-D(-1) bound state.

\item {\bf NS5-brane instantons}

The second type of branes is provided by the so called NS5-branes, magnetic dual to fundamental strings,
which have 6-dimensional world-volume and therefore can wrap the whole Calabi-Yau.
Again it is known on general ground that NS5-brane contributions behave as \cite{Becker:1995kb}
\be
\delta \de s^2\vert_{\text{NS5}}  \sim
e^{-2 \pi |k| \cV /g_s^2-\I\pi  k \sigma},
\label{couplNS5}
\ee
where $\cV$ is the Calabi-Yau volume. Thus, they are exponentially suppressed
comparing to D-instantons and characterized by a coupling to the NS-axion $\sigma$.
This last fact leads to breaking the last remaining continuous symmetry from the Heisenberg group
generated by $\kappa$-shifts in \eqref{Heisenb}.
As a result, the full non-perturbative HM moduli space does not have any continuous isometries,
but only a set of discrete ones which have been presented in section \ref{subsec_sym}.
\end{itemize}
\end{itemize}

\subsection{Where we are}

Until a few years ago, the tree level metric \eqref{hypmettree} and the general properties
of quantum corrections reviewed above exhausted all our knowledge about the HM moduli space.
But last years marked a remarkable progress in this direction.
It was initiated by the work \cite{Rocek:2005ij}, further elaborated in \cite{Neitzke:2007ke}, where the c-map
was formulated in terms of a projective superspace \cite{Karlhede:1984vr,Hitchin:1986ea,Lindstrom:1987ks}.
This description of QK geometry allows
to encode its metric in a holomorphic function, projective superspace Lagrangian, and is
closely related to the twistor space approach which we develop below.
But it works only in the case when the QK space has a sufficient number of commuting isometries
and therefore has a restricted area of applicability.
Nevertheless, this established a framework for studying quantum corrections and
the first new result obtained in this way
was the 1-loop contribution to the superspace Lagrangian \cite{Robles-Llana:2006ez}.
Then the one-loop corrected metric has been explicitly calculated in \cite{Alexandrov:2007ec}
performing a superconformal quotient \cite{deWit:2001dj}.
Moreover, it is believed now that this is the only perturbative correction
and all higher loop contributions can be removed by a field redefinition.

\begin{figure}[t!]
\setlength{\unitlength}{1cm}
\begin{center}
\begin{picture}(16,9.7)
\thicklines \put(4,9.){\makebox(2,0.6)[c]{Hypermultiplet sector $\cM_{\rm HM}$}}
\put(12.5,9.){\makebox(2,0.6)[c]{Vector multiplet sector $\cM_{\rm VM}$}}
\put(0.6,8.4){\makebox(2,0.6)[c]{IIA / $\CY$}}
\put(7,8.4){\makebox(2,0.6)[c]{IIB / $\CYm$}}
\put(10.75,8.4){\makebox(2,0.6)[c]{IIA / $\CYm$}}
\put(14.25,8.4){\makebox(2,0.6)[c]{IIB / $\CY$}}
\put(11,1.9){\makebox(2,0.6)[c]{\{$\alpha^\prime$\}}}
\put(14.25,1.9){\makebox(2,0.6)[c]{\{$-$\}}}
\put(14.4,2.2){\vector(-1,0){1.7}}
\put(13.125,2.2){\makebox(1,0.6)[c]{mirror}}
\put(11.3,2.2){\vector(-1,0){3.3}}
\put(9,2.2){\makebox(2,0.6)[c]{c-map}}
\put(6.3,1.9){\makebox(2,0.6)[l]{\{$\alpha^\prime$, $1\ell$\}}}
\put(7,2.6){\vector(0,1){1}}
\put(7,2.7){\makebox(2,0.6)[c]{$SL(2,\IZ)$}}
\put(0.7,3.6){\makebox(2,0.6)[l]{\{$1\ell$, A-D2\}}}
\put(5.6,3.6){\makebox(2,0.6)[l]{\{$\alpha^\prime$, D(-1), D1\}}}
 \put(1.55,4.3){\vector(0,1){1.1}}
\put(1.8,4.5){\makebox(2,0.6)[l]{e/m duality}}
\put(0,5.4){\makebox(2,0.6)[l]{ \{$1\ell$, A-D2, B-D2\} }}
\put(5.4,5.4){\makebox(2,0.6)[l]{\{$\alpha^\prime$, D(-1), D1, D3, D5\} }}
 \put(7,6.1){\vector(0,1){1.1}}
\put(7,6.3){\makebox(2,0.6)[c]{$SL(2,\IZ)$}}
\put(5.3,7.2){\makebox(2,0.6)[l]{\{$\alpha^\prime$, D(-1), D1, D3, D5, NS5\} }}
\put(0.25,7.2){\makebox(2,0.6)[l]{ \{$1\ell$, D2, NS5\} }}
\put(5.1,3.9){\vector(-1,0){1.7}}
\put(3.75,3.9){\makebox(1,0.6)[c]{mirror}}
\put(3.4,5.7){\vector(1,0){1.7}}
\put(3.75,5.7){\makebox(1,0.6)[c]{mirror}}
\put(5.1,7.5){\vector(-1,0){1.7}}
\put(3.75,7.5){\makebox(1,0.6)[c]{mirror}}
\put(0,8.3){\line(1,0){16}}
\put(10.6,2.9){\line(0,1){6.7}}
\end{picture}
\end{center}
\vspace{-2.4cm}
\parbox[c]{\textwidth}{\caption{\rm Prospective duality chain
for determining the quantum corrected low energy effective action of type II
strings compactified on a generic Calabi-Yau $\CY$ and its mirror partner $\CYm$ \cite{RoblesLlana:2007ae}.
\label{fig_scheme}}}
\end{figure}
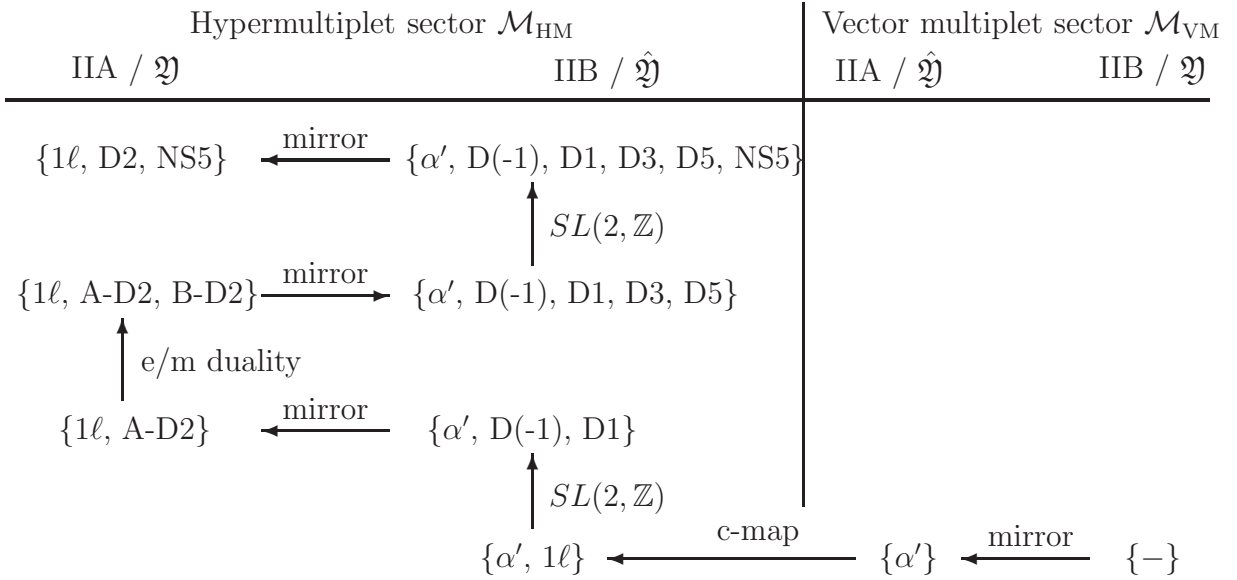

Furthermore, there is a procedure, suggested in \cite{RoblesLlana:2007ae} and shown in Fig. \ref{fig_scheme},
which allows, at least in principle, to restore all missing non-perturbative contributions by
using $SL(2,\IZ)$-duality, symplectic invariance and mirror symmetry.
In this duality chain, one can start even from the vector multiplet moduli space which in the type IIB
formulation does not receive {\it any} corrections. As we already mentioned several times,
the $\alpha'$-corrected metric on $\MV$ on the type IIA side can be obtained via mirror symmetry.
Through the c-map it gives rise to the tree level HM metric in type IIB, which should be
amended by the one-loop correction discussed just above. This correction together with $\alpha'$-contributions
break the $SL(2,\IR)$ invariance of the classical metric.
Enforcing its discrete version, $SL(2,\IZ)$, one obtains contributions due to D1-D(-1)-instantons
which are mirror symmetric to A-D2-instantons on the type IIA side.
All D2-brane contributions can be restored by symplectic transformations and
are mapped by repeated action of mirror symmetry to D3- and D5-brane instantons on the type IIB side.
Another application of S-duality then gives rise to NS5-brane and D5--NS5 bound state instantons.
Finally, applying mirror symmetry one last time produces the NS5-brane corrections in the type IIA formulation.

Where are we now on this route to the full non-perturbative description of the HM moduli space?
As described below, now we have control over all D-instanton corrections in the type IIA formulation.
Moreover, on the type IIB side we have a complete description of D(-1)-D1-instantons and D3-D5 and NS5-brane
contributions in the one-instanton approximation.
Although these results have been obtained by following the path presented in Fig.
\ref{fig_scheme}\footnote{The first step up towards non-perturbative contributions has been
done in \cite{RoblesLlana:2006is}, mirror symmetry has been applied in \cite{RoblesLlana:2007ae},
then all D-brane instantons have been included in \cite{Alexandrov:2008gh,Alexandrov:2009zh},
and one-instanton contributions due to NS5-branes have been found in \cite{Alexandrov:2010ca}.},
we will not repeat it in detail.
Instead, in chapter \ref{chap_Dbrane} we will jump immediately to the description of D2-instantons
on the type IIA side and present the final construction. Before that however we will have to develop
necessary mathematical tools (chapter \ref{chap_quatern}) and to reformulate the perturbative metric
using this new framework (chapter \ref{chap_pert}).

\section{Compactified gauge theory}
\label{sec_compgauge}

Let us recall that T-duality in three dimensions \eqref{Tdual3d} relates type IIA and type IIB theories,
compactified on the same Calabi-Yau and on circles of opposite radii, exchanging the moduli spaces coming
from HM and VM sectors of the four-dimensional effective theory.
Since this duality is expected to hold at the non-perturbative level,
the HM moduli space we considered in the previous section
can be realized also via circle compactifications of the vector multiplet sector
\be
\MH^{\rm B/A}(\CY)= \MH^{\rm B/A}(\CY\times S^1_{1/R})\simeq\MV^{\rm A/B}(\CY\times S^1_R).
\label{symmoduli}
\ee
The quantum corrections to $\MH$ are then mapped to the loop corrections from the Kaluza-Klein states
and to the non-perturbative corrections due to four-dimensional BPS black holes
with NUT charge $k$ whose world-line winds around the compactification circle.

Although this reinterpretation does not help in finding the non-perturbative description of $\MH$,
it provides a simplified version of our story which can be obtained in the limit where
gravity is decoupled and one remains with a $N=2$ supersymmetric gauge theory.
In four dimensions, in the low energy limit such a theory is described by a set of vector multiplets
which comprise $U(1)$ gauge fields and complex scalars.
Here we are concentrated on the Coulomb branch of the moduli space where the scalars acquire
vacuum expectation values $z^I$ and the initial (non-abelian) gauge group is broken to $U(1)^r$
with $r$ being the rank of the gauge group and giving the number of the vector multiplets.

The complex scalars span a moduli space $\Mgt$ which carries a special K\"ahler metric of {\it rigid} type
determining the low energy action. Such metric, as in the usual special K\"ahler case, is defined
by a holomorphic prepotential. But instead of \eqref{Kahlerpot_spec}, its K\"ahler potential is given by
\be
K=-2\Im\(\bz^I F_I(z)\).
\label{Kpot-gauge}
\ee
Then the bosonic part of the effective action reads as \cite{Seiberg:1988ur}
\be
S_{4d}=\frac{1}{4\pi}\int\Bigl[-\Im F_{IJ}\( \de z^I \wedge \star \de z^J + \cF^I\wedge \star\cF^J\)
+\Re F_{IJ}\,\cF^I\wedge \cF^J\Bigr].
\label{act-gauge}
\ee
Thus, as in the case of vector multiplets in $N=2$ supergravity, the low energy theory is
completely determined by the prepotential, which in this case is provided
by the Seiberg-Witten solution \cite{Seiberg:1994rs}.

However, we are interested really in a compactification of such theory on a circle of radius $R$ \cite{Seiberg:1996nz}.
Upon compactification, the vector multiplets give rise to the same number of hypermultiplets and
the low energy effective action is given by a non-linear $\sigma$-model with a \hk target space $\Mthree$,
which is another type of quaternionic geometries to be described in the next chapter.
This is a gauge theory analogue of the statement about the HM moduli space in string theory.

The space $\Mthree$ is parameterized by complex scalars $z^I$
and by Wilson lines of the gauge potential around the circle, which have ``electric" $\zeta^I$
and ``magnetic" components $\tzeta_I$ and are all periodic.
The perturbative metric on $\Mthree$ follows from the simple 3d truncation of the
4d vector multiplet Lagrangian \eqref{act-gauge} and thus is completely defined by the holomorphic prepotential $F(z)$.
This construction is known as {\it rigid c-map} \cite{Cecotti:1989qn}.
However, this metric gets instanton contributions from the massive spectrum due to
BPS particles going around the compactification circle.
For large compactification radius $R$ these contributions are exponentially suppressed
since for a particle of charge $\gamma=(q_I,p^I)$ they are weighted
by $e^{-2\pi R |Z_\gamma|}$ where $Z_\gamma$ is the same central charge function
as in \eqref{defZ} giving the mass of the BPS particle.

These instanton corrections look exactly the same as D-instanton corrections to $\MH$.
Therefore, one may expect that the D-instanton corrected HM moduli space appearing
in Calabi-Yau compactifications of type II string theory and the exact moduli space of
a 4d $N=2$ gauge theory compactified on a circle have similar descriptions.
We will see below that these expectations turn out to be correct and indeed, without
NS5-brane instantons, the constructions of $\MH$ and $\Mthree$ in the framework of the twistor approach
look almost identical. The NS5-brane corrections however are a distinguishing feature of
string theory, which does not have an analogue in supersymmetric gauge theories.
In a sense their presence marks the difference between QK and HK spaces and why the former
are more complicated than the latter.

\chapter{Twistor approach to quaternionic geometries}
\label{chap_quatern}

In the previous chapter we encountered two types of quaternionic geometries: \hk
and \qk spaces. The former are relevant for supersymmetric gauge theories and the latter
appear in Calabi-Yau compactifications of type II string theory.
The typical problem which one needs to solve in this context is to find a quantum corrected metric
on the quaternionic space in question.
On the other hand, it is hopeless to try to find quantum corrections in a gauge or string theory
directly to the components of the metric.
Therefore, we need to understand how such spaces can be conveniently
parametrized and in the best case to find an analogue of the holomorphic prepotential of
the special K\"ahler geometry, which would incorporate quantum corrections in an easy way.
This problem has been addressed in \cite{Alexandrov:2008ds,Alexandrov:2008nk} and here we present
the resulting construction.

\section{HK spaces}
\label{sec_HK}

\subsection{Twistorial construction}

A \hk manifold $\hkm$ of dimension $4n$ is defined as a Riemannian manifold whose holonomy group is contained in $Sp(n)$.
$\hkm$ is a particular case of K\"ahler manifold and moreover it is Ricci-flat.
In fact, in the $n=1$ case the Ricci flatness, supplemented by the self-duality of
the Weyl curvature, is the necessary and sufficient condition for a four-dimensional manifold to be \hk.
For arbitrary $n$, $\hkm$ admits a two-parameter family of integrable
complex structures with respect to which the metric is K\"ahler
\be
J(\varpi,\bar\varpi)=\frac{1-\varpi\bar\varpi}{1+\varpi\bar\varpi}\, J^3
+\frac{\varpi+\bar\varpi}{1+\varpi\bar\varpi}\, J^2
+ \I \frac{\bar\varpi-\varpi}{1+\varpi\bar\varpi}\, J^1 ,
\qquad
\varpi \in \IC \cup \infty=\CP,
\label{complstrHK}
\ee
where $J^i,\ i=1,2,3$, are three distinct complex structures which satisfy the algebra of quaternions
\be
J^i\, J^j = \eps^{ijk} J^k - \delta^{ij}
\label{Qalg}
\ee
and thus define the quaternionic structure of the HK manifold.
If $\omega^i$ are K\"ahler forms for the complex structures $J^i$ and $\omega^\pm=-\hf(\omega^1\mp \I\omega^2)$,
the K\"ahler form for the complex structure \eqref{complstrHK} can be written as
\be
\omega(\varpi,\bar\varpi)=\frac{1}{1+\varpi\bar\varpi}
\((1-\varpi\bar\varpi)\omega^3-2\I\bar\varpi \omega^+ +2\I\varpi\omega^-\).
\label{KformHK}
\ee

It is natural to consider the direct product $\hkt=\hkm\times \CP$ where the second factor
is associated to the sphere of complex structures. It is known as the {\it twistor space}
of $\hkm$. It is a K\"ahler manifold with the K\"ahler form \eqref{KformHK}
degenerate along the $\CP$ direction $\de\varpi$
and with the complex structure given by $J(\varpi,\bar\varpi)$ on the base $\hkm$ and by the standard
$\CP$ complex structure on the fiber.

It turns out that this twistor space carries two additional structures
which allow to encode the geometry of $\hkm$ in a very convenient way.
The first is given by a holomorphic 2-form $\Omega$
which defines a holomorphic symplectic structure
on each fiber of the holomorphic bundle $\pi:\ \hkt\to\CP$.
At finite $\varpi$ it can be expressed through $\omega^i$ as follows
\be
\Omega(\varpi)=\omega^+-\I\varpi \omega^3+\varpi^2\omega^- .
\label{holsymformHK}
\ee
Since $\omega^+$ is holomorphic in the complex structure $J^3$, the first two coefficients
of the expansion of $\Omega$ around $\varpi=0$ allow to read off both, the holomorphic coordinates
and the K\"ahler (1,1)-form. Together they are sufficient to reproduce the metric on $\hkm$.

The second additional structure provides a real structure on $\hkt$.
It is defined as a combined action of the complex conjugation and the antipodal map on $\CP$,
$\tau[\varpi]=-1/\bar\varpi$, and it must be compatible with all other structures of the twistor space.
In particular, it should preserve the holomorphic 2-form $\Omega$ up to a $\varpi$-dependent
holomorphic factor. It is easy to see that the expression \eqref{holsymformHK} satisfies this requirement.

Taken altogether, the holomorphic symplectic structure on the fibers and the real structure turn out
to be sufficient to ensure that $\hkt$ is the twistor space of a \hk manifold. Thus,
the problem of parametrization of HK manifolds can be reformulated as a similar problem for their twistor spaces
where it reduces to a characterization of consistent pairs of symplectic and real structures \cite{Hitchin:1986ea}.

\subsection{Holomorphic Darboux coordinates and transition functions}

Let us first concentrate on the holomorphic symplectic structure represented by
the holomorphic 2-form $\Omega$. More precisely, $\Omega$ appears to be a section
of the $\cO(2)$-twisted holomorphic vector bundle $\Lambda^2 T^*_F(2)$ where $T_F={\rm Ker}(\de \pi)$ is
the tangent bundle along the fibers of $\pi$.
A trivial extension of the Darboux theorem ensures that locally
it is always possible to choose holomorphic coordinates $\nui{i}^I,\mui{i}_I$
plus the coordinate on the fiber $\varpi$ where this section is given by
\be
\Omega^{[i]} =\de\nui{i}^I\wedge \de \mui{i}_I ,
\label{con1fo}
\ee
with $I=0,1,\dots,n-1$ and $4n=\dim_{\IR}\hkm$.
It is important however that this can be done only {\it locally}, which is indicated by the index $^{[i]}$
enumerating different open patches $\hU_i$ where \eqref{con1fo} is valid.
Extending \eqref{con1fo} out of the patch, the coordinates will eventually develop a singularity
and thus cannot be globally defined.
Therefore, in general, one has to cover the twistor space by such
patches and in each patch to choose the appropriate Darboux coordinates.

To ensure consistency with the real structure on $\hkt$,
it will be convenient to appropriately adapt the covering.
To this end, we assume that the antipodal map preserves the covering and the image of $\hU_i$
is another patch $\hU_{\bi}$. Then the compatibility of the symplectic and real
structures can be formulated as the following reality constraint
\be
\overline{\tau(\Omega^{[i]})}=\Omega^{[\bi]}
\label{relsympl}
\ee
so that the Darboux coordinates may be chosen to satisfy similar relations, all with sign plus, under
the combined action of the complex conjugation and the antipodal map.

Of course, on the overlap of two patches $\hU_i\cap\hU_j$ the two coordinate systems should be related
by a transformation, which is not however arbitrary but preserves
the holomorphic 2-form up to a $\varpi$-dependent factor
\be
\label{omij}
\Omega^{[i]}= f_{ij} ^2 \, \Omega^{[j]}  \  \mod\, \de\varpi.
\ee
Here $f_{ij}$ are transition functions of the $\cO(1)$ bundle on $\CP$ which are described
in detail in \cite{Alexandrov:2008ds}.
In particular, they satisfy
\be
f_{ij} f_{jk} = f_{ik} ,
\quad
f_{ii}=1 ,
\quad
\overline{\tau(f_{ij}^2)} = f_{\bi \bj}^2 ,
\ee
and the transition function between the two poles of $\CP$ explicitly reads as $f_{0\infty}=\varpi$.
Their appearance in \eqref{omij} is related to the fact that $\Omega$
is an $\cO(2)$-twisted section.

The condition \eqref{omij} implies that the local Darboux coordinates are related by ($\cO(2)$-twisted)
symplectic transformations. Such transformations are generated by holomorphic functions on $\hkt$,
which we call transition functions. To write them explicitly,
it is convenient to change variables and to work with $\etai{i}^I\equiv (\I\varpi)^{-1} f_{0i}^2\nui{i}^I$
instead of $\nui{i}^I$, where the index $\scriptstyle 0$ refers to some fixed patch $\hU_0$
which we choose to be the one around the north pole $\varpi=0$.
This allows to avoid the factor $f_{ij}$ in the most equations.
Then, taking transition functions to be functions of the initial ``position" and the final
``momentum" coordinates, $\Hij{ij}(\etai{i},\mui{j},\varpi)$, the gluing conditions look as follows
\be
\etai{j}^I =  \etai{i}^I -\p_{\mui{j}_I }\Hij{ij},
\qquad
\mui{j}_I =  \mui{i}_I
 + \p_{\etai{i}^I } \Hij{ij}.
\label{HKgluing}
\ee

It should be clear that the transition functions are not completely arbitrary, but must satisfy
various compatibility conditions. Their exact form can be found in \cite{Alexandrov:2008ds}, whereas
here we just briefly mention what they require:
\begin{itemize}
\item
$\Hij{ji}$ must be a generating function of the inverse of the symplectomorphism generated by $\Hij{ij}$,
which implies they are related by a Legendre transform;
\item
symplectomorphisms between different couples of patches should compose properly so that on the overlap
of three patches $\hU_i\cap\hU_j\cap\hU_k$, the transformation generated by $\Hij{ij}$ must
be the same as a composition of the transformations generated by $\Hij{ik}$ and $\Hij{kj}$;
\item
to be consistent with the reality conditions on Darboux coordinates in all patches,
the transition functions must satisfy
\be
\overline{\tau\(\Hij{ij}\)}=\Hij{\bi\bj}.
\label{realHij}
\ee
\end{itemize}
Besides, it is important to remember that the Darboux coordinates, and hence the transition functions,
are not uniquely defined. In each patch $\hU_i$, it is still possible to perform a local symplectomorphism
generated by a {\it regular} holomorphic function $\Gi{i}(\etai{i},\mui{i},\varpi)$. Such a gauge transformation
will affect all transition functions related to this patch by a regular contribution.

The transition functions are the main object of our interest because
it is these functions that play the role of the prepotentials we were looking for.
In particular, they will be used to encode quantum corrections to the moduli space metrics.
These transition functions contain all geometric information about the initial HK and its twistor spaces.
In other words, if one gives a covering of the twistor space and the associated system of
transition functions, then it is possible (at least in principle) to extract the metric from this information.

To achieve this goal, the first step is to find the Darboux coordinates
as functions of the fiber coordinate $\varpi$ and coordinates $q^\alpha$ on the base.
Such solutions, called twistor lines, can be found from the gluing conditions \eqref{HKgluing}.
Then substituting them into \eqref{con1fo} for $\Omega^{[0]}$ and expanding in $\varpi$,
the constraints of HK geometry ensure that the expansion takes the form \eqref{holsymformHK}.
From the first two coefficients one extracts the complex structure and the K\"ahler form so that
the metric can be easily found.

Of course, the most difficult problem on this way is to solve the gluing conditions.
Their discrete form is not very convenient for this purpose. Instead
they can be rewritten in an integral form \cite{Alexandrov:2009zh}
\be
\begin{split}
\etai{i}^I(\varpi,q^\alpha)& = x^I +
\varpi^{-1} u^I - \varpi \bu^I
-\frac12 \sum_j \oint_{C_j}\frac{\de\varpi'}{2\pi\I \varpi'}\,
\frac{\varpi'+\varpi}{\varpi'-\varpi}\, \p_{\mui{j}_I }\Hij{ij}(\varpi') ,
\\
\mui{i}_I(\varpi,q^\alpha)& = \vrh_I +
\half  \sum_j \oint_{C_j} \frac{\de \varpi'}{2 \pi \I \varpi'} \,
\frac{\varpi' + \varpi}{\varpi' - \varpi}
\, \p_{\etai{i}^I } \Hij{ij}(\varpi'),
\end{split}
\label{txiqline}
\ee
where $\varpi\in \cU_i$, which is the projection of $\hU_j$ on $\CP$, $C_j$ is a contour surrounding $\cU_j$
in the counterclockwise direction, whereas complex $u^I$ and real $x^I, \vrh_I$ are free parameters playing the role
of coordinates on the base HK manifold $\hkm$.
The sum in \eqref{txiqline} goes over all patches including those
which do not intersect with $\cU_i$. In that case the transition functions are defined by the
cocycle condition and by analytic continuation.

In terms of solutions of these equations, it is possible to give an explicit formula for
the K\"ahler potential in the complex structure $J^3$,
which can be expressed as an integral of transition functions \cite{Lindstrom:2008gs,Alexandrov:2009zh}
\be
\label{Kdefrew}
K_\hkm= \frac{1}{4\pi}\sum_j\oint_{C_j} \frac{\de\varpi}{\varpi}
\[\Hij{0j}-\etai{0}^I \p_{\etai{0}^I}\Hij{0j}
+\(\varpi^{-1} u^{I}-\varpi \bu^{I} \)\p_{\etai{0}^I} \Hij{0j} \].
\ee
In this complex structure the holomorphic coordinates on $\hkm$ are given by the coefficients of the leading
terms in the small $\varpi$ expansion of the Darboux coordinates $\etai{0}^I$ and $\mui{0}_I$
and therefore coincide with
\be
u^I
\qquad {\rm and} \qquad
w_I\equiv \frac{\I}{2}\,\mui{0}_\Lambda|_{\varpi=0}=\frac{\I}{2}\,\vrh_I +
\frac{1}{8\pi}  \sum_j \oint_{C_j} \frac{\de \varpi}{ \varpi}
\, \p_{\etai{0}^I } \Hij{0j}.
\label{defholcoor}
\ee

\subsection{$\cO(2)$ HK manifolds and their deformations}

The construction presented above becomes particularly simple and explicit
in the case of $\cO(2)$ HK manifolds where it coincides with the projective superspace approach.
In this case the twistor space $\hkt$ is supposed to have $n$ global $\cO(2)$ sections
which may be identified with the Darboux coordinates $\nu^I$.
The fact that they are globally well defined implies that the corresponding sections $\eta^I$
are the same in all patches. This is possible only if the transition functions $\Hij{ij}$
do not depend on the conjugate coordinates $\mui{j}_I$ what ensures that the HK space
has $n$ commuting isometries.
Taking into account this independence in \eqref{txiqline},
one immediately finds that\footnote{In \cite{Alexandrov:2008ds} also the case of $n$ $\cO(2k)$ global sections
has been described in this formalism. The corresponding transition functions are again $\mu_I$-independent
and the expansion of the Darboux coordinates $\eta^I$ comprise then $2k+1$ terms generalizing \eqref{etaO2}.
However, for $k>1$ the integral representation \eqref{txiqline} is not the convenient starting point
and should be appropriately modified. Here we restrict to the case $k=1$ due to its relevance for the
physical applications.}
\be
\eta^I(\varpi,q^\alpha) = x^I +\varpi^{-1} u^I - \varpi \bu^I.
\label{etaO2}
\ee
At the same time, the r.h.s. of the equation determining $\mui{i}_I$ becomes independent of them and thus
represents an explicit formula for these coordinates, and not an equation to be solved.

The independence of the transition functions on a half of coordinates simplifies many equations.
For example, the consistency conditions spelled above eq. \eqref{realHij} and the gauge transformations
generated by regular functions $\Gi{i}$ are written as
\be
\Hij{ji}=-\Hij{ij}, \qquad \Hij{ik}+\Hij{kj}=\Hij{ij},
\qquad
\Hij{ij}\mapsto \Hij{ij}+\Gi{i}-\Gi{j}.
\label{consist_condH}
\ee
Besides, the formula for the K\"ahler potential \eqref{Kdefrew} can be interpreted as a Legendre transform
of the following Lagrangian \cite{Karlhede:1984vr,Hitchin:1986ea}
\be
\label{tlag1}
\mathcal{L} (u^I, \bu^I,x^I)=\frac{1}{4\pi}\sum_j
\oint_{C_j} \frac{\de\varpi}{\varpi}\, \Hij{0j}(\eta^I,\varpi)
\ee
with respect to the coordinates $x^I$
\be
K_\hkm(u, \bu,w, \bw) = \langle \cL(u^I,\bu^I,x^I) - x^I  (w_I + \bw_I) \rangle_{x^I}  .
\label{K02}
\ee
This formula clearly shows that the K\"ahler potential and the metric do not depend on
the coordinates $\I\vrh_I=w_I-\bw_I$ so that $\p_{\vrh_I}$ provide the set of commuting Killing vectors on $\hkm$.

Assume now that our transition functions can be represented as
\be
\Hij{ij}(\etai{i},\mui{j},\varpi)=\Hnpij{ij}(\etai{i},\varpi)+\Hpij{ij}(\etai{i},\mui{j},\varpi) ,
\label{genf}
\ee
where $\Hpij{ij}$ is considered as a perturbation around $\Hnpij{ij}$.
Then one can try to find the metric on $\hkm$ by a perturbative expansion in powers of $\Hpij{ij}$.
The equations \eqref{txiqline} determining the holomorphic Darboux coordinates
are particularly suited for such perturbative analysis. Indeed, as was noted above, in the leading approximation
the coordinates $\etai{i}^I$ are globally well defined and given in \eqref{etaO2}, whereas $\mui{i}_I$
are obtained by integrating $\p_{\eta^I}\Hnpij{ij}(\eta,\varpi)$.
Then at the $n$th order of perturbative expansion, one first finds $\etai{i}^I$ by evaluating $\Hpij{ij}$
on the Darboux coordinates found at the previous step and gets $\mui{i}_I$ afterwards
by substituting $\etai{i}^I$ at $n$th order and $\mui{j}_I$ at $(n-1)$th order in the r.h.s. of
\eqref{txiqline}.

The linear approximation has been considered in detail in \cite{Alexandrov:2008ds}. In particular,
it has been shown that the K\"ahler potential is given by the same Legendre transform as in \eqref{K02}
where the Lagrangian is defined by the full perturbed transition functions. As a result, it becomes $\vrh_I$-dependent
and there are no isometries anymore.
The linear deformation of the K\"ahler potential is provided by the {\it Penrose transform} of the
perturbing transition functions evaluated on the unperturbed Darboux coordinates
\be
\label{kahlerdef}
K_{(1)}(u,\bu,w,\bw)=\frac{1}{4\pi}
\sum_j \oint_{C_j} \frac{\de\varpi}{\varpi}\,
\Hpij{0j}( \eta, \mui{j},\varpi) .
\ee

\subsection{Branch cuts and open contours}
\label{subsec_cuts}

So far we implicitly assumed that the singularities, developed by the Darboux
coordinates and canceled by transition functions going from one patch to another, are of the pole type.
The presence of branch cuts introduces additional complications.
If a branch cut does not belong to one patch, but goes through several patches,
in the formulas like \eqref{txiqline} one cannot use the usual contours $C_i$
because they are not closed anymore being split by the cut.
Fortunately, in \cite{Alexandrov:2008ds} it has been shown that the mutual compatibility
of transition functions ensures that one can always reconnect the contours in such a way
they the standard integral formulas as \eqref{txiqline} are still valid.
The new feature which arises here is that the transition functions
having branch cuts may be integrated along more complicated contours
than just the ones surrounding the patches.
What is important however is that these contours appear to be always {\it closed}.

\lfig{Logarithmic branch cuts and integration contours. The solid black line represents the
branch cut of $\mui{0}(\varpi)$, and the semi-dotted red lines are the branch cuts of $\Hij{0+}$.
On the left, the dotted line is an initial contour used to derive the integral representation \eqref{txiqline}.
On the right, the dotted line is the final figure-eight contour along which one integrates $\log \eta$.
The part of the contour between points 2 and 4 lies on a different Riemann sheet of
the logarithm. }{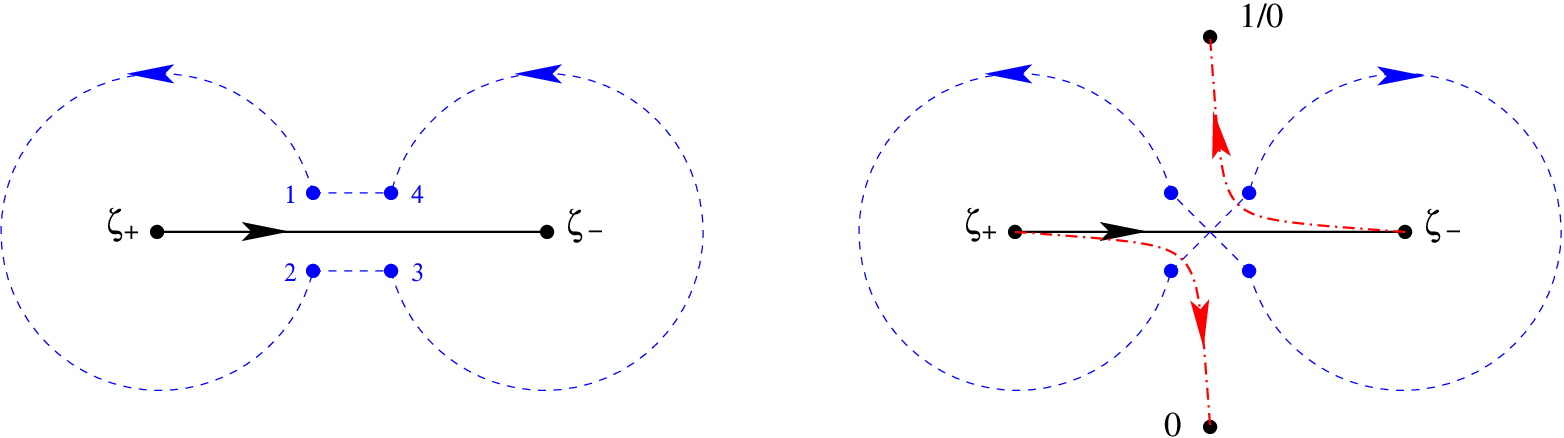}{13.5cm}{fig-eight}{-0.0cm}

A typical example is provided by the figure-eight contour encircling two
logarithmic singularities.
It appears, for example, in the situation when there are two transition functions
having a similar logarithmic term
\be
\Hij{0\pm}=\pm c\log\eta,
\ee
where $\cU_\pm$ are the patches surrounding the roots $\varpi=\zeta_\pm$ of $\eta(\varpi)=0$.
In this case a combination of two (non-closed) contours $C_+$ and $C_-$
can be reconnected to the figure-eight contour as shown on Fig. \ref{fig-eight}.

A very important remark is that the transition functions with branch cuts allow an effective
description in terms of {\it open contours}.
Namely, it is clear that the contribution of an integral of a singular transition function around
its branch cut, up to possible residues at the end points,
can be equally obtained by integrating its discontinuity along the cut. The same is true
for the figure-eight contours with the difference that the integral goes now along a contour
joining two branch singularities and representing a cut of a Darboux coordinate.
(In the above example, this is the line joining $\zeta_\pm$. Note that the cuts of Darboux coordinates
and transition functions do not necessarily coincide.)
Therefore, instead of working with singular transition functions and the corresponding complicated contours,
one can describe their effect on Darboux coordinates by their discontinuities associated to open contours.

Geometrically this result is obtained by shrinking the patches surrounding the branch points
($\cU_\pm$ in the above example) along a line, say $\ell$, joining them.
This allows to extend the Darboux coordinates from the bounding patches till $\ell$, whereas
the shrunken patches reduce just to small discs around the end points of $\ell$ and may be even
completely removed if the end points do not produce any residue contributions.
However, the analytical continuations from the left and right of $\ell$ do not match,
but differ by the discontinuity of the initial singular transition function across the branch cut
(more precisely, its appropriate derivative). The same effect is ensured by adding to
the integral representation \eqref{txiqline} a line integral along $\ell$ of the above discontinuity.

Thus, instead of assigning a set of open patches and transition functions,
a HK manifold can be characterized by providing a set of (closed or open) contours on $\CP$
and a set of associated holomorphic functions $\Hij{ij}$. We will see that most of the instanton corrections
to the gauge and string moduli spaces are introduced precisely in this way being associated
with open contours.

\section{QK spaces}
\label{sec_QK}

\subsection{HKC construction and twistor space}

Now we pass to the more complicated case of QK spaces appearing in supergravity \cite{MR664330}.
A QK manifold $\qkm$ is a $4n$-dimensional Riemannian manifold whose holonomy group is contained in $Sp(n)\times SU(2)$.
As in the HK case, for $n=1$ $\qkm$ can be described in a simpler way as a self-dual Einstein space
with a non-vanishing cosmological constant. Such spaces will be studied in detail in chapter \ref{chap-4d}
as they correspond to the important particular case of the universal hypermultiplet.

Similarly to HK spaces, a QK manifold carries a quaternionic structure represented by a triple of
almost complex structures $J^i$ satisfying the algebra \eqref{Qalg}.
These $J^i$ give rise to the quaternionic two-forms $\vec\omega_\qkm(X,Y)=g_\qkm(\vec J X,Y)$,
which are covariantly closed with respect to the $SU(2)$ part $\vec p$ of the Levi-Civita connection
and proportional to the curvature of $\vec p$,
\be \label{ompp}
\de\vec\omega_\qkm + \vec p \times
\vec\omega_\qkm = 0,
\qquad
\de\vec p+ \half\, \vec p \times \vec
p = \frac{\nu}{2}\, \vec \omega_\qkm ,
\ee
where the proportionality coefficient is related to the Ricci scalar curvature as $R=4n(n+2)\nu$.
In the limit of vanishing $\nu$ the \qk geometry becomes hyperk\"ahler. We are mainly interested in the case
of negative curvature.

A more complicated nature of the QK geometry compared to the HK one can be seen
from the fact that the former is not even K\"ahler (so that the name is somewhat misleading).
As a result, we do not have at our disposal such quantities like K\"ahler potential
which can be used to derive the metric in a simple way.
Fortunately, to each QK manifold
one can associate in a canonical way a HK space $\hkm_{\qkm}$ with some additional properties.
It is called hyperk\"ahler cone (HKC) or Swann bundle \cite{MR1096180}
and appears as a $\IC^2/\IZ_2$ bundle over $\qkm$.
In the physical parlance, when the HKC space is realized
as the target space of an $N=2$ supersymmetric $\sigma$-model \cite{deWit:2001dj},
the additional properties or, more precisely, restrictions can be interpreted
as conditions of superconformal symmetry.
This implies that there is a scaling symmetry and an SU(2) symmetry coming from the RR-sector
of the $\sigma$-model. This gives precisely 4 degrees of freedom to be factored out to go from
the HKC space down to the QK base.

Not only this construction allows to work with an easier type of geometry, but
the additional symmetries make it even more special. In particular, they lead to the existence of
the so called \hk potential $\chi$ such that the metric on $\hkm_{\qkm}$, in {\it real} coordinates,
satisfies
\be
g_{MN}=D_M\p_N\chi.
\ee
In any complex structure $\chi$ reduces to a K\"ahler potential. Thus, this is a very convenient quantity
to work with, encoding the geometry of the Swann bundle.

However, this construction does not solve yet the problem of parametrization of QK manifolds.
Nevertheless, it suggests a natural strategy:
since the parametrization problem of HK spaces is solved by considering their twistor spaces,
one should start from such twistor space and impose the conditions ensuring that this is the twistor space of
a \hk cone. After factoring out the auxiliary degrees of freedom,
this procedure leads to the twistor space $\qkt$ of the initial QK manifold, which in fact appears
as an intermediate step in the Swann construction. Namely, it is a $\CP$ bundle over $\qkm$, whereas
$\hkm_{\qkm}$ is a $\IC^\times$ bundle over $\qkt$.
Thus, one has the following chain of bundles
\be
\cZ_{\hkm_{\qkm}} \rightarrow \hkm_{\qkm} \rightarrow \qkt \rightarrow \qkm,
\ee
where at each step the real dimension is decreased by 2.

It turns out that in terms of transition functions on $\cZ_{\hkm_{\qkm}}$, the condition that
$\hkm_{\qkm}$ is a HKC requires that these functions do not depend explicitly on $\varpi$ and
are (quasi-)homo\-geneous\footnote{Here ``quasi" refers to the possibility
of a mild modification of the homogeneity condition due to logarithmic terms in the transition functions
of the kind considered in section \ref{subsec_cuts}.
The modification is characterized by a set of constant parameters
called ``anomalous dimensions". Although they affect some of the following equations,
in this subsection we assume that they all vanish (see \cite{Alexandrov:2008nk} for complete results).
However, we introduce them back in \eqref{txiqlineQK}
because they are responsible for physically important effects.\label{foot_quasi}}
of first degree in $\eta^I$ \cite{Alexandrov:2008nk}.
One can then impose these conditions in the formulas of the previous section
and extract the corresponding equations describing the initial QK manifold.
In particular, due to the homogeneity condition, $\eta^I$ appear now as projective coordinates.
Singling out one of them, say $\eta^\alpha$, one ends up with an odd number of inhomogeneous ones
given by $(I=(\Lambda,\alpha),\ \Lambda=0,\dots, n-1)$
\be
\xi^\Lambda=\frac{\eta^\Lambda}{\eta^\alpha},
\qquad
\txi_\Lambda=\mu_\Lambda,
\qquad
\alpha=\mu_\alpha.
\ee
These are holomorphic coordinates on the twistor space $\qkt$. In particular,
since none of them coincides with the coordinate on the $\CP$ fiber, this demonstrates
that, in contrast to the twistor space of a HK manifold, $\qkt$ is a non-trivial bundle
with fibers which are not holomorphic.

The most important for us is what is happening with the symplectic structure represented by
a set of holomorphic 2-forms $\Omega^{[i]}$. It turns out that it descends to
a holomorphic {\it contact structure} on $\qkt$. The latter is provided by a set of holomorphic one-forms
$\cX^{[i]}$, defined up to rescaling by a nowhere vanishing holomorphic smooth function, such
that $\cX^{[i]}\wedge (\de \cX^{[i]})^n$ is the non-vanishing holomorphic top form.
To obtain $\cX^{[i]}$ from $\Omega^{[i]}$, it is sufficient to write the holomorphic 2-form
as a differential of the Liouville form
\be
\Omega^{[i]}=\de\Theta^{[i]},
\qquad
\Theta^{[i]}=\nui{i}^I\de\mui{i}_I,
\ee
and divide the latter by $\nui{i}^\alpha$,
\be
\cX^{[i]}=(\nui{i}^\alpha)^{-1}\Theta^{[i]} =\de \ai{i} +\xii{i}^\Lambda \de\txii{i}_\Lambda.
\label{formcX}
\ee
This shows that $\xii{i}^\Lambda,\txii{i}_\Lambda$ and $\ai{i}$ are local Darboux coordinates for
the holomorphic contact one-form $\cX^{[i]}$, which can be thought as an odd-dimensional counterpart
of the symplectic structure.

The existence of the holomorphic contact structure is the characteristic feature
of the twistor space of a QK manifold \cite{MR664330,MR1327157}.
Moreover, it is known to be proportional to a certain canonical (1,0)-form $D\varpi$
defined by the $SU(2)$ connection $\vec p$
\be
\label{contact}
\hCX\ui{i} =   \frac{4}{\I\varpi}\,e^{\Phi\ui{i}} D\varpi,
\qquad
D\varpi=\de\varpi + p^+ -\I\varpi p^3  + \varpi^2 p^-  ,
\ee
where $\varpi$ again denotes the coordinate on the $\CP$ fiber of the twistor space.
The proportionality coefficient is given in terms of a function $\Phi\ui{i}\equiv\Phi\ui{i}(q^\alpha,\varpi)$,
holomorphic in $\varpi$, which we refer as ``contact potential".
Note that in general it is defined only locally what is reflected by the patch index.
Nevertheless, this is a very important quantity for several reasons.
First of all, its real part provides the \kahler potential on the twistor space $\qkt$
\be
\label{Knuflat}
K_{\qkt}\ui{i} = \log\frac{1+\varpi\bar \varpi}{|\varpi|}
+ \Re\Phi\ui{i}.
\ee
Secondly, it gives rise to a certain function on the QK base manifold
\be
\label{phipot}
\phi \equiv \Re\left[  {\Phi^{[+]}}(\varpi=0)   \right]  ,
\ee
where $\scriptstyle{[+]}$ denotes the patch around $\varpi=0$.
This new function turns out to be closely related to the \hk potential on the HKC
\be
\chi=\frac{e^\phi}{4\rf} ,
\label{chiphi}
\ee
where $\rf$ is a certain function invariant under the $SU(2)$ isometric action on $\hkm_\qkm$,
with weight one under dilations (see \cite{Alexandrov:2008nk} for more details).
Thus, $\phi$ is a natural potential to be associated with a QK space.
Moreover, in physical applications it can be identified with the dilaton field and
acquires nice transformation properties under action of symmetries.

Finally, note that the contact one-form and the Darboux coordinates are subject to the reality conditions
written as
\be
\overline{\tau(\hCX\ui{i})}= \hCX\ui{\bi}.
\label{relcontact}
\ee

\subsection{Transition functions}

As in the HK case, the geometry of the twistor space is encoded in
the gluing conditions between different coordinate systems given by $(\xii{i}^\Lambda,\txii{i}_\Lambda,\ai{i})$,
which provide the canonical representation of the contact one-form \eqref{formcX}.
However, now the gluing conditions
become a bit more complicated since they describe local contact transformations.
Such transformations are generated by holomorphic functions of $\xii{i}^\Lambda$ in one patch and
$\txii{j}_\Lambda,\ai{j}$ in another patch, which we denote by the same symbol $\Hij{ij}$ as in the previous section.
Similarly to \eqref{omij}, the contact one-form is rescaled by a holomorphic factor
\be
\CX\ui{i} =  \hat f_{ij}^{2} \, \CX\ui{j},
\label{glue2}
\ee
but this factor turns out to be completely determined by the transition functions
\be
\hat f_{ij}^2=1-\p_{\ai{j} }\Hij{ij}.
\label{eqhf}
\ee
The gluing conditions for the Darboux coordinates read as \cite{Alexandrov:2009zh}
\be
\begin{split}
\xii{j}^\Lambda &=  \xii{i}^\Lambda -\p_{\txii{j}_\Lambda }\Hij{ij}+\xii{j}^\Lambda \, \p_{\ai{j} }\Hij{ij},
\\
\txii{j}_\Lambda &=  \txii{i}_\Lambda + \p_{\xii{i}^\Lambda } \Hij{ij},
\\
\ai{j} &=  \ai{i}+\Hij{ij}- \xii{i}^\Lambda \p_{\xii{i}^\Lambda}\Hij{ij},
\end{split}
\label{QKgluing}
\ee
and can be rewritten in the following integral form
\beq
\xii{i}^\Lambda(\varpi,q^\alpha)& =& A^\Lambda +
\varpi^{-1} Y^\Lambda - \varpi \bY^\Lambda
-\frac12 \sum_j \oint_{C_j}\frac{\de\varpi'}{2\pi\I \varpi'}\,
\frac{\varpi'+\varpi}{\varpi'-\varpi}\( \p_{\txii{j}_\Lambda }\Hij{ij}
-\xii{j}^\Lambda \, \p_{\ai{j} }\Hij{ij} \) ,
\nonumber \\
\txi_\Lambda^{[i]}(\varpi,q^\alpha)& = &  B_\Lambda +
\half  \sum_j \oint_{C_j} \frac{\de \varpi'}{2 \pi \I \varpi'} \,
\frac{\varpi' + \varpi}{\varpi' - \varpi}
\, \p_{\xii{i}^\Lambda } \Hij{ij}-2\I  \ci{+}_\Lambda \log \varpi ,
\label{txiqlineQK}
\\
\ai{i}(\varpi,q^\alpha)& = & B_\alpha +
\half  \sum_j \oint_{C_j} \frac{\de \varpi'}{2 \pi \I \varpi'} \,
\frac{\varpi' + \varpi}{\varpi' - \varpi}
\( \Hij{ij}- \xii{i}^\Lambda \p_{\xii{i}^\Lambda}\Hij{ij}\)-2\I\(  \ci{+}_\alpha  \log \varpi
+\cij{+}_\Lambda\(Y^\Lambda \varpi^{-1} + \bY^\Lambda \varpi\)\).
\nonumber
\eeq
Here complex $Y^\Lambda$ and real $A^\Lambda,B_\Lambda,B_\alpha$
parametrize the space of solutions of the gluing conditions and
can be chosen as coordinates on $\qkm$. However, they contain $4n+1$ variables, one more than is needed
for a $4n$-dimensional QK manifold. One auxiliary coordinate can be absorbed by a phase
rotation of the $\CP$ coordinate $\varpi$, which allows to choose, for example, $Y^0$ to be real.
On the other hand, $c_\Lambda$ and $c_\alpha$ are just numerical parameters.
They provide the so called ``anomalous dimensions" mentioned in footnote \ref{foot_quasi}.
They arise due to the possibility to have transition functions on the HKC $\hkm_{\qkm}$
which are not simply homogeneous, but quasi-homogeneous and characterize this
``anomalous" behavior.

As is the case for the K\"ahler potential of HK spaces, there is an integral
representation of the contact potential introduced above in terms of
solutions of the equations \eqref{txiqlineQK}.
It is given by the following formula \cite{Alexandrov:2009zh}
\be
\Phi\ui{i}(\varpi,q^\alpha)=\phi-\frac12 \sum_j \oint_{C_j} \frac{\de \varpi'}{2 \pi \I \varpi'}
\,\frac{\varpi' + \varpi}{\varpi' - \varpi}
\, \log\(1-\p_{\ai{j}}\Hij{ij}(\varpi') \),
\label{eqchip}
\ee
where the $\varpi$-independent part reads as
\be
e^\phi=\frac{\frac{1}{8\pi} \sum_j\oint_{C_j}\frac{\de\varpi}{\varpi}
\(\varpi^{-1} Y^{\Lambda}-\varpi \bY^{\Lambda} \)
\p_{\xii{i}^\Lambda } \Hij{ij}+{c_\Lambda} A^\Lambda+{\cij{+}_\alpha}}
{2\cos\[\frac{1}{4\pi}\sum_j\oint_{C_j}\frac{\de\varpi}{\varpi}\,\log\(1-\p_{\ai{j}}\Hij{ij} \)\]}.
\label{contpotconst}
\ee

Finally, note that the conclusions of section \ref{subsec_cuts} are equally applied to the given case
and in all above equations some of the contours may be taken to be open. This is a sign of
the presence of branch cuts and there is always a possibility to reformulate such a construction
in terms of the usual open patches and closed contours only.
But the latter is usually much more involved than the one in terms
of open contours.

\subsection{The metric}
\label{subsec_QKmetric}

As in the previous section, the first and the most complicated step on the way to the metric is to solve
the gluing conditions \eqref{txiqlineQK} and to find the holomorphic Darboux coordinates
as functions of coordinates on $\qkm$ and the $\CP$ coordinate $\varpi$.
Once this step is completed, the procedure to extract the metric is straightforward,
although more involved than in the HK case.
It involves the following steps \cite{Alexandrov:2008nk,Alexandrov:2008gh}:
\begin{enumerate}
\item
compute the components of the $SU(2)$ connection from
\eqref{formcX}, \eqref{contact}, \eqref{eqchip} and \eqref{contpotconst};
\item
compute the triplet of quaternionic 2-forms $\vec\omega_\qkm$ using the second equation in \eqref{ompp};
\item
expand the holomorphic one-forms
$\de\xi^\Lambda,\ \de\txi_\Lambda$ and $\de\alpha$
around $\varpi=0$ and project them along the base $\qkm$, producing
local one-forms on $\qkm$ of Dolbeault type $(1,0)$,
which allows to get the almost complex structure $J^3$;
\item
obtain the metric via $g = \omega_\qkm^3 \cdot J^3$.
\end{enumerate}
We refer to \cite{Alexandrov:2008gh} for further details where all elements of the
construction are expressed in terms of first few coefficients of the expansion of Darboux coordinates
around $\varpi=0$.

\subsection{Toric QK geometries, deformations and isometries}
\label{subsec_toricQK}

There exists a QK analogue of $\cO(2)$ HK spaces where the integral equations \eqref{txiqlineQK}
provide explicit expressions for the Darboux coordinates and we do not have any equations to solve anymore.
This is the case where the transition functions are independent of $\txii{j}_\Lambda$ and $\ai{j}$,
which describes QK manifolds with $n+1$ commuting isometries. We will call them {\it toric} geometries.
The corresponding Killing vectors are given by $\p_{B_\Lambda}$ and $\p_{B_\alpha}$
and their moment maps provide global sections on the twistor space which coincide
with the unit function and a half of Darboux coordinates
\be
\label{gxi}
\xi^\Lambda =\varpi^{-1}Y^\Lambda  + A^\Lambda - \varpi\bY^\Lambda .
\ee
The HKC of a toric manifold is an $\cO(2)$ HK space whose K\"ahler potential
appears as a Legendre transform \eqref{K02}.
Thus, the toric manifolds are those QK spaces which can be described using the projective superspace approach
and this fact has been extensively used in the study of quantum corrections to the HM moduli space in
\cite{Rocek:2005ij,Robles-Llana:2006ez,RoblesLlana:2006is}.

Generic QK spaces can be viewed as perturbations around the toric case if the transition functions
can be represented as in \eqref{genf}. Again, the integral equations \eqref{txiqlineQK}
can be used to provide a perturbative solution expanded in powers of $\Hpij{ij}$.
The linear approximation was studied in detail in \cite{Alexandrov:2008nk}.

A generic perturbation breaks all continuous isometries of $\qkm$.
However, if an isometry has survived, it is known that it can be lifted to a {\it holomorphic} isometry of
the twistor space $\qkt$. Moreover, even if a continuous symmetry is broken by perturbations to
a discrete subgroup, the remaining discrete isometry should still be realizable on $\qkt$
holomorphically.
This property was instrumental in our study of string moduli spaces
because a combination of discrete symmetries with holomorphicity turns out to be sufficiently powerful
to generate all quantum corrections to the classical metric.

Besides the toric geometries, there is another special case which is worth to consider in detail.
Let us assume that $\qkm$ has one continuous isometry. One can always choose the Darboux coordinates
such that the corresponding holomorphic Killing vector on $\qkt$ is $\p_{\ai{i}}$.
As a result, our transition functions are independent on the coordinates $\ai{j}$
which leads to crucial simplifications. First of all, in this case the functions $\hat f_{ij}$ \eqref{eqhf}
become trivial so that the contact one-form becomes globally defined.
Besides, the contact potential \eqref{eqchip}
is also real, globally defined and independent on $\varpi$, $\Phi\ui{i}=\phi(q^\alpha)$.
But more importantly is that such $\qkm$ can be identified locally with
a HK space of the {\it same} dimension.
This identification was called {\it QK/HK correspondence} \cite{Alexandrov:2011ac}
and plays an important role for moduli spaces in theories with $N=2$ supersymmetry.

\subsection{QK/HK correspondence}
\label{subsec_QKHK}

The QK/HK correspondence is a one-to-one map between a QK space with a quaternionic isometry
and a HK space of the same dimension also having an isometry and
equipped with a hyperholomorphic line bundle $\lb$ \cite{Alexandrov:2011ac}
(see also \cite{Haydys}).
This map can be constructed quite explicitly if one proceeds as follows
\be
\begin{split}
& \hkm_\qkm^{(4n+4)}
\\
\nearrow & \qquad \searrow
\\
\qkm^{(4n)} \quad & \qquad\qquad \hkm^{(4n)}
\end{split}
\label{dualQKHK}
\ee
It means that, first, one associates a HKC $\hkm_\qkm$ to the initial QK manifold.
In particular, the isometry on $\qkm$ gives rise
to a tri-holomorphic isometry on $\hkm_\qkm$. Then at the second step this isometry can be used to perform
a hyperk\"ahler  quotient which produces a HK space of the same dimension as the initial QK space.

Following this procedure, it is possible to give explicit formulae for the dual HK metric
and the hyperholomorphic connection of $\lb$. Let the initial QK metric and
the $SU(2)$ connection are represented as
\be
\label{dsqkgen}
\de s^2_{\qkm} =  \cff \(\de\sigp+\Theta\)^2 + \de s^2_{\qkm/\p\sigp},
\qquad
p_3 = - \frac{1}{\rho} \( \de \sigp + \Theta\) + \pTheta
\ee
where $\p_\sigp$ generates the isometry, $\Theta$ is a connection one-form,
$\cff$ is a function on $\qkm$ invariant under $\p_\sigp$, and
we chose a $SU(2)$ frame such that the QK moment map $\vec \mu_{\qkm}$
for $\p_\sigp$ is aligned along the third axis and described by a function $\rho$.
Then the dual metric is found to be \cite{Alexandrov:2011ac}
\be
\label{dsPhk}
\de s^2_{\hkm} =\frac{\de\rho^2}{\rho} -2\rho \, \de s^2_{\qkm/\p_\sigp} + 4 \rho \, |p_+|^2   +
(\cfff+\rho) \(\de\phip+\pTheta\)^2,
\ee
where we traded $\cff$ for a new function $\cfff$ defined by
\be
\cff=\frac{\cfff+\rho}{2\rho^2\cfff}.
\ee
As is clear from \eqref{dsPhk}, the dual metric possesses an isometry generated by $\p_\phip$.
It is also possible to show that the curvature of the connection
\be
\lambda_{\lb} =\cfff \( \de\phip+\pTheta\) + \Theta
\label{lambdlbTheta}
\ee
is a (1,1)-form in all complex structures and thus defines a hyperholomorphic line bundle.
It is very important to consider the HK space together with this line bundle, because
there are many QK spaces which are mapped to the same HK manifold and which differ only by the form of
the hyperholomorphic connection $\lambda_{\lb}$.

In fact, the QK/HK correspondence becomes particularly simple in our twistor approach.
Indeed, as indicated in the previous subsection, one can always adjust Darboux coordinates such that
the transition functions are $\alpha$-independent, which implies that
the coordinate $\sigp$ should be identified with $B_\alpha$.
Then the gluing conditions \eqref{QKgluing} for $\xi^\Lambda$ and $\txi_\Lambda$
reduce to the usual symplectomorphisms identical to \eqref{HKgluing} so that
the equations determining the twistor lines in HK and QK cases become equivalent.
From this fact one immediately concludes that, to describe the dual spaces, $\qkm$ and $\hkm$,
in the twistor approach, it is sufficient to take the same covering of their twistor spaces and
the same transition functions
\be
\Hij{ij}_{\hkm}(\eta,\mu,\varpi)=\Hij{ij}_{\qkm}(\eta,\mu),
\label{QKHK-tr}
\ee
which implies also a simple relation between the Darboux coordinates
\be
\eta^\Lambda(\varpi)=\xi^\Lambda(e^{-\I\phip}\varpi),
\qquad
\mu_\Lambda(\varpi)=\txi_\Lambda(e^{-\I\phip}\varpi)
\ee
provided we identify
\be
u^\Lambda=Y^\Lambda e^{\I\phip},
\qquad
x^\Lambda=A^\Lambda,
\qquad
\vrh_\Lambda=B_\Lambda.
\label{identrealcoor}
\ee
The relation \eqref{QKHK-tr} shows that the image of the duality map \eqref{dualQKHK}
consists of those HK spaces whose transition functions can be taken independent on
the $\CP$ coordinate $\varpi$. This independence is the origin of the isometry along $\phip$.

This construction leaves aside so far the remaining Darboux coordinate $\alpha$.
It is precisely this holomorphic coordinate that is responsible for the existence
of the hyperholomorphic line bundle $\lb$ on $\hkm$. It defines a holomorphic section
on $\cZ_\hkm$ and appears as a potential for the connection $\lambda_{\lb}$
\be
\label{Khhgen}
K_{\lb} = \Im\alpha,
\qquad
\lambda_{\lb}=\frac{1}{2\I}(\p-\bar\p)K_{\lb}.
\ee
It is interesting that the contact potential $e^\phi$ \eqref{contpotconst} of the QK geometry
also plays an important role:
it is proportional to the moment map $\rho$ (see \eqref{dsqkgen}) \cite{Alexandrov:2008nk}
and relates the K\"ahler potentials for the HK metric and the hyperholomorphic curvature
\be
4e^\phi=K_\hkm+K_{\lb}.
\ee

The above formulae have been written ignoring the possibility to have non-vanishing anomalous dimensions
which appear explicitly in \eqref{txiqlineQK}. However, it is easy to incorporate them
and one finds that their variation induces just a coordinate transformation on the dual HK manifold \cite{Alexandrov:2011ac},
whereas both, the metric and the hyperholomorphic curvature, stay invariant.
Thus, it seems that the QK/HK correspondence erases any information about them.
But this is not the case as the anomalous dimensions affect the hyperholomorphic connection $\lambda_{\lb}$
changing it by an exact term. This is a manifestation of the fact, the QK/HK correspondence
produces $\lambda_{\lb}$ in a fixed K\"ahler gauge. This is clear from \eqref{lambdlbTheta}
as K\"ahler transformations involving $\de\phip$ change the function $\cfff$ and, as a consequence,
the QK metric \eqref{dsqkgen}.

Why is this pure mathematical duality relevant for us?
The point is that the QK/HK correspondence relates the local and rigid c-map metrics based on the same prepotential $F$.
Besides, the same relation continues to hold upon inclusion of the one-loop correction
on the QK side: as we will see in section \ref{chap_pert}.\ref{sec_pertspace}, in the twistor description
it appears as a non-vanishing anomalous dimension $c_\alpha$ and hence does not affect the dual HK metric.
Thus, the QK/HK correspondence provides a nice connection between perturbative metrics on
the HM moduli space $\MH$ of compactified Type II string theory and the moduli space $\Mthree$
of circle compactification of a $N=2$ gauge theory, considered
in section \ref{chap_moduli}.\ref{sec_compgauge}.\footnote{It should be noted however that
the holomorphic prepotentials relevant for gauge and string theories are of different types.
In particular, the image of the local c-map metric under
the QK/HK correspondence is a HK metric of Lorentzian signature, whereas the metric on $\Mthree$
should be positive definite.}
It shows that, although the HM moduli space is \qk, it can be viewed as coming
from a more simple \hk space.

Moreover, as was discussed in section \ref{chap_moduli}.\ref{subsec_quantcor},
all D-instanton corrections to the metric on $\MH$
preserve the isometry along the NS-axion. Therefore, in the approximation where the instanton corrections
from NS5-branes are ignored, $\MH$ is still subject to the QK/HK correspondence.
A remarkable fact is that the image of this map is again given by $\Mthree$, now corrected by BPS instantons,
so that the latter can be identified with the D-instantons in string theory.
This provides an explanation why their twistor constructions, which we are going to present in chapter \ref{chap_Dbrane},
look very similar.
At the same time, we see that the inclusion of NS5-brane contributions destroys the correspondence and makes
the moduli space $\MH$ truly \qk.

\chapter{Tree level and perturbative corrections}
\label{chap_pert}

\section{Tree level moduli space in the twistor description}
\label{sec_treetwist}

The tree level metric on the HM moduli space is obtained by the c-map procedure
as described in section \ref{chap_moduli}.\ref{subsec_cmap} and is given explicitly in \eqref{hypmettree}.
It possesses several continuous isometries which, in particular, include the Heisenberg symmetry
transformations \eqref{Heisenb} shifting the RR-fields and the NS-axion.
This implies that at tree level $\MH$ is a toric QK manifold.
Its twistor description has been proposed in \cite{Alexandrov:2008nk} heavily relying on
the results of \cite{Rocek:2005ij,Neitzke:2007ke} which used the projective superspace formalism. The latter
essentially coincides with our twistor description of HK spaces which are constrained to be
\hk cones with $n+1$ commuting isometries. Here we present only the final construction.

As we learnt in the previous chapter, to define a QK manifold, it is sufficient to provide
its covering and the associated set of transition functions.
In practice, we will always work locally in the moduli space. As a result, we should care
only about the covering of the $\CP$ fiber and the Darboux coordinates develop singularities
only in $\varpi$.

Taking this comment into account, we consider the Riemann sphere and introduce a covering by three patches,
$\cU_+,\cU_-$ and $\cU_0$: the first patch surrounds the north pole $(\varpi=0)$,
the second is around the south pole $(\varpi=\infty)$, and the third one covers the region around
the equator of the sphere, as is shown on Fig. \ref{fig-pertsphere}.
It is clear that in this case there are only two independent transition functions defined on the intersection
of $\cU_0$ with $\cU_\pm$.
It turns out that this construction produces the HM moduli space provided these functions are chosen as follows
\be
\label{symp-cmap}
\Hij{+0}= F(\xi),
\qquad
\Hij{-0}=\bF(\xi).
\ee
Since we are describing a toric QK manifold, the transition functions depend only on the Darboux coordinates
$\xi^\Lambda$. For these coordinates we do not have to specify the patch because they are globally
defined as in \eqref{gxi}. The full set of Darboux coordinates can be easily found using \eqref{txiqlineQK}
and in the patch $\cU_0$ they are given by \cite{Neitzke:2007ke}
\be
\label{gentwi}
\begin{split}
\xii{0}^\Lambda &= \zeta^\Lambda + \frac{\tau_2}{2}
\left( \varpi^{-1} z^{\Lambda} -\varpi \,\bz^{\Lambda}  \right) ,
\\
\txii{0}_\Lambda &= \tzeta_\Lambda + \frac{\tau_2}{2}
\left( \varpi^{-1} F_\Lambda(z)-\varpi \,\bF_\Lambda(\bz) \right),
\\
\tai{0}&= \sigma + \frac{\tau_2}{2}
\left(\varpi^{-1} W(z)-\varpi \,\bar W(\bz) \right) ,
\end{split}
\ee
where
\be
W(z) \equiv  F_\Lambda(z) \zeta^\Lambda - z^\Lambda \tzeta_\Lambda
\label{defWz}
\ee
and used the following redefinition of coordinates
\be
A^\Lambda=\zeta^\Lambda ,
\quad
B_\Lambda=\tzeta_{\Lambda}-\zeta^\Sigma \Re F_{\Lambda\Sigma}(z) ,
\quad
B_\alpha =- \hf\(\sigma   +\zeta^\Lambda B_\Lambda\) ,
\quad
Y^\Lambda = \frac{\tau_2}{2}\, z^\Lambda .
\label{relABzeta}
\ee
The new coordinates\footnote{Note that $z^0\equiv 1$ and therefore
$Y^0=\tau_2/2$ is real in agreement with the comment below \eqref{txiqlineQK}.}
$\zeta^\Lambda,\tzeta_\Lambda,\sigma,z^a$ turn out to coincide with the physical fields
of the compactified string theory in the type IIA formulation. Comparing with Table \ref{HMfields},
one observes that we are missing the dilaton $\phi$. It can be obtained from the additional coordinate $\tau_2$,
which is in fact identified with the ten-dimensional string coupling.
More precisely, the dilaton appears as the contact potential, which does not depend
on the $\CP$ coordinate due to the isometries
and therefore can be considered as a coordinate on the moduli space. Using \eqref{contpotconst}, one finds
\be
\label{phipertA}
e^{\Phi}=e^\phi
= \frac{\tau_2^2}{16}\,K(z,\bz).
\ee

\lfig{Covering of $\CP$ by three patches and the associated transition functions.}{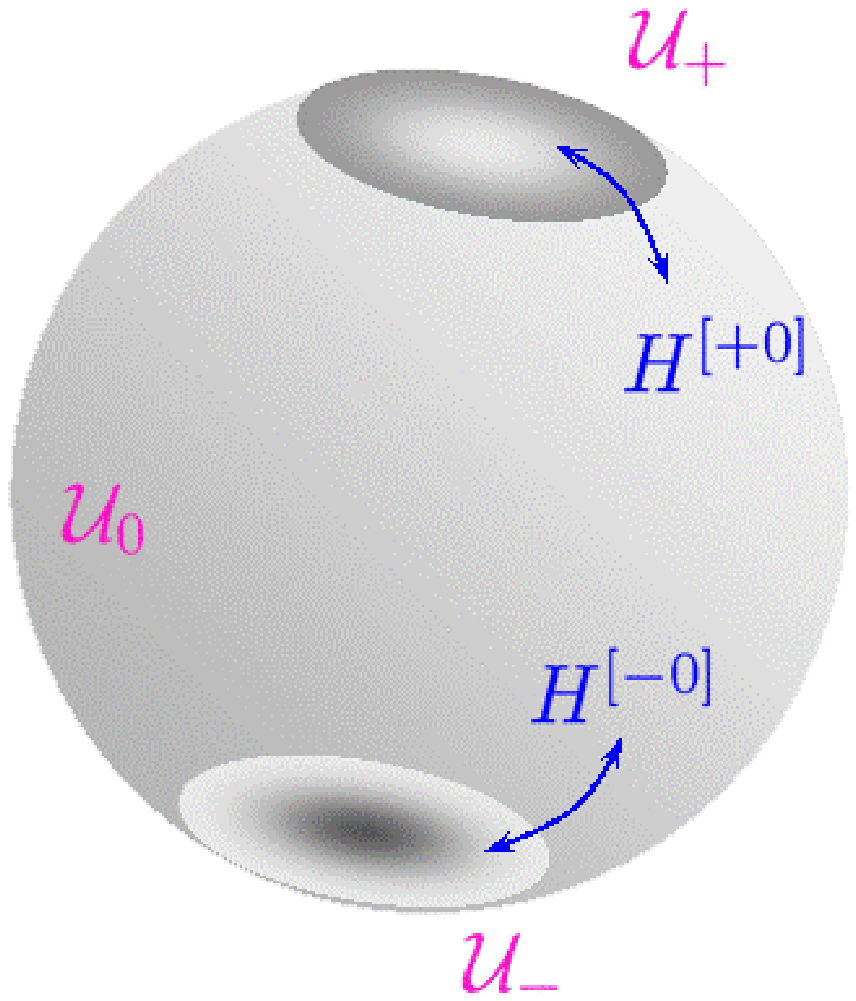}{7cm}{fig-pertsphere}{-0.8cm}

The last thing which we need to explain regarding \eqref{gentwi} is $\tai{i}$.
This is a shifted version of $\ai{i}$ defined as
\be
\tai{i}=-2\ai{i}- \xii{i}^\Lambda\txii{i}_\Lambda.
\label{deftalph}
\ee
It is more convenient because $\tai{0}$ is explicitly invariant under symplectic transformations \eqref{emag}.
At the same time, $(\xii{0}^\Lambda,\txii{0}_\Lambda)$ form a symplectic vector. This property
ensures the symplectic invariance of the whole construction. Besides, it demonstrates
the distinguished role of the patch $\cU_0$. In other patches the Darboux coordinates can be obtained
by using the gluing conditions \eqref{QKgluing} with the transition functions \eqref{symp-cmap}.
In $\cU_\pm$ the poles
at $\varpi=0$ and $\varpi=\infty$, respectively, are canceled, but
the explicit symplectic invariance is broken.

From this twistor construction and the generic behavior
of the instanton corrections described in section \ref{chap_moduli}.\ref{subsec_quantcor},
one can already learn a few lessons on how these corrections should be incorporated.
Comparing the dependence of the Darboux coordinates \eqref{gentwi} on the RR-fields and the NS-axion
with the leading form of the D-brane and NS5-brane instantons, \eqref{d2quali} and \eqref{couplNS5},
one finds that the transition functions encoding them should scale, respectively, as
\be
H_\gamma\sim e^{-2\pi \I\(q_\Lambda \xi^\Lambda -p^\Lambda \txi_\Lambda\)},
\qquad
H_k \sim e^{-\pi \I k \tilde\alpha}h_k(\xi,\txi).
\label{asympH}
\ee
Thus, the Darboux coordinates $\xi^\Lambda,\txi_\Lambda$ and $\tilde\alpha$ can be viewed as
a twistor space lift of $\zeta^\Lambda,\tzeta_\Lambda$ and $\sigma$.
This also implies that if one introduces only D2-instantons associated with A-cycles, i.e. one puts $p^\Lambda=0$,
then the Darboux coordinates $\xi^\Lambda$ remain unmodified and given by the tree level result \eqref{gentwi}.

Following the procedure outlined in section \ref{chap_quatern}.\ref{subsec_QKmetric}, one can
derive the metric on the QK manifold defined by the above covering and transition functions.
Being expressed in terms of the type IIA physical fields, the metric coincides
with the well known result for the tree level metric on the HM moduli space \eqref{hypmettree},
which confirms the validity of our construction.
Note that the transition functions \eqref{symp-cmap} are completely
defined in terms of the holomorphic prepotential.
This is essentially the content of the c-map. Thus, in our formalism it gets a simple and nice
realization by means of the identification \eqref{symp-cmap}.

\section{Perturbative moduli space}
\label{sec_pertspace}

\subsection{One-loop correction in the twistor space}
\label{subsec_perttwist}

A way to incorporate the one-loop $g_s$-correction has been suggested in \cite{Robles-Llana:2006ez}.
The proposal has been formulated in the projective superspace framework based on symmetry considerations
and on the previous results of
\cite{Antoniadis:1997eg,Gunther:1998sc,Antoniadis:2003sw,Anguelova:2004sj}.
Namely, it was found that the one-loop correction gives rise to an additional term
in the Lagrangian \eqref{tlag1} of the following form
\be
\cL_{\mbox{\scriptsize 1-loop}}=\frac{c}{\pi\I}\int_{C_8}\frac{\de\varpi}{\varpi}\, \eta^\alpha\log\eta^\alpha,
\label{oneloopL}
\ee
where $C_8$ is the figure-eight contour encircling the two zeros of $\eta^\alpha(\varpi)$ and
$c$ is a deformation parameter which encodes the one-loop correction.
This parameter was found in \cite{Robles-Llana:2006ez} by matching the microscopic string calculations
of \cite{Antoniadis:1997eg,Antoniadis:2003sw} and is completely determined
by the Euler characteristic of the Calabi-Yau,\footnote{While most of considerations in this review
are independent of the precise value of $c$, this value is important to have agreement with perturbative string theory.
Furthermore, it is correlated with the one-loop $\alpha'$-correction to the holomorphic prepotential \eqref{lve},
which is also determined by the Euler characteristics.
These two corrections are mixed by S-duality discussed below in chapter \ref{chap_IIB}.}
\be
c=-\frac{\chi_\CY}{192 \pi} .
\label{cchi}
\ee
The deformation term \eqref{oneloopL} precisely corresponds to the situation described in section \ref{chap_quatern}.\ref{subsec_cuts}
and is generated by a quasi-homogeneous transition function $\eta^\alpha\log\eta^\alpha$.
As we know, such contributions do not descend to non-trivial transition functions on the twistor space
of the initial QK manifold, but instead they give rise to non-vanishing anomalous dimensions.
Indeed, as was shown in \cite{Alexandrov:2008nk}, to obtain the one-loop corrected HM moduli space,
it is sufficient to supplement the transition functions \eqref{symp-cmap} by one non-vanishing anomalous
dimension
\be
\ci{+}_\alpha=-2c=\frac{\chi_\CY}{96 \pi}, \qquad \ci{+}_\Lambda=0.
\label{anomdim}
\ee
As was argued in \cite{Robles-Llana:2006ez,Alexandrov:2008nk}
on the basis of symmetries and topological considerations, the moduli space metric does not get higher loop corrections
so that this simple modification incorporates {\it all} perturbative effects.

This correction affects the last Darboux coordinate in \eqref{gentwi} and the contact potential.
From \eqref{txiqlineQK} and \eqref{contpotconst}, one finds
\be
\tai{0}= \sigma + \frac{\tau_2}{2}
\left(\varpi^{-1} W(z)-\varpi \,\bar W(\bz) \right)+\frac{\I\chi_\CY}{24\pi}\,\log\varpi,
\qquad
e^\phi
= \frac{\tau_2^2}{16}\,K(z,\bz)+\frac{\chi_\CY}{192\pi}.
\label{oneloopres}
\ee
Thus, the only effect of the one-loop correction is the introduction of logarithmic singularities
at the north and south poles of $\CP$ and a shift of the dilaton by a constant term.
Of course, its effect on the metric is much more significant.
It will be the subject of the next subsection.

\subsection{Moduli space metric}
\label{subsec_pertmet}

First, the one-loop corrected metric has been found in \cite{Robles-Llana:2006ez}, however,
in a somewhat inexplicit form. A much more simple and explicit expression has been obtained in \cite{Alexandrov:2007ec}
by performing the standard superconformal quotient \cite{deWit:2001dj} of the HKC
defined by the tree level Lagrangian and its correction \eqref{oneloopL}.
Later the same result has been reproduced from the twistor formulation of the previous subsection \cite{Alexandrov:2008nk}.
As a result, the one-loop corrected metric was found to be
\be
\begin{split}
\de s_{\MH^{\rm pert}}^2=& \, \frac{r+2c}{r^2(r+c)}\,\de r^2
-\frac{1 }{2r}\, (\Im\cN)^{\Lambda\Sigma}\(\de\tzeta_\Lambda -\cN_{\Lambda\Lambda'}\de\zeta^{\Lambda'}\)
\(\de\tzeta_\Sigma -\bar\cN_{\Sigma\Sigma'}\de\zeta^{\Sigma'}\)
\\ &
+\frac{2c}{r^2 K}\left| z^\Lambda\de\tzeta_\Lambda-F_\Lambda\de\zeta^\Lambda\right|^2
+ \frac{r+c}{16 r^2(r+2c)}\, D\sigma^2
+\frac{4(r+c)}{r}\,\cK_{a\bar b}\de z^a \de \bz^{\bar b}  ,
\end{split}
\label{hypmetone}
\ee
where $D\sigma$ is the one-form
\be
\label{Dsigone}
D\sigma = \de \sigma + \tzeta_\Lambda \de \zeta^\Lambda -  \zeta^\Lambda \de \tzeta_\Lambda
+ 8 c \, \cA_K
\ee
corrected at one-loop by the \kahler connection
\be
\cA_K=\frac{\I}{2}\(\cK_a\de z^a-\cK_{\ba}\de \bz^{\ba}\).
\ee
Thus, we observe that the effect of the one-loop correction, described by the numerical coefficient $c$ \eqref{cchi},
is to modify the dilaton-dependent coefficients of various terms
and to add an additional term to the connection \eqref{Dsigone}. The latter describes the circle bundle
of the NS-axion over the torus of the RR-fields. This bundle and the one-loop correction to its curvature
will play an important role in the study of NS5-brane contributions in chapter \ref{chap_NS5}.

More importantly, the one-loop correction introduces a curvature singularity at $r=-2c$ for $c<0$.
One expects that it should be resolved by instanton corrections.
Such a resolution can only happen once all corrections, including those
which come from NS5-branes, are taken into account at all orders. Since this goal has not been
achieved yet, this resolution problem remains open.
The other two apparent singularities at $r=0$ and $r=-c$ can be shown to be
of coordinate nature \cite{Alexandrov:2009qq}.

\chapter{D-brane instantons}
\label{chap_Dbrane}

\section{D-instanton corrected twistor space}

The metric \eqref{hypmetone} presented in the previous chapter provides the complete
perturbative description of the HM moduli space. The next step is to include
non-perturbative corrections. As we saw in section \ref{chap_moduli}.\ref{subsec_quantcor},
they come either from D-branes wrapping non-trivial cycles or NS5-branes.
In this chapter we will concentrate on the former type of corrections and postpone the latter
till chapter \ref{chap_NS5}. This implies that we are working in the region of the moduli space
where the string coupling determined by the dilaton is small and therefore
the NS5-brane corrections are exponentially suppressed comparing to the D-instanton ones.
Besides, we restrict ourselves to the type IIA formulation so that our results
should explicitly respect symplectic invariance. The mirror type IIB formulation
will be considered in the next chapter.

Thus, here our aim is to present the construction of the HM moduli space
in the type IIA formulation which includes contributions from D-brane instantons.
We do not go through its derivation, which followed the chain of dualities shown in Fig. \ref{fig_scheme},
but give just the final picture \cite{Alexandrov:2008gh,Alexandrov:2009zh}.
However, for reader's convenience, we approach it in several steps. First, we explain
how one can incorporate contributions coming from D-branes wrapping {\it one} particular cycle
of the Calabi-Yau. Then we superpose contributions from different cycles
obtaining a description valid in the {\it one-instanton approximation}. And finally we explain
how non-linear effects can be taken into account, which produces the final construction
encoding all D-instanton corrections.

\subsection{BPS rays and transition functions in the one-instanton approximation}

Let us start by picking up a non-contractible cycle on the Calabi-Yau which can
give rise to non-trivial instanton corrections to the HM metric.
Since we are working in the type IIA formulation, D-instantons arise only from D2-branes
and thus one should consider a three-dimensional cycle.
More precisely, this should be a special Lagrangian submanifold in a homology class
$q_\Lambda \gamma^\Lambda-p^\Lambda \gamma_\Lambda\in H_3(\CY,\IZ)$ \cite{Aspinwall:2004jr}.
The vector $\gamma=(p^\Lambda,q_\Lambda)$ can be seen as the charge associated to the brane,
with $-\gamma$ being the charge of the anti-brane.
Our first goal is to show how one can generate non-perturbative contributions to the HM metric
coming from these particular BPS objects.

The construction we are going to present crucially relies on the twistor methods developed
in chapter \ref{chap_quatern} and can be viewed as an enrichment of the twistor space describing
the perturbative HM moduli space from the previous chapter.
Thus, what we need is to provide a refinement of the covering,
used in that description and shown in Fig. \ref{fig-pertsphere}, and to give transition functions
associated with the new patches. Once this task is completed, the D-instanton corrected
metric can be found following the recipe of section \ref{chap_quatern}.\ref{subsec_QKmetric}.

To describe the new covering, let us recall that
every charge vector gives rise to the central charge function $Z_\gamma(z)$ \eqref{defZ}.
For fixed complex structure moduli $z^a$, the phase of this function
can be used to define the so called ``BPS ray'' $\ellg{\gamma}$.
This is a line on $\CP$ going between the north and south poles along the direction determined by $\arg Z_\gamma(z)$:
\be
\ellg{\gamma}= \{ \varpi :\,  Z_\gamma(z)/\varpi \in \I\IR^{-} \} .
\label{rays}
\ee
It is clear that the ``BPS ray'' associated with the anti-brane of the opposite charge $-\gamma$
is its antipodal image. Together they form a circle which splits the patch $\cU_0$
into two patches. It turns out that the D-instanton contributions related to D2-branes of charge $\pm\gamma$
are generated by introducing discontinuities of the Darboux coordinates
across the circle.
This implies that there should be a non-trivial transition function
which can be shown to be defined in terms of
the dilogarithm function $\Li_2(x)=\sum_{n=1}^\infty {x^n}/{n^2}$.
Across the BPS ray $\ellg{\gamma}$ it is given by \cite{Alexandrov:2008gh}
\be
H_\gamma(\xi,\txi)=\frac{\hnkl}{(2\pi)^2}\,
\Li_2\left(\sigma_{\text{D}}(\gamma) e^{-2\pi \I \Xikl} \right),
\label{prepH}
\ee
where $\Xikl=q_{\Lambda}\xi^\Lambda- p^\Lambda \txi_\Lambda$, whereas $\hnkl$ and $\sigma_{\text{D}}(\gamma)$
are numerical factors to be discussed below. Across the opposite ray
$\ellg{-\gamma}$ the transition function is given by a similar expression where the sign in the exponent
and in the argument of $\sigma_{\text{D}}$ is flipped,
which is nothing else but $H_{-\gamma}$ provided $\hnkl$ do not depend on the sign of the charge,
$\gnkl{\gamma}=\gnkl{-\gamma}$.

Few comments about the transition functions \eqref{prepH} are in order:
\begin{itemize}
\item
The function \eqref{prepH} is consistent with the expected scaling \eqref{asympH},
whereas the sum in the dilogarithm encodes multi-covering effects.
\item
The direction of the BPS ray $\ellg{\gamma}$ is chosen such that the function $H_\gamma$ evaluated on
the tree level Darboux coordinates \eqref{gentwi} is exponentially decreasing as $\varpi\to 0$ or $\infty$
along $\ellg{\gamma}$.
It is clear that this property continues to hold for all contours which lie in the half-plane centered at $\ellg{\gamma}$.
\item
The coefficients $\hnkl$ in \eqref{prepH} are some numbers which carry a topological information
about the Calabi-Yau manifold.
For $p^\Lambda=0$ they coincide with the genus zero Gopakumar-Vafa invariants determining the instanton
part of the holomorphic prepotential \eqref{lve} in the mirror formulation,
\be
\gnkl{0,q_\Lambda}=n_{q_a}^{(0)}\quad\mbox{\rm for}\quad \{ q_a \} \ne 0 ,
\qquad
\gnkl{0,(0,q_0)}
= \chi_\CY .
\ee
For generic charge they are expected to be integer and were related to the generalized Donaldson--Thomas
invariants introduced in \cite{ks} and satisfying the so called KS wall-crossing formula which will be explained
in section \ref{subsec_wallcros}.
\item
The other numerical factor $\sigma_{\text{D}}(\gamma)$ in \eqref{prepH}
is a ``quadratic refinement of the intersection form"
on the charge lattice $\Gamma$. It is defined as a homomorphism
$\sigma: \Gamma\to U(1)$, i.e. a phase assignment such that
\be
\label{qrifprop}
\sigma_{\text{D}}(\gamma+\gamma') =
(-1)^{\langle \gamma, \gamma' \rangle}\, \sigma_{\text{D}}(\gamma)\, \sigma_{\text{D}}(\gamma'),
\ee
where we used the symplectic scalar product defined in \eqref{SymplecticPairing}.
The general solution of \eqref{qrifprop} can be parametrized as \cite{Belov:2006jd}
\be
\sigma_{\text{D}}(\gamma) = \expe{-\frac12 q_\Lambda p^\Lambda
+ q_\Lambda \theta_{\text{D}}^\Lambda
- p^\Lambda \phi_{{\text{D}},\Lambda}}\equiv \sigma_{\Theta_{\text{D}}}(\gamma),
\label{quadraticrefinementpq}
\ee
where $\Theta=(\theta^\Lambda,\phi_\Lambda)$ are the so called characteristics and
we introduced a very convenient notation $\expe{z}=\exp\[2\pi\I z\]$.
In the case when the characteristics are half-integer, the quadratic refinement becomes just
a sign factor. Although it seems to be harmless, its presence is important for consistency with
wall-crossing \cite{Gaiotto:2008cd} and affects the global structure of the moduli space
\cite{Alexandrov:2010np,Alexandrov:2010ca}.
\item
In \eqref{prepH} we did not specify the patch indices of the Darboux coordinates entering
the arguments of $H_\gamma$. This is because this form of the transition function is valid
strictly speaking only in the one-instanton approximation. In this approximation it is sufficient to use
for $\xi^\Lambda,\txi_\Lambda$ their perturbative expressions in the patch $\cU_0$ \eqref{gentwi}.
At the full non-linear level one should evaluate $\xi^\Lambda$ on the left of the BPS ray and $\txi_\Lambda$
on the right, which would naively spoil the symplectic invariance of the construction.
For the moment we ignore this complication and will address it below in section \ref{subsec_exact}.
\label{issuepatch}
\item
\lfig{Two coverings of $\CP$. The left picture shows a regular covering by open patches
which generates the HM twistor space affected by one D-instanton. The covering on the right
is obtained in the limit where the strips $\cU_\pm$ go to
zero width along the BPS rays $\ellg{\pm\gamma}$, while maintaining a non-zero size at
the north and south pole.}{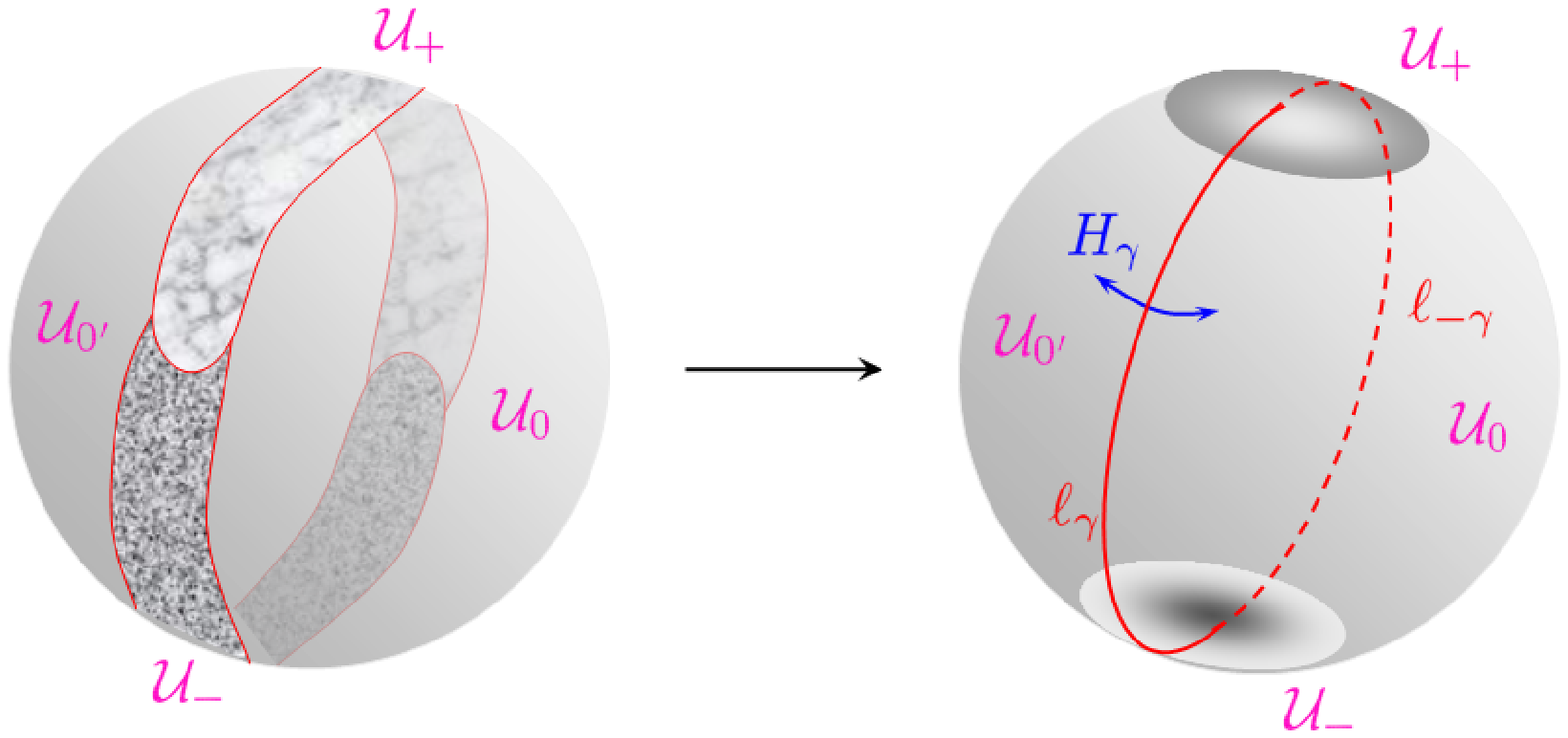}{17cm}{fig_sphere}{-0.8cm}
The presented construction of the instanton corrected HM moduli space is an example of
the description based on the use of open contours. We discussed such possibility
in section \ref{chap_quatern}.\ref{subsec_cuts}. Here the open contours are given by the BPS rays $\ellg{\gamma}$
and, as expected, introduce branch cuts in the Darboux coordinates. This fact leads to
the necessity to modify also the transition functions between $\cU_0$ and
the patches $\cU_\pm$ around the poles because the tree level transition functions \eqref{symp-cmap}
are not sufficient to remove the cuts.
This goal can be achieved by considering
\be
\label{gensymp}
\Hij{+0}=  F(\xii{+})
+\Gg
,
\qquad
\Hij{-0} =   \bF(\xii{-})
-\Gg
,
\ee
where the instanton part is described by the function
\be
\Gg(\Xikl)=\frac{\I\hnkl}{4\pi^3}
\int_0^{-\I\infty} \frac{\Xi\,\de\Xi}{\Xikl^2-\Xi^2}\,
\Li_2\left(\sigma_{\text{D}}(\gamma) e^{-2\pi \I\, \Xi} \right).
\label{funGg}
\ee
It has two branch cuts from $\varpi=0$ and $\varpi=\infty$ to the two roots of $\Xikl(\varpi)=0$.
The discontinuity along these cuts is precisely given by $H_\gamma$ which allows to get
regular coordinates in $\cU_\pm$.
\item
Moreover, the transition functions \eqref{gensymp} can be used to provide a description of the same moduli space
which uses only open patches and closed contours. For this purpose, one extends the patches $\cU_\pm$
down to the equator with a non-vanishing mutual intersection as shown in Fig. \ref{fig_sphere}.
This gives a completely regular description of the twistor space where the functions \eqref{gensymp}
and the anomalous dimension \eqref{anomdim} are the only input data. The previous picture
with the BPS rays arises upon shrinking $\cU_\pm$ along $\ellg{\pm\gamma}$ while retaining
a finite size around the north and south pole \cite{Alexandrov:2009zh}.
\end{itemize}

\lfig{``Octopus-like" and ``melon-like" coverings of $\CP$ and
the associated transition functions.}{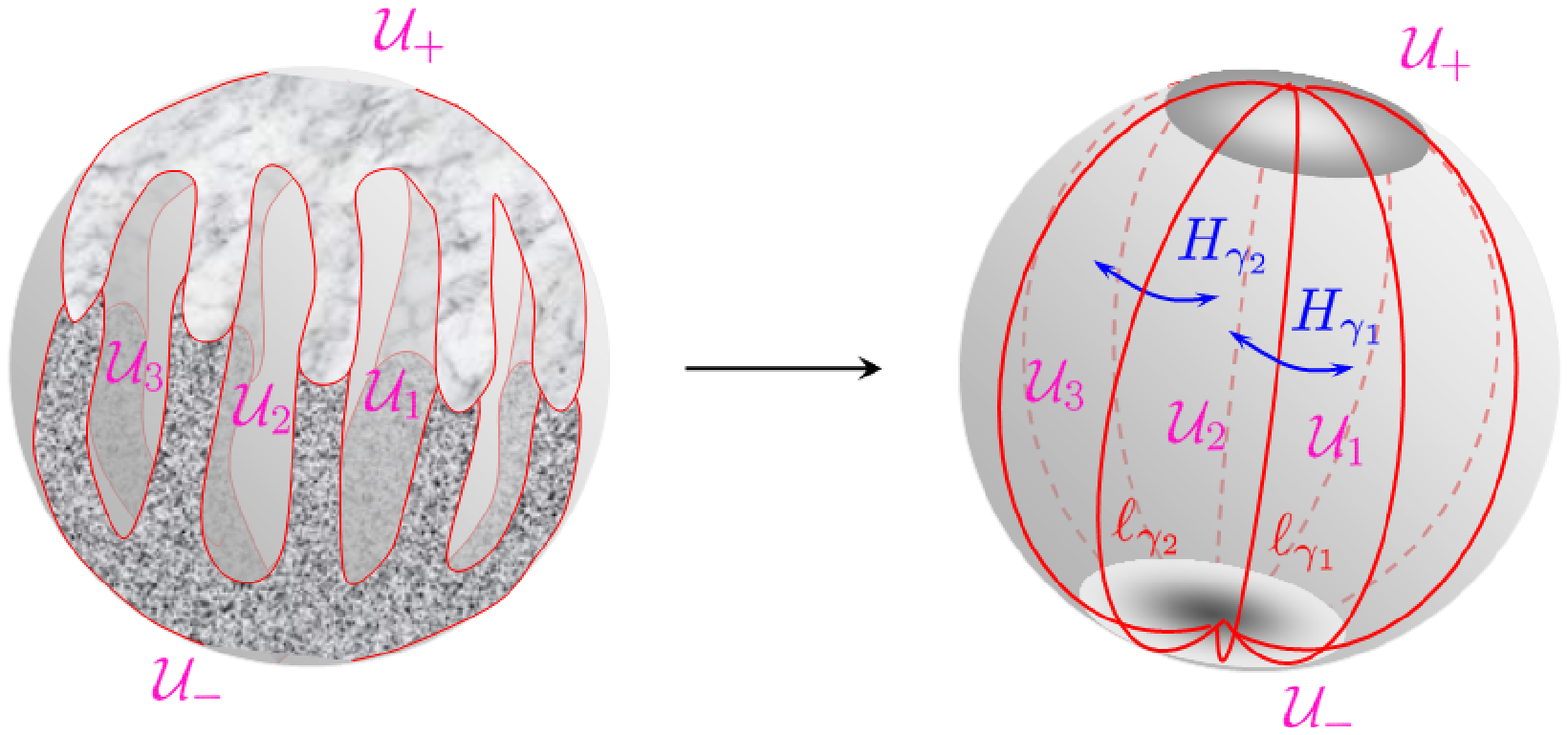}{17cm}{fig-coverD}{-0.8cm}

To include D-branes wrapping other cycles, one should repeat the above procedure associating with each charge
a new BPS ray and introducing across it a discontinuity described by the transition function
given by the same formula \eqref{prepH}.
As a result, one obtains the ``melon-like" picture (Fig. \ref{fig-coverD})
where the covering consists of sectors of $\CP$ divided by the contours $\ellg{\gamma}$
and the usual patches $\cU_\pm$.

Similarly one can generalize the formulation based on open patches. This leads to the picture shown
on the left of Fig. \ref{fig-coverD} which resembles two octopuses with touching fingers.
The corresponding transition functions are given by \eqref{gensymp}
with the functions $\Gg$ summed over different charges.

Although it is straightforward to evaluate the Darboux coordinates in the one-instanton
approximation \cite{Alexandrov:2008gh},
we refrain from giving them explicitly. Instead, in section \ref{subsec_alleqs} we provide
the full result which includes all D-instanton corrections to all orders in the instanton expansion.
From this result the one-instanton approximation will trivially follow.

\subsection{All orders construction}
\label{subsec_exact}

Let us go beyond the one-instanton approximation which was used so far.
It turns out that for this purpose one should just adjust the transition functions \eqref{prepH}
in the way that ensures the symplectic invariance and consistency with wall-crossing.

Let us order all charges $\gamma\in\Gamma$ (including those for anti-branes) according to decreasing
of the phase of the central charge function and enumerate them by the index $a$.
Furthermore, the sector bounded by $\ellg{\gamma_{a-1}}$ and $\ellg{\gamma_{a}}$ on Fig. \ref{fig-coverD}
will be denoted by $\cU_a$ and we define
\be
\Xiijg{ab}{c} \equiv
q_{c,\Lambda}\xii{a}^\Lambda- p_c^\Lambda \txii{b}_\Lambda,
\qquad
\Xigi{a}\equiv\Xiijg{aa}{a}.
\label{Xarggem}
\ee
Then the above two requirements uniquely fix the gluing conditions for
the Darboux coordinates $\xii{a}^\Lambda,\txii{a}_\Lambda$ across the BPS rays
which must be given by
\be
\begin{split}
\xii{a+1}^\Lambda &= \xii{a}^\Lambda
-\frac{\gnkl{\gamma_a}}{2\pi \I}\,p_a^\Lambda \log\(1-\sigma_{\text{D}}(\gamma_a) e^{-2\pi \I \Xigi{a}}\),
\\
\txii{a+1}_\Lambda &= \txii{a}_\Lambda
-\frac{\gnkl{\gamma_a}}{2\pi \I}\,q_{a,\Lambda} \log\(1-\sigma_{\text{D}}(\gamma_a) e^{-2\pi \I \Xigi{a}}\).
\end{split}
\label{KSgl}
\ee
It is easy to check that these transformations are symplectomorphisms in the $(\xi,\txi)$-subspace
and therefore there should be a Hamiltonian function generating them. This function would coincide
with \eqref{prepH} if not the fact, mentioned on page \pageref{issuepatch},
that its arguments should belong to different patches. The correct transition function
can nevertheless be found \cite{Alexandrov:2009zh}, although it is given only implicitly by
\be
\Hij{a\,a+1}(\xii{a},\txii{a+1})=H_{\gamma_a}-\hf\,q_{a,\Lambda} p_a^\Lambda (H_{\gamma_a}')^2 ,
\label{transellg}
\ee
where $H_{\gamma}(\Xikl)$ is defined in \eqref{prepH} and $\Xigi{a}$
should be understood as a function of $\Xiijg{a\,a+1}{a}$ (and thus as a function of the arguments of $\Hij{a\,a+1}$)
defined through the transcendental equation
\be
\Xiijg{a\, a+1}{a} =\Xigi{a} - q_{a,\Lambda} p_a^\Lambda H_{\gamma_a}'(\Xigi{a}) .
\label{Xargsim}
\ee
This complicated definition is designed to ensure that
\be
\p_{\xii{a}^\Lambda}\Hij{a\,a+1}=q_{a,\Lambda}H_{\gamma_a}',
\qquad
\p_{\txii{a+1}_\Lambda}\Hij{a\,a+1}=-p_a^\Lambda H_{\gamma_a}'
\ee
and thus the gluing conditions \eqref{KSgl} are indeed satisfied.

Similarly, the transition functions to the poles \eqref{gensymp} acquire non-linear terms.
Their exact expressions can be found in \cite{Alexandrov:2009zh} and are quite involved.
However, the instanton terms in $\Hij{\pm 0}$ do not affect the evaluation of the metric and
are needed only for consistency of the construction.
Altogether the above data completely define the HM moduli space in the type IIA formulation
with inclusion of {\it all} D-brane instantons.

\subsection{Darboux coordinates and contact potential}
\label{subsec_alleqs}

Although the transition functions uniquely define
the twistor space and the underlying quaternionic manifold, to actually compute the metric
one needs to solve the integral equations \eqref{txiqlineQK} for the twistor lines.
In our case, they can be shown to reduce to the following system \cite{Gaiotto:2008cd,Alexandrov:2009zh}
\be
\label{eqXigi}
\begin{split}
\Xigi{}(\varpi)=&\, \Thkli{}+\frac{\tau_2}{2}\(\varpi^{-1}Z_{\gamma}-\varpi\bZ_{\gamma}\)
\\
& \qquad\qquad +
\frac{1}{8\pi^2}\sum_{\gamma'} \gnkl{\gamma'}\<\gamma,\gamma'\> \int_{\ellg{\gamma'}}\frac{\d \varpi'}{\varpi'}\,
\frac{\varpi+\varpi'}{\varpi-\varpi'}\,
\log\(1-\sigma_{\text{D}}(\gamma') e^{-2\pi \I \Xi_{\gamma'}(\varpi')}\),
\end{split}
\ee
where
\be
\Thkl
\equiv q_\Lambda \zeta^\Lambda - p^\Lambda\tzeta_\Lambda.
\label{THg}
\ee
These equations encode all non-trivialities of the problem.
Once the functions $\Xigi{}(\varpi)$ are known, the Darboux coordinates can be easily given
in terms of the integral which appears also on the r.h.s. of \eqref{eqXigi}
\be
\Igg{}(\varpi)=\int_{\ellg{\gamma}}\frac{\d \varpi'}{\varpi'}\,
\frac{\varpi+\varpi'}{\varpi-\varpi'}\,
\log\(1-\sigma_{\text{D}}(\gamma) e^{-2\pi \I \Xigi{}(\varpi')}\).
\label{newfun}
\ee
Indeed, one finds the following result \cite{Alexandrov:2009zh}
\beq
\xii{a}^\Lambda &=& \zeta^\Lambda + \frac{\tau_2}{2}
\left( \varpi^{-1} z^{\Lambda} -\varpi \,\bz^{\Lambda}  \right)+
\frac{1}{8\pi^2}\sum_{b} \gnkl{\gamma_b} p_b^\Lambda \cJ_{\gamma_b}(\varpi) ,
\nn\\
\txii{a}_\Lambda &=&
\tzeta_\Lambda + \frac{\tau_2}{2}
\left( \varpi^{-1} F_\Lambda-\varpi \,\bF_\Lambda \right)
+\frac{1}{8\pi^2}\sum_{b} \gnkl{\gamma_b} q_{b,\Lambda}\cJ_{\gamma_b}(\varpi)  ,
\label{exline}\\
\tai{a} &= &\sigma
+\frac{\tau_2}{2}\(\varpi^{-1} \cW-\varpi \bar \cW\) +\frac{\I\chi_X}{24\pi} \log \varpi
+ \frac{\I}{8\pi^3}\sum_b \gnkl{\gamma_b}
\int_{\ellg{\gamma_b}}\frac{\d \varpi'}{\varpi'}\,
\frac{\varpi+\varpi'}{\varpi-\varpi'}\,
L_{\gamma_b}(\varpi') ,
\nn
\eeq
where $\varpi\in \cU_a$,
\be
\cW=W-\frac{1}{8\pi^2}\sum_b \gnkl{\gamma_b} \, Z_{\gamma_b} \,\Igg{b}(0),
\ee
and in the last Darboux coordinate we used
\beq
L_\gamma(\varpi)&=& L\(\sigma_{\text{D}}(\gamma) e^{-2\pi \I \Xigi{}(\varpi)}\)
-\hf\,\log \sigma_{\text{D}}(\gamma)\log\(\sigma_{\text{D}}(\gamma) e^{-2\pi \I \Xigi{}(\varpi)}\),
\\
\label{defrogers}
L(z) &=& \Li_2(z)+\frac12\, \log z \log(1-z).
\eeq
The latter function is known as the Rogers dilogarithm.
To complete the set of geometric data, we also give a formula for the contact potential, which
can still be identified with the dilaton.
Computing \eqref{contpotconst} in our situation gives the following result \cite{Alexandrov:2009zh}
\be
\begin{split}
e^{\phi} =&\, \frac{\tau_2^2}{16}\, K(z,\bz)+\frac{\chi_X}{192\pi}
\\
& \qquad
-\frac{ \I\tau_2}{64\pi^2}\sum\limits_{a} \gnkl{\gamma_a}
\int_{\ellg{\gamma_a}}\frac{\d \varpi}{\varpi}\,
\( \varpi^{-1}Z_{\gamma_a} -\varpi\bZ_{\gamma_a}\)
\log\(1-\sigma_{\text{D}}(\gamma_a) e^{-2\pi \I \Xigi{a}(\varpi)}\).
\end{split}
\label{phiinstmany}
\ee

In the limit of small string coupling, the equations \eqref{eqXigi} are well suited
for perturbative treatment. Regarding the last term as a perturbation, one can easily generate
an instanton expansion which is equivalent to the expansion in powers of $\hnkl$.
In particular, to get the one-instanton approximation, one should take
\be
\Xikl(\varpi)= \Thg{}+\frac{\tau_2}{2}\(\varpi^{-1}Z_{\gamma}-\varpi\bZ_{\gamma}\)
\label{resX}
\ee
and substitute it into \eqref{newfun} and subsequent formulae.
It is interesting to note that if one restricts the attention to a sector in the charge lattice
where all charges satisfy the following condition
\be
\label{restr}
\<\gamma_a,\gamma_b\>=0,
\ee
the last term in \eqref{eqXigi} vanishes and the linear instanton approximation becomes exact.
A particular set of such ``mutually local states" is given, for example,
by D2-branes wrapping only A-cycles which all have $p^\Lambda=0$.

\subsection{Symplectic invariance}

Although the Darboux coordinates \eqref{exline} and the contact potential \eqref{phiinstmany}
perfectly respect the symplectic symmetry, one can wonder why this is not the case for the transition functions
\eqref{transellg} and whether one can find a more invariant formulation?
It is clear that the answer to the first question is hidden in the asymmetry in the arguments
of $\Hij{a\,a+1}$: they belong to different patches and therefore do not form a symplectic vector.
Precisely this fact is also the reason for the presence of the non-symplectic invariant quadratic term in \eqref{transellg}.

Therefore, if the second question has an affirmative answer, it should be possible to write
the transition functions generating gluing conditions across BPS rays
as functions of Darboux coordinates in one patch.
However, not every function of such form can be considered as a generating
function of symplectomorphisms in the $(\xi,\txi)$-subspace. It is easy to check that
a function $\hHij{ij}(\xii{i},\txii{i})$ generates a symplectomorphism
if it satisfies the following condition
\be
\p_{\xii{i}^\Lambda}\hHij{ij}\, d\(\p_{\txii{i}_\Lambda}\hHij{ij}\) =
\p_{\txii{i}_\Lambda}\hHij{ij}\, d\( \p_{\xii{i}^\Lambda}\hHij{ij}\).
\label{condsymple}
\ee
In other words, this property ensures that the transformation
\be
\xii{j}^\Lambda=  \xii{i}^\Lambda-\p_{\txii{i}_\Lambda}\hHij{ij},
\qquad
\txii{j}_\Lambda=\txii{i}_\Lambda+\p_{\xii{i}^\Lambda}\hHij{ij}
\label{explicit-symtr}
\ee
preserves the symplectic form $\de\xi^\Lambda\wedge\de\txi_\Lambda$.
Moreover, it can easily be lifted to a contact transformation provided the Darboux coordinate $\tilde\alpha$
transforms as
\be
\tilde\alpha^{[j]} =\tilde\alpha^{[i]}-2\hHij{ij}+\xii{i}^\Lambda\p_{\xii{i}^\Lambda}\hHij{ij}
+\txii{i}_\Lambda\p_{\txii{i}_\Lambda}\hHij{ij}.
\label{glutildealp}
\ee

Remarkably, the condition \eqref{condsymple} is identically satisfied if the function $\hHij{ij}$
depends on its arguments only through their linear combination. This is precisely our case where
the linear combination is given by $\Xikl$! Thus, it is legal to take $\hHij{a\, a+1}=H_{\gamma_a}$
and one easily verifies that the transformations generated by \eqref{explicit-symtr}
coincide with the ones given in \eqref{KSgl}, whereas \eqref{glutildealp} explicitly shows how the
dilogarithm function of \eqref{prepH} turns into the Rogers dilogarithm appearing in \eqref{exline}.

\section{Gauge theories and wall-crossing}

\subsection{BPS instantons}

As was discussed in section \ref{chap_moduli}.\ref{sec_compgauge},
$N=2$ supersymmetric gauge theories compactified on a circle give rise to moduli spaces
carrying a \hk geometry. At the perturbative level the metric on such moduli space is obtained
via rigid c-map and is determined by the holomorphic prepotential of the four-dimensional theory.
But there are also instanton contributions which come from the massive BPS states of the spectrum
and look similar to the D-instanton corrections in superstring theory considered above.

The complete non-perturbative description of these BPS instanton contributions has been developed
in \cite{Gaiotto:2008cd}. It is based on the twistor space approach presented
in section \ref{chap_quatern}.\ref{sec_HK} and the resulting picture is essentially identical
to the description of D-instantons from the previous section. Namely, with each BPS state
one associates a BPS ray whose position is defined by the phase of the central charge
and the Darboux coordinates on the opposite sides of the ray are related
by the symplectomorphism \eqref{KSgl}.
In particular, the Darboux coordinates are given by solution of the same system of
non-linear integral equations \eqref{eqXigi}.
A mathematical explanation of this similarity is provided by the QK/HK correspondence presented
in section \ref{chap_quatern}.\ref{subsec_QKHK}
and can be traced back to the presence of continuous isometries on the two sides of the duality.
In our case they originate from R-symmetry and due to ignorance of NS5-brane instantons, respectively.

Although the description of \cite{Gaiotto:2008cd} is sufficient to get the metric on the moduli space,
in the following we will need an explicit formula for the K\"ahler potential.
It was derived in \cite{Alexandrov:2010pp} from the general formula \eqref{Kdefrew}
specified for our covering and transition functions.
The final result reads as follows
\beq
K_{\Mthree}&= & \frac{R^2}{8}\, K(z,\bz)+\frac{1}{4}\, N_{IJ}\(\zeta^I\zeta^J+
\frac{1}{64\pi^4}\,\sum_{a,b}\gnkl{\gamma_a}\gnkl{\gamma_b}p_a^I p_b^J
\int_{\ellg{\gamma_a}}\!\!\cD_a\varpi
\int_{\ellg{\gamma_b}}\!\!\cD_b\varpi' \)
\nn\\
&&\hspace{-2cm}
-\frac{1}{16\pi^3}\, \sum_{a}\gnkl{\gamma_a}\int_{\ellg{\gamma_a}}\!\! \frac{\d \varpi}{\varpi}
\[\Li_2 \(\sigma(\gamma_a)e^{-2\pi \I \Xigi{a}}\)
-2\pi \I \zeta^I\(q_{a,I}-p_a^J\Re F_{IJ}\)\log\(1-\sigma(\gamma_a)e^{-2\pi \I \Xigi{a}}\)\]
\nn\\
&&\hspace{-2cm}
+\frac{\I}{128\pi^4 }\, \sum_{a\ne b} \gnkl{\gamma_a}\gnkl{\gamma_b}\<\gamma_a,\gamma_b\>
\int_{\ellg{\gamma_a}}\!\!\cD_a\varpi \int_{\ellg{\gamma_b}}\!\!\cD_b\varpi'\,
\frac{\varpi+\varpi'}{\varpi-\varpi'},
\label{Kahlerinst}
\eeq
where we abbreviated
\be
\cD_a\varpi=\frac{\d \varpi}{\varpi}\,\log\(1-\sigma(\gamma_a)e^{-2\pi \I \Xigi{a}(\varpi)}\).
\label{measure}
\ee
Here the K\"ahler potential is written in terms of the real coordinates on the moduli space.
To be applied for the computation of the metric, it should be reexpressed
as a function of the holomorphic
coordinates \eqref{defholcoor} which in our case can be taken as
\be
z^I
\qquad {\rm and}
\qquad
w_I\equiv \tzeta_I-F_{IJ}\zeta^J
-\frac{1}{8\pi^2}\sum_a \gnkl{\gamma_a}\(q_{a,I}-F_{IJ}(z)p_a^J\)\int_{\ellg{\gamma_a}}\!\! \cD_a\varpi.
\label{defholw}
\ee
Note that the quantity \eqref{Kahlerinst} is not explicitly symplectic invariant. Instead,
$K_{\Mthree}$ can be shown to transform by a K\"ahler transformation \cite{Alexandrov:2010pp}
so that it is perfectly consistent with the symplectic symmetry.

\subsection{Wall-crossing}
\label{subsec_wallcros}

So far we were working over a fixed point in the moduli space. Let us assume now that we are slowly changing
the moduli $z^I$. As a result, one changes also the phases of the central charges $Z_\gamma(z)$ and,
consequently, the position of BPS rays. It is natural to expect that at some point two BPS rays
may align and then exchange their order. On the other hand, it is easy to see that the symplectomorphisms
across the BPS rays are not commutative provided they correspond to non-mutually local charges,
$\langl\gamma,\gamma'\rangl\ne 0$. Thus, two different orders of BPS rays lead to two different
solutions for Darboux coordinates and, therefore, to two different metrics. But how is this possible
given the fact that the metric on our moduli space must be continuous?

To achieve the continuity of the metric, one should take into account the wall-crossing phenomenon
and, in fact, the behavior we have just observed is its nice geometric manifestation.
Indeed, it has been known for long time \cite{Argyres:1995gd,Ferrari:1996sv,Denef:2000nb}
that in both gauge theories and supergravity there are certain
walls of codimension 1 in the moduli space, called lines of marginal stability,
where the spectrum of single-particle BPS states discontinuously jumps.
This means that, crossing the wall, some BPS states either decay or on the contrary recombine to form a bound state.
For example, it is known that the distance between two centers in
a two-centered black hole solution is given by \cite{Denef:2000nb}
\be
r_{12}=\hf\, \frac{\langl\gamma_1,\gamma_2\rangl |Z_{\gamma_1}+Z_{\gamma_2}|}{\Im\(Z_{\gamma_1}\bZ_{\gamma_2}\)}.
\label{distance}
\ee
Due to the presence of the denominator,
the hypersurface in the moduli space where the two central charges, $Z_{\gamma_1}$ and $Z_{\gamma_2}$,
align splits the moduli space into two parts. In one part $r_{12}$ is positive and in the second it is negative.
But a negative distance is meaningless which indicates that in this region of the moduli space the two-centered
solution is unstable and decays.

This example illustrates the general fact that the lines of marginal stability appear where the phases
of two or more central charge functions coincide.
This is precisely the condition for the alignment of BPS rays in our twistor construction!
Thus, already at this point we see that the two phenomena are related two each other.

Furthermore, the information about the single-particle spectrum is contained in the BPS indices $\hnkl$,
which appear explicitly in our transition functions \eqref{prepH}. The wall-crossing implies
that, besides the charge, they carry also a piecewise-constant dependence on the moduli $z^I$ and jump across the lines
of marginal stability. Thus, there are two origins of discontinuity in our story:
the jump of $\hnkl$ and the reordering of BPS rays.
This gives a chance that they can cancel each other and the resulting metric would be continuous.
As was shown in \cite{Gaiotto:2008cd}, this is indeed the case.

In fact, this result is a direct consequence of a formula
suggested by Kontsevich and Soibelman \cite{ks} which determines $\hnkl$ on one side of the wall
if they are known on the opposite side.
Let us introduce operators
\be
U_\gamma=\exp\(\sum_{n=1}^\infty \frac{e_{n\gamma}}{n^2}\)
\ee
constructed from the generators of the Lie algebra of infinitesimal symplectomorphisms of the complex torus
which satisfy
\be
\[ e_\gamma, e_{\gamma'}\]=(-1)^{\langl\gamma,\gamma'\rangl}
\langl\gamma,\gamma'\rangl e_{\gamma+\gamma'}.
\ee
Then given two charges, $\gamma_1$ and $\gamma_2$, whose central charges align,
the KS formula states that the product of symplectomorphisms $U_\gamma^{\gnkl{\gamma;z}}$,
ordered according to decreasing of the phase of $Z_\gamma(z)$,
stays the same on the two sides of the wall, i.e.
\be
\label{wc}
\prod^\curvearrowleft_{\substack{\gamma=n \gamma_1+m \gamma_2\\m>0, n>0} }
U_{\gamma}^{\Omega^-(\gamma)} =
\prod^\curvearrowright_{\substack{\gamma=n \gamma_1+m \gamma_2\\m>0, n>0} }
U_{\gamma}^{\Omega^+(\gamma)}.
\ee
Identifying $e_\gamma=\sigma_{\text{D}}(\gamma) e^{-2\pi \I \Xikl}$, one recognizes in the operator $U_\gamma^{\gnkl{\gamma;z}}$
the exponentiated transition function \eqref{prepH}, whereas its adjoint action on $e_{\gamma'}$
reproduces the symplectomorphisms \eqref{KSgl}. Given these identifications, the KS formula is nothing else
but the statement that the wall-crossing does not change the symplectic structure of the twistor space.
As a result, the two effects mentioned above compensate each other and the resulting metric on the moduli space is continuous.

Although originally this interpretation of the KS wall-crossing formula has been developed
in the context of gauge theories and symplectic geometry, it can also be lifted to contact geometry
associated with \qk spaces \cite{Alexandrov:2011ac}. To this end, it is enough to show that the transformations
of the Darboux coordinate $\tilde\alpha$ induced by $U_\gamma$ satisfy a property similar to \eqref{wc}.
These transformations can be read off from \eqref{exline} or \eqref{glutildealp} and are given by the Rogers dilogarithm.
Then the corresponding wall-crossing formula becomes equivalent to one of the dilogarithm identities
and one concludes that our twistor construction consistently implements
the wall-crossing in the context of superstring theory.

\section{Integrability}

\subsection{Relation to TBA}
\label{subsec_releq}

It turns out that the instanton corrected moduli spaces described above,
in both cases of gauge and string theory,
are intimately related to an integrable structure.
This relation is based on the observation of \cite{Gaiotto:2008cd} that
the equations \eqref{eqXigi}, which encode geometric information about the twistor space
and its base manifold, turn out to coincide with the equations of Thermodynamic Bethe Ansatz,
very well known in the theory of two-dimensional integrable models \cite{Zamolodchikov:1989cf}.
To establish this relation, for each BPS particle of type $a$ one defines
\begin{itemize}
\item
rapidity parameter $\theta$ as
$\varpi=\I e^{\I \psi_a +\theta}$ where $\psi_a=\arg Z_{\gamma_a}$,

\item
spectral density
$\epsilon_a(\theta)=2\pi \I \(\Xigi{a}(\I e^{\I \psi_a +\theta})-\Theta_{\gamma_a}\)$,

\item
mass $\beta m_a =2\pi \tau_2|Z_{\gamma_a}|$ with $\beta$ being the inverse temperature,

\item
chemical potential
$\beta\mu_a=-2\pi \I \Theta_{\gamma_a}+\log(-\sigma(\gamma_a))$,

\item
kernel
\be
\phi_{ab}(\theta)=-\frac{\I}{2}\,\<\gamma_a,\gamma_b\>
\frac{e^{\theta}+e^{\I \(\psi_{b}-\psi_a\)}}{e^{\theta}-e^{\I \(\psi_{b}-\psi_a\)}},
\label{ident}
\ee
\end{itemize}
and denotes $n_a=\gnkl{\gamma_a}$.
Then all BPS rays are mapped to the real axis
and the integral equations \eqref{eqXigi} take the following form
\be
m_a \beta\cosh\theta=\epsilon_a(\theta)+\frac{1}{2\pi}\sum_{b}n_b \int_{-\infty}^\infty\d \theta'\,
\phi_{ab}(\theta-\theta')\log\(1+e^{\beta\mu_b-\epsilon_b(\theta')}\).
\label{Bethe}
\ee
These equations are identical to the standard TBA equations except for the factors $n_a$ which
can be interpreted as weights of particles of type $a$.

Although this is a curious observation, one can ask whether this is just a coincidence or
a reflection of some deep relation between the non-perturbative moduli spaces and integrability.
In the following subsections we will give two observations which put the observed relation on a firmer ground
and thus support the second option \cite{Alexandrov:2010pp}.

\subsection{Potentials, free energy and Yang--Yang functional}
\label{subsec_relpot}

First of all, if the above observation is more than a coincidence then
it should be possible to relate not only equations but also some physically or mathematically
interesting quantities.
Let us consider first the case of gauge theories. In this case the moduli space is HK
and the most natural quantity to consider is the K\"ahler potential.
Its full non-perturbative expression was given in \eqref{Kahlerinst}.
In \cite{Alexandrov:2010pp} it was found that if one extracts its symplectic invariant, instanton part,
it turns out to coincide
with an important quantity in the theory of TBA which is known as Yang--Yang functional.
This is a functional which generates the action principle for the TBA equations following
from it by varying with respect to the spectral densities.
It can be conveniently written as \cite{Nekrasov:2009rc}
\be
\begin{split}
\cW[\vph,\rho]=&\, \frac{1}{8\pi^2}\sum_{a,b}\int \de\theta\int \de\theta' \,
\phi_{ab}(\theta-\theta')\rho_a(\theta)\rho_b(\theta')
\\
&\qquad
+\frac{1}{2\pi}\sum_a \int\de \theta \[\rho_a(\theta)\vph_a(\theta)
-n_a\Li_2\(e^{\lambda_a(\theta)-\vph_a(\theta)}\)\].
\end{split}
\label{YYfun}
\ee
Here $\vph_a(\theta)$ is the interacting part of the spectral density and $\lambda_a(\theta)$ encodes
its free part together with the chemical potential
\be
\vph_a(\theta)=\eps_a(\theta)-m_a\beta\cosh\theta,
\qquad
\lambda_a(\theta) =\beta(\mu_a-m_a\cosh\theta)-\pi\I.
\label{identYY}
\ee
Varying \eqref{YYfun} with respect to $\vph_a$ and $\rho_a$ and using \eqref{identYY},
one indeed gets the TBA equations \eqref{Bethe}.
The critical value of the Yang--Yang functional is then given by
\beq
\cW_{\rm cr}&=&
-\frac{1}{8\pi^2}\sum_{a,b}n_a n_b\int \de\theta\int \de\theta' \,
\phi_{ab}(\theta-\theta')\log\(1-e^{\lambda_a(\theta)-\vph_a(\theta)}\)\log\(1-e^{\lambda_b(\theta')-\vph_b(\theta')}\)
\nn\\
&& -\frac{1}{2\pi}\sum_a n_a\int\de \theta\, \Li_2\(e^{\lambda_a(\theta)-\vph_a(\theta)}\).
\label{YYfuncrit}
\eeq
Comparison of this expression with the exact K\"ahler potential \eqref{Kahlerinst} of the gauge
theory moduli space reveals that it reproduces two instanton symplectic invariant terms of the latter.
Thus, one has the following relation
\be
\label{relYYK}
\begin{split}
\vphantom{\frac{A^A}{A_A}}
K_{\Mthree}= & \frac{R^2}{8}\, K(z,\bz)
-\frac{1}{4}\, N^{IJ}(w_I-\bw_I)(w_J-\bw_J)
\\
&
+\frac{1}{64\pi^4}\,\sum_{a,b}\gnkl{\gamma_a}\gnkl{\gamma_b}\cQ_{ab}
\int_{\ellg{\gamma_a}}\!\!\cD_a\varpi
\int_{\ellg{\gamma_b}}\!\!\cD_b\varpi'
+\frac{1}{8\pi^2}\,\cW_{\rm cr},
\vphantom{\frac{A^A}{A_A}}
\end{split}
\ee
where $N^{IJ}$ is the inverse of $N_{IJ}=-2\Im F_{IJ}$, $\cQ_{ab}$ is constructed
in terms of charges as
\be
\cQ_{ab}=\frac{1}{4}\, N_{IJ}p_a^I p_b^J
+N^{IJ}\(q_{a,I}-p_a^K\Re F_{IK}\)\(q_{b,J}-p_b^L\Re F_{JL}\),
\ee
and we rewrote several contributions in terms of the holomorphic coordinates $w_I$ defined in \eqref{defholw}.

Let us now turn to string theory. In this case the moduli space is QK and does not have
a K\"ahler potential or any other function which gives the metric in an easy way.
The most ``close" quantity of that kind is provided by the contact potential $\phi$.
We already saw that it plays an important role being, for example, associated with the four-dimensional
string coupling. Besides, it is also closely
related to the hyperk\"ahler potential on the corresponding HKC (see \eqref{chiphi}).
Due to these reasons, it is natural to expect
that it will play also a certain role in the correspondence with integrable systems.
And this indeed turns out to be the case:
it is trivial to see that the instanton part of the contact potential \eqref{phiinstmany}
coincides with the free energy of the integrable system given by
\be
\cF(\beta)=\frac{\beta}{2\pi}\sum_{a} m_a \int_{-\infty}^\infty\d \theta\,
\cosh\theta\,\log\(1+e^{\beta\mu_a-\epsilon_a(\theta)}\),
\label{grstener}
\ee
so that one has the following relation
\be
\label{identener}
e^{\phi}=  \frac{\tau_2^2}{16}\, K(z,\bz)+\frac{\chi_X}{192\pi}-\frac{1}{32\pi^2}\,\cF(\beta).
\ee

Thus, in both cases we have a relation between fundamental quantities characterizing the geometry
and physics of the model. Moreover, given the form of \eqref{relYYK} and \eqref{identener} and the fact that
the free energy is usually considered as one of the most important quantities in the context of TBA,
the string theory relation looks somewhat
simpler and more fundamental than the gauge theory one.
This indicates that may be the full string theory result will still be consistent with integrability.

\subsection{$S$-matrix}
\label{subsubsec_Smat}

The second observation concerns the integrability of the $S$-matrix corresponding to the TBA \eqref{Bethe}.
In integrable two-dimensional models the $S$-matrix is factorizable and is completely determined
by its two-particle sector. The two-particle $S$-matrix in turn
can be obtained from the kernel of the integral TBA equation because
$\phi_{ab}(\theta)$ is known to be given by its logarithmic derivative:
$\phi_{ab}(\theta)=-\I\frac{\p\log S_{ab}}{\p\theta}$.
Integrating \eqref{ident}, one finds the following result
\be
S_{ab}(\theta)\sim\[\sinh\(\hf\(\theta+\I(\psi_a-\psi_b)\)\)
\]^{\<\gamma_a,\gamma_b\>}.
\label{Smat}
\ee
As was noticed in \cite{Gaiotto:2008cd},
this $S$-matrix is non-unitary. However, it is not necessarily a problem
since integrability does not require unitarity.
On the other hand, it imposes on the $S$-matrix a set of severe conditions.
This is their complete list:
\begin{itemize}
\item {\it Lorentz invariance} --- It implies that the $S$-matrix should depend only on the rapidity difference
of two particles $\theta=\theta_a-\theta_b$.
\item {\it Zamolodchikov algebra} --- It means that the particle creation operators must satisfy
$\Phi_a(\theta)\Phi_b(\theta')=S_{ab}(\theta-\theta')\Phi_{b}(\theta')\Phi_a(\theta)$. Applying this identity twice,
one gets the following restriction on the $S$-matrix
\be
S_{ab}(\theta)S_{ba}(-\theta)=1.
\label{ZamalgS}
\ee

\item {\it Crossing symmetry} --- It relates the scattering in the $s$- and $t$-channels and requires that
\be
S_{b\ba}(\pi \I-\theta)= S_{ab}(\theta),
\ee
where $\ba$ denotes the antiparticle of a particle of type $a$.

\item {\it Yang-Baxter equation} ---
It means that the order in which the particles are scattered should not matter and can be depicted as
follows
\\
\unitlength .3mm 
\linethickness{0.4pt}
\ifx\plotpoint\undefined\newsavebox{\plotpoint}\fi 
\begin{picture}(310,152)(-100,30)
\thinlines
\put(20,145){\vector(1,-1){120}}
\put(190,145){\vector(1,-1){120}}
\put(139.5,144.75){\vector(-1,-1){120}}
\put(309.5,144.75){\vector(-1,-1){120}}
\put(47.5,145){\vector(0,-1){120}}
\put(282.5,145){\vector(0,-1){120}}
\put(18.75,151){\makebox(0,0)[cc]{$a$}}
\put(188.75,151){\makebox(0,0)[cc]{$a$}}
\put(47,152){\makebox(0,0)[cc]{$b$}}
\put(282,152){\makebox(0,0)[cc]{$b$}}
\put(139,151){\makebox(0,0)[cc]{$c$}}
\put(309,151){\makebox(0,0)[cc]{$c$}}
\thinlines
\qbezier(47.5,111.25)(50.75,110.375)(52,113)
\qbezier(282,111.25)(278.75,110.375)(277.5,113)
\qbezier(47.5,45.25)(42.875,42.875)(42.75,48)
\qbezier(282.5,45.25)(287.125,42.875)(287.25,48)
\qbezier(75.75,80.75)(80,76.25)(84.25,80.75)
\qbezier(245.75,80.75)(250,76.25)(254.25,80.75)
\put(53.5,100.75){\makebox(0,0)[cc]{$\theta$}}
\put(277.5,100.75){\makebox(0,0)[cc]{$\theta'$}}
\put(43.5,37.5){\makebox(0,0)[cc]{$\theta'$}}
\put(288.5,37.5){\makebox(0,0)[cc]{$\theta$}}
\put(80,63.75){\makebox(0,0)[cc]{$\theta+\theta'$}}
\put(250,63.75){\makebox(0,0)[cc]{$\theta+\theta'$}}
\put(164.75,84.75){\makebox(0,0)[cc]{$=$}}
\end{picture}
\be
S_{a b}(\theta)S_{a c}(\theta+\theta')S_{b c}(\theta')
=S_{b c}(\theta')S_{a c}(\theta+\theta')S_{a b}(\theta),
\label{YBeq}
\ee

\item {\it Bootstrap identity} --- This is the most non-trivial requirement on the $S$-matrix which relates
its singularity structure to the spectrum. Namely, it demands that if $S_{ab}(\theta)$ has
a pole in the physical strip, {\it i.e.}, at $\theta=\I u_{ab}^{c}$ where $u_{ab}^c\in(0,\pi)$, then
the spectrum should contain a bound state $\bc$
with the mass
\be
m_{\bc}^2=m_a^2+m_b^2+2m_a m_b \cos u_{ab}^c,
\label{massc}
\ee
appearing in the fusing process $a+b\to \bc$.
But the most important condition is that it does not matter whether an additional particle, say $d$,
scatters with the bound state $\bc$ or consequently with the two particles $a,b$,
which leads to the following constraint
\\
\unitlength .3mm 
\linethickness{0.4pt}
\ifx\plotpoint\undefined\newsavebox{\plotpoint}\fi 
\begin{picture}(310,152)(-100,30)
\thinlines
\put(70,145){\line(2,-3){30}}
\put(200,145){\line(2,-3){30}}
\put(130,145){\line(-2,-3){30}}
\put(260,145){\line(-2,-3){30}}
\put(100,80){\vector(2,-3){30}}
\put(230,80){\vector(2,-3){30}}
\put(100,80){\vector(-2,-3){30}}
\put(230,80){\vector(-2,-3){30}}
\put(50,145){\vector(2,-1){100}}
\put(180,115){\vector(2,-1){100}}
\put(100,100){\line(0,-1){20}}
\put(230,100){\line(0,-1){20}}
\put(68.75,151){\makebox(0,0)[cc]{$a$}}
\put(198.75,151){\makebox(0,0)[cc]{$a$}}
\put(47,152){\makebox(0,0)[cc]{$d$}}
\put(180,122){\makebox(0,0)[cc]{$d$}}
\put(260,152){\makebox(0,0)[cc]{$b$}}
\put(130,152){\makebox(0,0)[cc]{$b$}}
\put(108,94){\makebox(0,0)[cc]{$\bc$}}
\put(238,94){\makebox(0,0)[cc]{$\bc$}}
\thinlines
\put(164.75,84.75){\makebox(0,0)[cc]{$=$}}
\end{picture}
\be
S_{da}(\theta-\I \bu_{ca}^b)S_{db}(\theta+\I \bu_{bc}^a)=S_{d\bc}(\theta),
\qquad
\bu_{ab}^c=\pi - u_{ab}^c.
\label{bootstr}
\ee

\end{itemize}

In \cite{Alexandrov:2010pp} it was shown that all these conditions are satisfied.
In particular, note that the poles of the $S$-matrix \eqref{Smat} correspond to
$u_{ab}^c= \psi_b-\psi_a$.
The mass formula \eqref{massc} together with its expression in terms of the central charge
$Z_\gamma$ yields
\be
\beta m_{\bc}=2\pi \tau_2|Z_{\gamma_a+\gamma_b}|,
\ee
which implies that the bound state $\bc$ has charge $\gamma_a+\gamma_b$ consistently with the physical
interpretation. Moreover, upon changing the moduli, the fusing angles $u_{ab}^c$ may
penetrate from the physical strip to a non-physical one or {\it vice versa}. This happens at
$\psi_a=\psi_b$ which precisely corresponds to a line of marginal stability!
Thus, there is a direct link between stability of BPS states and positions of fusing angles on the complex plane.

We believe that all these observations strongly indicate in favor of a hidden integrable structure
in the geometry of the instanton corrected moduli spaces.

\subsection{Y-system and unusual features of TBA}

TBA equations are integral equations on spectral densities.
Quite often it is useful to rewrite them in the form of finite-difference equations
which are known as Y-system (see \cite{Kuniba:2010ir} for a review).
For this purpose one introduces the Y-functions
\be
Y_a(\theta)=-e^{\beta\mu_a-\epsilon_a(\theta)}
\ee
and, using $\psi_{\ba}=\psi_a+\pi$ and $\Theta_{\ba}=-\Theta_a$,
one easily verifies that our TBA leads to the following relations \cite{Alexandrov:2010pp}
\be
Y_a \(\theta+\frac{\pi \I}{2}\)Y_{\ba} \(\theta-\frac{\pi \I}{2}\)
=\prod_{\ba<b<a}\Bigl[1-Y_b\bigl(\theta+\I\({\textstyle{\pi\over 2}}+\psi_a-\psi_b\)\bigr)\Bigr]^{\hng{b}\<\gamma_a,\gamma_b\>}.
\label{Ysys}
\ee
This is the Y-system for the case at hand. For a general configuration of charges it is extremely complicated
and possesses a few unusual features:
\begin{itemize}
\item First, on the l.h.s. of \eqref{Ysys} one multiplies functions associated with a particle
and its antiparticle, whereas usually one has only one function.

\item
Second, on the r.h.s. the Y-functions are all evaluated at different points, whereas usually their arguments
do not contain any shifts. Moreover, usually the fusing angles and the only shifts appearing in functional relations
are rational multiples of $\pi$. Here they are completely arbitrary and vary continuously with
the moduli $z^I$.

\item Third, usually the power of each factor in the product on the r.h.s is related to the incidence matrix of a
graph which structure is severely constrained by the periodicity of the Y-system.
This incidence matrix is equivalently described by the matrix
$N_{ab}=\int^{+\infty}_{-\infty} \phi_{ab}(\theta)d\theta$. In our case it is just not defined
because the kernel is not integrable.

\end{itemize}

In fact, these features can be traced back to the following properties of the original TBA:
\begin{itemize}
\item
The spectral densities and Y-functions satisfy somewhat unusual reality conditions
\be
\bar\epsilon_a(\theta)=\epsilon_{\ba}(-\theta),
\qquad
\bY_a(\theta)=Y_{\ba}(-\theta).
\label{realY}
\ee
As a result, there is no parity symmetry and the Y-functions are not necessarily
real on the real axis of the spectral parameter.
\item
The kernel $\phi_{ab}(\theta)$ \eqref{ident} is not decaying at infinity.
\item
The system is supplied by arbitrary imaginary chemical potentials $\mu_a$.
\end{itemize}
Altogether, these features make our TBA much more complicated than one usually encounters
in two-dimensional integrable models. In particular, the techniques developed to extract
the free energy and Y-functions in the high-temperature limit, which corresponds either
to a small compactification radius in gauge theory or to the strong coupling limit in string theory,
become inapplicable. To generalize these techniques to the present context
represents an interesting mathematical problem.

\chapter{Quantum mirror symmetry and S-duality}
\label{chap_IIB}

\section{Type IIB formulation}
\label{sec_IIBform}

So far we considered the HM moduli space in type IIA string theory. Let us now turn to the
type IIB formulation compactified on the mirror CY $\CYm$. Although these two formulations
should be equivalent according to mirror symmetry, it is of great importance to have both of them at our disposal.
There are many issues which are difficult to deal with in one framework, but which become easy
in another. Besides, the properties of their D-branes and their explicit symmetries are also quite different.
In particular, type IIB theory provides us with the powerful S-duality, which was indispensable
in getting many of the results presented in this review. Finally, at the full quantum level,
mirror symmetry itself is so far only a conjecture and needs to be proven.

Let us therefore, before going into details,
summarize the main differences between the two mirror formulations.
The first difference is the physical nature of coordinates on the moduli space.
From Table \ref{HMfields} one observes that in type IIB the complex structure moduli $z^a$ are replaced
by the complexified K\"ahler moduli $v^a$ \eqref{coorCK}\footnote{We use the indices $a,b$
to label coordinates on $\KK$ because we work on the mirror Calabi-Yau $\CYm$ for which $h^{1,1}(\CYm)=h^{2,1}(\CY)$.
Thus, this is not in contradiction with our conventions spelled in footnote \ref{foot_indices}.},
whereas the periods of the RR 3-form potential are replaced by periods
of even-dimensional RR forms (the precise relation between the periods
and the fields $c^0, c^a, c_a, c_0$ will be given in \eqref{RRiib}).
Note that it is convenient to combine one of these potentials, $\tau_1\equiv c^0$,
with the ten-dimensional string coupling $\tau_2\equiv 1/g_s$
into the ten-dimensional axio-dilaton field $\tau = \tau_1+\I \tau_2$.

The second difference is the symmetry which characterizes each particular formulation.
Whereas type IIA theory is explicitly invariant under symplectic transformations,
the characteristic feature of type IIB is the presence of $SL(2,\IZ)$ duality.
In particular, the physical fields should form a representation of this duality group.
Therefore, if mirror symmetry holds at quantum level, our constructions
done in the type IIA framework must have a hidden $SL(2,\IZ)$ symmetry which
should become explicit after passing to the type IIB formulation.

Finally, recall that D-branes in type IIB have even-dimensional world-volumes
and therefore they give rise to instanton effects by wrapping even-dimensional cycles of Calabi-Yau.
Thus, there is a series of instanton contributions to the moduli space metric
coming from D(-1), D1, D3 and D5-branes as well as from the NS5-brane pertaining to both formulations.
Although all D-instanton corrections should follow by mirror symmetry from the construction of
the previous chapter, to write them explicitly in the type IIB framework is an extremely difficult problem.
The main obstacle is that it requires the knowledge of the mirror map at the non-perturbative level.
As we will show, this problem has been solved only partially.
But before entering this subject, we need to understand better some of the type IIB features mentioned above.

\subsection{D-branes and charge quantization}
\label{sec_chargeQ}

The first problem we are going to discuss is how to characterize
D-brane instantons on the type IIB side. It turns out that the picture
where some $n$-dimensional objects wrap $n$-dimensional non-trivial cycles
and are classified by integers enumerating these cycles is too naive.
The correct mathematical description associates
BPS D-instantons to elements in
the derived category of coherent sheaves $\mathcal{D}(\CYm)$
\cite{MR1403918,Douglas:2000gi}.
In particular, for non-vanishing D5-brane charge $p^0$, a D5-D3-D1-D(-1)-instanton
configuration can be represented
as a coherent sheaf $E$ on $\CYm$, of rank $\rk(E)=p^0$.
The charge vector classifying such configurations is given by
the expansion of the so called generalized Mukai vector in the basis
of even-dimensional cohomology \cite{Minasian:1997mm}
\be
\label{mukaimap}
\gamma' \equiv \ch(E)\sqrt{\Td(\CYm)} = p^0 + p^a \omega_a - q'_a \omega^a + q'_0 \omega_{\CYm} ,
\ee
where $\ch(E)$ and $\Td(\CYm)$ are the Chern character of $E$ and Todd class of
$T\CYm$. This leads to the following expressions for the charges
\be
\begin{split}
& p^a =\int_{\gamma^a} c_1(E) \, ,
\qquad
q'_a =-\left( \int_{\gamma_a} \ch_2(E) +   \frac{p^0}{24}  c_{2,a}\right),
\\
& \qquad\quad q'_0 = \int_{\CYm}\left(\ch_3(E)+\frac{1}{24}\, c_1(E)\,c_2({\CYm})
\right) ,
\end{split}
\label{ElectricChargesD5}
\ee
which explicitly shows that the charges $q'_\Lambda$ are not integer.
This becomes problematic regarding applications to mirror symmetry.
Indeed, the charges in the type IIA formulation are classified by $H_3( X,\IZ)$ and hence integer.
Thus, there seems to be a clash between the two charge lattices and one needs to understand
how they are related to each other.

This problem has been solved in \cite{Alexandrov:2010ca}.
The idea is to compare the central charge associated to a D-instanton and given in type IIB by
\be
Z_{\gamma'}=\int_{\CYm} e^{-v^a\omega_a}\gamma',
\ee
with the central charge \eqref{defZ} known from type IIA theory where the prepotential should be
restricted to its large volume limit
\be
\label{prepcl}
\Fcl(X)=-\kappa_{abc}\frac{X^a X^b X^c}{6 X^0}+ \frac12\, A_{\Lambda\Sigma} X^\Lambda X^\Sigma.
\ee
We emphasize that it is important to keep the quadratic term despite it does not affect the K\"ahler potential on $\KK$.
Then, upon identification of the moduli $z^a=v^a$, the two expressions coincide provided
\be
q'_\Lambda = q_\Lambda -A_{\Lambda\Sigma} p^\Sigma.
\label{chargeshift}
\ee
Thus, this is the matrix $A_{\Lambda\Sigma}$ that makes possible the conversion of rational charges into
integer ones.
However, this conversion works only if the matrix satisfies a set of conditions.
Their full list reads as follows \cite{Alexandrov:2010ca}:
\be
\begin{split}
i)&\ A_{00}\in \IZ,
\\
ii)&\ A_{0a}\in \frac{c_{2,a}}{24} + \IZ ,
\\
iii)&\ A_{ab}  p^b- \frac12\, \kappa_{abc} p^b p^c  \in \IZ\quad {\rm for}\ \forall p^a\in\IZ ,
\\
iv)&\  \frac{1}{6}\,\kappa_{abc}p^a p^b p^c+\frac1{12}\,c_{2,a}p^a\in\IZ \quad {\rm for}\ \forall p^a\in\IZ .
\end{split}
\label{condA1}
\ee
Although the last two conditions look completely non-trivial, remarkably
they are indeed fulfilled in few examples where $A_{\Lambda\Sigma}$ has actually been computed.
For example, on the quintic Calabi-Yau one has
\be
\kappa_{111}= 5,
\qquad
A_{11}=-\frac{11}{2},
\qquad
c_{2,1}=50
\ee
and it is amusing to check that \eqref{condA1} does hold. In general, these conditions can be considered
as a consequence of mirror symmetry. They ensure that the charge vector
$\gamma=(p^\Lambda,q_\Lambda)$ is valued in  $H^{\rm even}(\cX,\IZ)$
and can be meaningfully identified with the homology class of a special Lagrangian submanifold on the mirror
type IIA side. In contrast, the primed charges \eqref{ElectricChargesD5} belong to a shifted lattice
\be
q'_a\in \mathbb{Z} -\frac{p^{0}}{24}\, c_{2,a} - \frac12 \kappa_{abc} p^b p^c,
\qquad
q'_0\in \mathbb{Z}-\frac{1}{24}\, p^{a} c_{2,a} .
\label{fractionalshiftsD5}
\ee

\subsection{S-duality and mirror symmetry}
\label{subsec_Sdualmir}

We had to start with the subtle issue of quantization of D-brane charges, even
before considering the type IIB formulation at the perturbative level, because it
turns out that the shift by the matrix $A_{\Lambda\Sigma}$ \eqref{chargeshift}
revealed above has important implications on the mirror map and S-duality transformations.

To see how this comes about, let us first establish how mirror symmetry relates
the periods of RR potentials on type IIA and type IIB sides. All these periods
appear linearly in the imaginary part of the D-instanton action whose real part is determined by the central charge.
Thus, it is natural to find the relation by equating D-instanton actions in type IIA and in type IIB theories.
In the former case the action is given in \eqref{d2quali}
and in the latter case it is known to be \cite{Marino:1999af,Aspinwall:2004jr}
\be
S_{\text{D-IIB}}=2\pi g_s^{-1} |Z_{\gamma'}|
+2\pi\I\int_{\CYm}\gamma'\wedge \hA_{\rm even} e^{-\hB_2}.
\label{Dbraneact}
\ee
Decomposing the RR potential $\hA_{\rm even}$ as
\be
\label{ABze}
\hA_{\rm even} e^{-\hB_2}
= \zeta^0 - \zeta^a \omega_a - \tzeta'_a \omega^a-\tzeta_0' \omega_{\CYm}
\ee
and using the identification \eqref{chargeshift} between the charges,
one finds the same axionic couplings as in \eqref{d2quali} provided the primed periods are shifted
in a way similar to the charges
\be
\tzeta'_\Lambda = \tzeta_\Lambda - A_{\Lambda\Sigma} \zeta^\Sigma.
\label{RRshift}
\ee

This result indicates that all symplectic vectors experience such a shift under mirror symmetry.
This shift corresponds to a symplectic transformation whose meaning is to remove
the quadratic term from the prepotential \eqref{lve} (or \eqref{prepcl})
\be
F'(X) = F(X) - \frac12 A_{\Lambda\Sigma} X^\Lambda X^\Sigma.
\ee
Thus, in terms of primed variables, in the small coupling, large volume limit the metric on the HM moduli space
is obtained from \eqref{hypmettree} by plugging in the cubic prepotential \eqref{prepcubic}.

However, this is not yet what we want. We need a formulation which is explicitly $SL(2,\IR)$ invariant.
It can be achieved by a further change of variables which provides the {\it classical mirror map}
\cite{Bohm:1999uk}
\be
\label{symptob}
\begin{split}
z^a&=b^a+\I t^a,
\qquad\qquad
\zeta^0=\tau_1 ,
\qquad\qquad
\zeta^a = - (c^a - \tau_1 b^a) ,
\\
\tzeta'_a &=  c_a+ \frac{1}{2}\, \kappa_{abc} \,b^b (c^c - \tau_1 b^c) ,
\qquad
\tzeta'_0 =\, c_0-\frac{1}{6}\, \kappa_{abc} \,b^a b^b (c^c-\tau_1 b^c) ,
\\
\sigma &= -2 (\psi+\frac12  \tau_1 c_0) + c_a (c^a - \tau_1 b^a)
-\frac{1}{6}\,\kappa_{abc} \, b^a c^b (c^c - \tau_1 b^c).
\end{split}
\ee
The advantage of the variables appearing on the r.h.s. of \eqref{symptob} is that they are adapted
to the action of $SL(2,\IR)$ symmetry.
Under a transformation
\be
\label{Sdualde2}
\gsl = \begin{pmatrix} a & b \\ c & d \end{pmatrix}
\ee
with $ad-bc=1$, they transform as
\be
\label{SL2Z}
\begin{split}
&\quad \tau \mapsto \frac{a \tau +b}{c \tau + d} ,
\qquad
t^a \mapsto t^a |c\tau+d| ,
\qquad
c_a\mapsto c_a- c_{2,a}\eps(\gsl)
,
\\
&\quad
\begin{pmatrix} c^a \\ b^a \end{pmatrix} \mapsto
\begin{pmatrix} a & b \\ c & d  \end{pmatrix}
\begin{pmatrix} c^a \\ b^a \end{pmatrix} ,
\qquad
\begin{pmatrix} c_0 \\ \psi \end{pmatrix} \mapsto
\begin{pmatrix} d & -c \\ -b & a  \end{pmatrix}
\begin{pmatrix} c_0 \\ \psi \end{pmatrix}.
\end{split}
\ee
It is straightforward to verify that the classical HM metric obtained by the change of variables \eqref{symptob}
is indeed invariant under the action \eqref{SL2Z}.

Some comments are in order:
\begin{itemize}
\item
The continuous $SL(2,\IR)$ symmetry holds only classically. Already $\alpha'$-corrections
to the holomorphic prepotential given by the last two terms in \eqref{lve} break it.
However, the discrete subgroup $SL(2,\IZ)$ can be restored by including D(-1) and D1-instanton corrections
\cite{RoblesLlana:2006is} and the full non-perturbative metric on $\MH$ is also expected
to preserve this symmetry.
\item
Combining the classical mirror map \eqref{symptob} with the expansion \eqref{ABze},
one finds the following definitions of the RR scalars in terms of period integrals
\cite{Louis:2002ny,Alexandrov:2008gh}
\be
\label{RRiib}
\begin{split}
& c^0= \hA_0,
\qquad
c^a=\int_{\gamma^a} \hA_2,\qquad
c_a=- \int_{\gamma_a} (\hA_4 - \frac{1}{2}\, \hB_2 \wedge \hA_2) ,
\\
& \qquad
c_0=-\int_{\CYm} (\hA_6 - \hB_2 \wedge  \hA_4 + \frac{1}{3}\, \hB_2 \wedge \hB_2 \wedge \hA_2 ).
\end{split}
\ee
However, these expressions are meaningful only at the classical level.
In general, we define the type IIB physical fields as those which transform under
$SL(2,\IZ)$ according to the law \eqref{SL2Z}.
Of course, this makes possible that the mirror map \eqref{symptob} receives quantum
corrections and one of our aims is to show how they can be obtained.
\item
The S-duality transformation \eqref{SL2Z} differs from the standard one \cite{Gunther:1998sc,Bohm:1999uk}
by the presence of a constant shift in the transformation law of the D3-axion $c_a$.
The shift is taken to be proportional to the multiplier system $\eps(\gsl)$ of the Dedekind eta function $\eta(\tau)$
which is a fractional number defined, up to the addition of integers, by the ratio
\be
\label{multeta}
\eta\left(\frac{a\tau+b}{c\tau+d}\right)\slash \eta(\tau)
=e^{2\pi\I \eps(\gsl)}(c\tau+d)^{1/2} .
\ee
In particular, the phase factor $e^{2\pi\I \eps(\gsl)}$ is a 24th root of unity
(cf. fractional terms in the quantization condition \eqref{fractionalshiftsD5}).
The necessity to introduce this shift can be understood from two reasons which are both related
to the effects induced by the matrix $A_{\Lambda\Sigma}$ \cite{Alexandrov:2010ca}.
First, S-duality and the Heisenberg symmetry \eqref{Heisenb} are not completely uncorrelated.
The transformation generated by $\gsl_b={\scriptsize \begin{pmatrix} 1 & b \\ 0 & 1  \end{pmatrix}}$
should actually be the same as the Heisenberg shift with parameter $\eta^0=b$.
But without transforming $c_a$, the mismatch \eqref{RRshift} between primed (used in S-duality)
and unprimed (used in Heisenberg symmetry) variables makes these two transformations different.
Although this coincidence might seem to be not crucial, the second issue shows that it is really needed.
The problem is that, in the absence of the shift of $c_a$, the axionic couplings coming with D-instanton
contributions are not invariant under the same transformation $\gsl_b$ due to the fractional nature of
the primed charges $q'_\Lambda$. If this was the case, this would spoil the $SL(2,\IZ)$ symmetry.
Fortunately, the correction of the S-duality transformation of $c_a$ cures both these problems.
\end{itemize}

\subsection{Monodromy invariance and spectral flow symmetry}

There is another important symmetry which must be taken into account in
the non-perturbative analysis of the HM moduli space. This is monodromy invariance
originating from the possibility to go around singularities in the moduli space of
complexified K\"ahler deformations $\KK(\CYm)$.
Since the metric on $\MH$ must be regular, such monodromies should leave it invariant.
Here we are particularly interested in the monodromy around the large volume point which
is induced by a large gauge transformation of the B-field acting as
\be
\label{monb}
b^a\mapsto b^a+\epsilon^a\ ,\qquad \epsilon^a\in \IZ.
\ee
Moreover, due to the presence of the B-field in the definition of the RR scalars \eqref{ABze},
the monodromy is accompanied by the symplectic rotation
\be
\label{bjacr}
\begin{split}
\zeta^a\mapsto \zeta^a + \epsilon^a \zeta^0,
\qquad
\tzeta'_a\mapsto \tzeta'_a -\kappa_{abc}\zeta^b \epsilon^c
-\frac12\,\kappa_{abc} \epsilon^b \epsilon^c \zeta^0 ,
\\
\tzeta'_0\mapsto \tzeta'_0 -\tzeta'_a \epsilon^a+\frac12\, \kappa_{abc}\zeta^a \epsilon^b \epsilon^c
+\frac16\,\kappa_{abc} \epsilon^a \epsilon^b \epsilon^c \zeta^0.
\qquad
\end{split}
\ee
Note that being rewritten in terms of the unprimed RR-fields,
i.e. composed with the symplectic transformation \eqref{RRshift},
the symplectic matrix becomes integer valued. Indeed, an easy calculation yields \cite{Alexandrov:2010ca}
\be
\label{monosymp}
\( {\zeta^\Lambda\atop \tzeta_\Lambda} \)\mapsto\rho(M)\cdot\( {\zeta^\Lambda\atop \tzeta_\Lambda} \),
\qquad
\rho(M)=\(\begin{array}{cccc}
1\ & 0 & 0 & 0
\\
\epsilon^a & {\delta^a}_b & 0 & 0
\\
-L_a(\epsilon)\ & -\kappa_{abc}\epsilon^c & {\delta_a}^b & 0
\\
\ L_0(\epsilon) & L_b(\epsilon)+2A_{bc}\epsilon^c & \ -\epsilon^b\  & 1
\end{array}\) ,
\ee
where we introduced two functions
\be
\label{defL0La}
L_0(\epsilon^a)=  \frac{1}{6}\,\kappa_{abc}\epsilon^a \epsilon^b \epsilon^c+\frac1{12}\,c_{2,a}\, \epsilon^a ,
\qquad
L_a(\epsilon^a)= \frac12\, \kappa_{abc} \epsilon^b \epsilon^c - A_{ab}  \epsilon^b.
\ee
Due to the conditions \eqref{condA1}, both these functions are integer valued for $\epsilon^a\in \IZ$
which ensures our statement.

A nice fact is that the D-instanton action \eqref{Dbraneact} stays invariant under the above monodromy transformation
if it is supplemented by a similar transformation of charges
\be
\gamma\mapsto \gamma[\epsilon]=\rho(M)\cdot\gamma.
\label{chargesymp}
\ee
Since the symplectic matrix is integer valued, the transformation preserves the charge lattice
and is therefore admissible. In the following we call it {\it spectral flow}.

Due to the invariance of the D-instanton action, the combined action of the monodromy and the spectral flow
preserves also the BPS indices $\gnkl{\gamma;v}$, where we indicated explicitly
their piecewise constant dependence on the K\"ahler moduli. This dependence is important since
the spectral flow alone in general does not leave the indices invariant \cite{Manschot:2009ia}:
the transformation can change the position of walls of marginal stability and a given
point in the moduli space may turn out to fall in a different chamber.

Given this ``spectral flow symmetry",
it is convenient also to introduce combinations of charges which are invariant under the spectral flow.
Such combinations are easily computed and are given by \cite{Alexandrov:2010ca,amp-in-progress}
\be
\begin{split}
p^0\ne 0: & \qquad
\hat q_a= q'_a + \frac12\, \kappa_{abc} \frac{p^b p^c}{p^0},
\qquad
\hat q_0 =
q'_0 +\frac{p^a q'_a}{p^0}+ \frac13\, \kappa_{abc} \frac{p^a p^b p^c}{(p^0)^2},
\\
{p^0=0 \atop p^a\ne 0}: &\qquad
\hat q_0 =
q_0' -\frac12 \kappa^{ab} q'_a q'_b,
\end{split}
\label{invcharge}
\ee
where we distinguished two cases and introduced the matrix $\kappa_{ab}\equiv \kappa_{abc} p^c$
together with its inverse $\kappa^{ab}$.
If not the dependence on the moduli, one could say that the BPS indices are functions of these
charges only. Finally, it is important to note that for $p^0=0$ the invariant charge is bounded from above,
\be
\label{boundqhat0}
\hat q_0 \leq \hat q_0^{\rm max}=
\frac{1}{24} \( \kappa_{abc} p^a p^b p^c + c_{2,a} p^a\).
\ee

\section{S-duality in twistor space}
\label{sec_Sdual}

The characteristic feature of the type IIB formulation is the S-duality symmetry.
In our terms it is generated by the transformation \eqref{SL2Z} of the type IIB physical fields
parametrizing the HM moduli space.
However, as we saw in the previous chapters, quantum corrections to the HM metric
are naturally formulated at the twistor space level.
Therefore, it is natural to ask how S-duality acts on the twistor space and which
constraints it imposes on the twistor construction.
In this section we answer these questions and moreover provide a reformulation of the D-instanton
corrected twistor space of chapter \ref{chap_Dbrane} which explicitly
respects S-duality.

\subsection{Classical S-duality}

We start from the classical twistor space which is described by the Darboux coordinates \eqref{gentwi}
with the classical prepotential \eqref{prepcl}. As usual, it is convenient to define the primed
variables
\be
\txi'_\Lambda = \txi_\Lambda - A_{\Lambda\Sigma} \xi^\Sigma,
\qquad
\alpha'=\alpha-\hf\, A_{\Lambda\Sigma}\xi^\Lambda\xi^\Sigma.
\label{txishift}
\ee
In the patch $\cU_0$ they are given by the same expressions but with $\tzeta_\Lambda$ replaced by $\tzeta'_\Lambda$ and
with the prepotential given by ${\Fclp}=-\frac{1}{6}\,\frac{\kappa_{abc}X^a X^b X^c}{X^0}$.

As was mentioned in section \ref{chap_quatern}.\ref{subsec_toricQK},
all isometries of a QK manifold can be lifted to a holomorphic action on its twistor space.
In particular, this implies that S-duality, represented in the classical approximation by
the $SL(2,\IR)$ symmetry group, should have a holomorphic representation on the Darboux coordinates
$\xi^\Lambda,\txi_\Lambda$ and $\alpha$. This indeed turns out to be the case provided
the fiber coordinate $\varpi$ transforms as
\be
\varpi  \mapsto  \frac{1+\tzm\varpi}{\tzm-\varpi}=
-\frac{\tzp-\varpi}{1+\tzp\varpi} ,
\label{trans-t}
\ee
where $\tzpm$ denote the two roots of $c\xi^0(\varpi)+d=0$
\be
\tzpm = \frac{ c \tau_1 + d \mp | c\tau + d |}{c \tau_2} ,
\qquad
\tzp \tzm = -1.
\label{poles}
\ee
Then the transformations \eqref{SL2Z}, the mirror map \eqref{symptob}
and the explicit expressions for the Darboux coordinates in the classical limit in terms of the type IIA fields
lead to the following non-linear holomorphic action in the patch $\cU_0$ \cite{Alexandrov:2008gh}
\be
\label{SL2Zxi}
\begin{split}
&
\xi^0 \mapsto \frac{a \xi^0 +b}{c \xi^0 + d} , \qquad
\xi^a \mapsto \frac{\xi^a}{c\xi^0+d} , \qquad
\txi'_a \mapsto \txi'_a +  \frac{ c}{2(c \xi^0+d)} \kappa_{abc} \xi^b \xi^c-c_{2,a}\eps(\gsl)
,
\\
&
\begin{pmatrix} \txi'_0 \\ \alpha' \end{pmatrix} \mapsto
\begin{pmatrix} d & -c \\ -b & a  \end{pmatrix}
\begin{pmatrix} \txi'_0 \\  \alpha' \end{pmatrix}
+ \frac{1}{6}\, \kappa_{abc} \xi^a\xi^b\xi^c
\begin{pmatrix}
c^2/(c \xi^0+d)\\
-[ c^2 (a\xi^0 + b)+2 c] / (c \xi^0+d)^2
\end{pmatrix} .
\end{split}
\ee
Moreover, one finds that
\be
e^\phi \mapsto \frac{e^\phi}{|c\tau+d|},
\qquad
K_{\cZ}\mapsto  K_\cZ - \log(|c\xi^0+d| ),
\qquad
\hCX\ui{i}\mapsto \frac{\hCX\ui{i}}{c\xi^0+d}
\label{SL2phi}
\ee
so that the \kahler potential varies by a \kahler transformation
and the complex contact structure stays invariant.
This ensures that S-duality is indeed a symmetry of the classical twistor space.

\subsection{Inclusion of instantons: general construction}
\label{subsec-inclgener}

The action of S-duality in the classical approximation \eqref{SL2Zxi}
has been obtained using explicit expressions for the Darboux coordinates.
The only non-trivial step was to find the correct transformation of the fiber coordinate \eqref{trans-t}.
However, once we include instanton corrections, we do not have such expressions at our disposal anymore.
Instead, the Darboux coordinates are determined by complicated integral equations and in the best case
can be expressed through integrals of type \eqref{newfun}. Therefore, the first natural step
would be to understand how such integrals, or more generally the integrals of type
\be
\sum_j \oint_{C_j}\frac{\de\varpi'}{\varpi'}\,
\frac{\varpi'+\varpi}{\varpi'-\varpi}\, H(\xi,\txi),
\label{integral}
\ee
transform under S-duality.

To discuss this issue, we restrict ourselves to the instanton corrections
with the vanishing five-brane charge $p^0=0$. As is clear from \eqref{exline},
this ensures that the Darboux coordinate $\xi^0$ remains uncorrected and is given globally
by the same classical expression \eqref{gentwi}. This implies in turn that
the transformation of $\varpi$ should also be unchanged and one can use the formula \eqref{trans-t}.

But now we arrive at a problem: the measure and the kernel in the integral \eqref{integral}
transform in a complicated way which does not allow for a simple geometric interpretation.
Fortunately, this problem can be cured by a modification of the integration kernel
which amounts just to a coordinate redefinition on the moduli space \cite{Alexandrov:2012bu}.
Let us define
\be
\label{modker}
K(\varpi,\varpi') =
\frac12\(\frac{\varpi'+\varpi}{\varpi'-\varpi}+\frac{1/\varpi'-\varpi'}{1/\varpi'+\varpi'}\)
=\frac{(1+\varpi\varpi')}{(\varpi'-\varpi)(1/\varpi'+\varpi')}.
\ee
Then it is easy to check that $K(\varpi,\varpi')\,\frac{\de\varpi'}{\varpi'}$ is invariant
under S-duality!\footnote{To understand the origin of the new kernel, it might be helpful
to change the fiber coordinate to \cite{Alexandrov:2012bu}
$$
z=\frac{\varpi+\I}{\varpi-\I}.
$$
In its terms one has
$$
\varpi=-\I \frac{1+z}{1-z},
\qquad
\xi^0 = \tau_1 + \I \tau_2\, \frac{1/z+z}{1/z-z}.
$$
The new coordinate has the advantage of having a simpler transformation under S-duality
$$
\label{ztrans}
z\mapsto \frac{c\bar\tau+d}{|c\tau+d|}\, z.
$$
Moreover, it provides a very simple representation for the new kernel
$$
K(\varpi,\varpi') \frac{\de \varpi'}{\varpi'}
=\frac12\,\frac{z'+z}{z'-z}\, \frac{\de z'}{z'},
$$
which makes its invariance transparent.
}
Since $K(\varpi,\varpi')$ differs from the original kernel by a $\varpi$-independent term,
this additional contribution can be absorbed into a redefinition of the constant terms in the $\varpi$-expansion
of Darboux coordinates. Thus, to write them in terms of the new kernel,
it is sufficient to redefine $\zeta^\Lambda,\tzeta_\Lambda$ and $\sigma$.
After this redefinition, we are ready to give our main result.

The following theorem not only shows how S-duality is realized to {\it all} orders in the instanton expansion
in the absence of five-brane instantons, but also gives conditions on the transition functions
ensuring the $SL(2,\IZ)$ symmetry.\footnote{To avoid cluttering of notations, we omit primes on the RR fields,
Darboux coordinates and holomorphic prepotential. Thus we ignore the effects due to the matrix $A_{\Lambda\Sigma}$.
But they can be easily restored by performing the corresponding symplectic transformation.
The primes on sums over $m,n$ will denote that the value $m=n=0$ is omitted.}

{\bf Theorem \cite{Alexandrov:2012bu}:} {\it Let the twistor space $\cZ$ is defined by the covering
\be
\cZ=\cU_0\cup\cU_+ \cup \cU_- \cup\(\cup_{m,n}' \cU_{m,n}\),
\label{th-cover}
\ee
where the patches $\cU_\pm$ cover the north and south poles of $\CP$, $\cU_{m,n}$ are mapped to each other
under the antipodal map and $SL(2,\IZ)$-transformations
\be
\tau\[\cU_{m,n}\]= \cU_{\tau(m),\tau(n)},
\ee
\be
\cU_{m,n}\mapsto \cU_{m',n'},
\qquad
\( m'\atop n'\) =
\(
\begin{array}{cc}
a & c
\\
b & d
\end{array}
\)
\( m \atop n \) ,
\ee
and $\cU_0$ covers the rest.
The associated transition functions are taken to be
\be
\label{transfun}
\Hij{+0}=\Fcl(\xii{+}^\Lambda) ,
\qquad
\Hij{-0}=\bFcl(\xii{-}^\Lambda) ,
\qquad
\Hij{(m,n)0}= G_{m,n}(\xii{m,n}^\Lambda,\txii{0}_a),
\ee
where
the holomorphic functions $G_{m,n}$ are required to transform under $SL(2,\IZ)$ as
\be
G_{m,n}
\mapsto \frac{G_{m',n'}}{c\xi^0+d}
-\frac{c}{2}\, \frac{\kappa_{abc}\p_{\txii{0}_a}G_{m',n'}\p_{\txii{0}_b}G_{m',n'}}{(c\xi^0+d)^2}
\(\xii{m',n'}^c-\frac{2}{3}\,\p_{\txii{0}_c}G_{m',n'}\)
+\mbox{reg.}
\label{StransG}
\ee
where $+ \mbox{reg.}$  denotes equality up to terms regular in the patch $\cU_{m',n'}$.

Then the following statements hold:
\begin{enumerate}
\item The Darboux coordinates in the patch $\cU_0$, as functions of the fiber coordinate $\varpi$,
are given by the following expressions:
\beq
\hspace{-1cm}
\xi^0 &=& \zeta^0+\frac{\tau_2}{2}\(\varpi^{-1}-\varpi\),
\nn
\\
\hspace{-1cm}
\xii{0}^a &=& \zeta_{\rm cl}^a +\varpi^{-1}Y^a -\varpi \bY^a+
 {\sum_{m,n}}' \oint_{C_{m,n}} \frac{\de \varpi'}{2 \pi \I \varpi'} \,
K(\varpi,\varpi') \p_{\txii{0}_a}G_{m,n},
\nn
\\
\hspace{-1cm}
\txii{0}_\Lambda& =& \tzeta^{\rm cl}_\Lambda +\varpi^{-1}\Fcl_\Lambda(Y)-\varpi\bFcl_\Lambda(\bY)
- {\sum_{m,n}}' \oint_{C_{m,n}} \frac{\de \varpi'}{2 \pi \I \varpi'} \,
K(\varpi,\varpi')\p_{\xii{m,n}^\Lambda}G_{m,n},
\label{twistlines}
\\
\hspace{-1cm}
\ai{0}&= &  -\hf(\tsigma+\zeta^\Lambda \tzeta_\Lambda)^{\rm cl}
-\(\varpi^{-1}+\varpi\)\(\varpi^{-1} \Fcl(Y)+\varpi \bFcl(\bY)\)
-\zeta_{\rm cl}^\Lambda\( \varpi^{-1}\Fcl_\Lambda(Y)-\varpi\bFcl_\Lambda(\bY)\)
\nn \\
&&
-{\sum_{m,n}}' \oint_{C_{m,n}} \frac{\de \varpi'}{2 \pi \I \varpi'} \,
\[
K(\varpi,\varpi') \(1-\xii{m,n}^\Lambda(\varpi')\p_{\xii{m,n}^\Lambda}\)G_{m,n}
\right.
\nn \\
&& \left.\qquad\qquad\qquad\qquad\qquad
+\frac{ (\varpi \varpi')^{-1} \Fcl_a(Y)+\varpi \varpi'\bFcl_a(\bY) }{1/\varpi'+\varpi'}\,\p_{\txii{0}_a}G_{m,n} \],
\nn
\eeq
where $C_{m,n}$ are contours on $\CP$ surrounding $\cU_{m,n}$ in the counter-clockwise direction and
$\tau_2,\zeta^0,\zeta_{\rm cl}^a,\tzeta^{\rm cl}_\Lambda,\tsigma^{\rm cl},Y^a$ are some coordinates on the QK base.

\item
Provided these coordinates are identified with the following combinations of the physical type IIB fields
\be
\label{mirror-map}
\begin{split}
\zeta^0 & =\tau_1,
\qquad
\zeta_{\rm cl}^a = - (c^a - \tau_1 b^a),
\\
\tzeta^{\rm cl}_a &=  c_a+ \frac{1}{2}\, \kappa_{abc} \,b^b (c^c - \tau_1 b^c) ,
\qquad
\tzeta^{\rm cl}_0 =\, c_0-\frac{1}{6}\, \kappa_{abc} \,b^a b^b (c^c-\tau_1 b^c),
\\
\tsigma^{\rm cl} &= -2 (\psi+\frac12  \tau_1 c_0) + c_a (c^a - \tau_1 b^a)
-\frac{1}{6}\,\kappa_{abc} \, b^a c^b (c^c - \tau_1 b^c)
+\frac{\tau_2^2}{3}\, \kappa_{abc}b^a b^b b^c,
\\
Y^0 & =\frac{\tau_2}{2},
\qquad
Y^a =\frac{\tau_2}{2}\( b^a+\I t^a\)+{\sum_{m,n}}' \oint_{C_{m,n}} \frac{\de \varpi}{2 \pi \I \varpi^2} \,
\frac{\p_{\txii{0}_a}G_{m,n}}{\(1/\varpi+\varpi\)^2},
\end{split}
\ee
the Darboux coordinates \eqref{twistlines} transform under $SL(2,\IZ)$ transformations \eqref{SL2Z} and \eqref{trans-t}
according to the classical laws \eqref{SL2Zxi}.
\item
The same transformation rules hold also in the patches $\cU_{m,n}$,
with the understanding that the Darboux coordinates appearing on the r.h.s. are those
attached to the patch $\cU_{m',n'}$, e.g. $\xii{m,n}^a \mapsto \xii{m',n'}^a/(c\xi^0+d)$.
\item
The contact potential is given by
\beq
e^{\Phi}& =&\frac{2}{3\tau_2}\,\kappa_{abc} \Im Y^a\Im Y^b \Im Y^c
\label{contpotIIB}
\\
&& -\frac{1}{16\pi}{\sum_{m,n}}'
\oint_{C_{m,n}}\frac{\de\varpi}{\varpi}\[
\(\varpi^{-1} Y^{\Lambda}-\varpi \bY^{\Lambda} \)\p_{\xii{m,n}^\Lambda}G_{m,n}
+\(\varpi^{-1} \Fcl_{a}(Y)-\varpi \bFcl_a(\bY) \)\p_{\txii{0}_a}G_{m,n}\]
\nn
\eeq
and transforms under $SL(2,\IZ)$ as in \eqref{SL2phi}
\end{enumerate}
As a result, the \qk manifold associated to the above twistorial data carries an isometric action of
$SL(2,\IZ)$.
}

\vspace{0.2cm}

Some comments are in order:
\begin{itemize}

\item
The patches $\cU_{m,n},\cU_\pm$ are not necessarily all different. Typically, $\cU_{m,n}$ are associated
to zeros of $m\xi^0+n$ denoted above by $\tzpmcd{m,n}$.
Then one might have $\cU_{km,kn}=\cU_{m,n}$ for $k\in \Nint$
and $\cU_\pm$ coinciding with $\cU_{0,n}$ for $n>0$ and $n<0$, respectively.
If this is the case, the functions $G_{m,n}$ should be defined for $m,n$ coprime only.

\item
Another possibility is to include open contours similar to the BPS rays appearing in the type IIA description.
In fact, all previous statements hold for an arbitrary set of contours
$C_{m,n}$ mapped unto each other by $SL(2,\IZ)$.
For example, $C_{m,n}$ can be an open contour from $\tzpcd{m,n}$ to $\tzmcd{m,n}$.
In this case however the structure of patches might be more complicated.
The only requirement on the covering is that
it is invariant under the antipodal map and $SL(2,\IZ)$ transformations so that
a patch $\cU_i$ is mapped to some patch $\cU_{i_{c,d}}$. The Darboux coordinates in each patch
are given by the formulas \eqref{twistlines} and transform under S-duality
by \eqref{SL2Zxi} where the coordinates on the r.h.s. are from the patch $\cU_{i_{c,d}}$.

\item
If the transition function $G_{m,n}$ is associated with such an open contour, the regular terms
in its $SL(2,\IZ)$ transformation are not allowed anymore.
In other words, the transformation \eqref{StransG} should be exact,
because in this case the transition function already represents
a discontinuity through a logarithmic branch cut.

\item
If not the quadratic and cubic terms in $G_{m,n}$, the transformation law \eqref{StransG}
would simply mean that the transition functions are similar to modular functions of weight $(-1,0)$.
More precisely, such a modular function would arise as a formal sum of $G_{m,n}$ over all patches.
However, the non-linear terms spoil the simple modular transformation.
They appear due to the fact that the arguments of transition functions are taken in different patches
and are completely similar to the quadratic terms in \eqref{transellg}.
But, in contrast to the situation in type IIA theory, it is not known whether there is a simplified
description which allows to get rid of them.

\item
Note that the coordinate change \eqref{mirror-map} is almost identical to the classical mirror map \eqref{symptob}.
There are only two differences. First, $Y^a$ (usually related to $z^a$) acquires an additional
instanton contribution. Second, the coordinate $\tilde\sigma^{\rm cl}$ differs from $\sigma$ in \eqref{symptob} by
the presence of the last term. Although it is a classical contribution,
it was absorbed into $\tsigma^{\rm cl}$ to simplify the expression for $\alpha$ in terms of $Y^a$.

\item
This theorem assumes that the functions $G_{m,n}$ are independent on $\txi_0$ and $\alpha$, which is equivalent to
the ignorance of five-brane instanton corrections. At the moment, it is known how to generalize the above
results to include such dependence only in the one-instanton approximation.
As one could suspect, in this approximation the S-duality invariance is again ensured
by the requirement that the transition functions transform with the modular weight $(-1,0)$, i.e.
their transformation should be given by the first term in \eqref{StransG}.\label{ext-theor}

\item
Finally, note that all anomalous dimensions are taken to vanish.
On the other hand, on the type IIA side the anomalous dimension $\ci{+}_\alpha$ is non-zero and encodes
the one-loop $g_s$-correction to the HM metric (see section \ref{chap_pert}.\ref{subsec_perttwist}).
This implies that in the type IIB twistor space the one-loop correction appears from a completely different origin.

\end{itemize}

These results provide a general framework for the twistorial construction of the HM moduli space
in the type IIB formulation. All D(-1), D1 and D3-instanton corrections should fit the conditions
of the above theorem. In particular, the transition functions generating these corrections
should transform according to \eqref{StransG}. But this theorem does not answer to
the main question: what are these transition functions?
The theorem does not specify them, but only imposes conditions on their transformation properties.
On the other hand, if mirror symmetry does hold at quantum level, the resulting moduli space should agree
with the one constructed in the type IIA formulation.
Thus, there should exist a resummation procedure which, starting from the description of chapter \ref{chap_Dbrane},
allows to find the transition functions relevant
in type IIB theory and they should automatically satisfy the required S-duality constraint \eqref{StransG}.
This might be considered as a very non-trivial test of consistency of the type IIA construction.
Below we report on the results in this direction.

This reasoning does not apply to D5-brane instantons. They are not invariant under S-duality and therefore
should be treated separately. In fact, S-duality mixes them with NS5-branes. This fact will be used
in chapter \ref{chap_NS5} to generate the latter corrections and thus to get an access to the complete
non-perturbative description of $\MH$.

\subsection{D(-1)-D1-instantons}

Let us first consider D(-1)-D1-instanton corrections, i.e. those which have vanishing magnetic charge $p^\Lambda$.
In this case one has great simplifications. First of all, as indicated in section \ref{chap_Dbrane}.\ref{subsec_alleqs},
in the sector with vanishing $p^\Lambda$
the linear instanton approximation for Darboux coordinates becomes exact so that they are not given
by complicated integral equations, but known explicitly. Moreover, since the transition functions should be
$\txi_\Lambda$-independent, all non-linear terms in the transformation law for $G_{m,n}$ \eqref{StransG}
vanish and the transition functions in the type IIB formulation are expected to originate from a simple modular form.

How the type IIB description of this sector, fitting the theorem of the previous subsection,
arises from the type IIA construction of D-instantons has been shown in
\cite{Alexandrov:2009qq}.\footnote{In fact, D(-1) and D1-instanton contributions have been obtained before
the invention of the type IIA construction based on the BPS rays and played an important role
on the way to it \cite{RoblesLlana:2006is}. However, these contributions were found only in the form of corrections
to the hyperk\"ahler potential \eqref{chiphi} on the associated HKC.
Neither the complete twistor formulation, nor its relation to the type IIA twistor space were known.}
The idea is that one should perform a Poisson resummation of the Darboux coordinates with respect to
D-brane charge $q_0$.
Then the couple of integers $(m,n)$ from the theorem originates as follows:
$n$ appears in the process of Poisson resummation as the integer conjugate to $q_0$ and $m$ comes from
the expansion of the dilogarithm in the type IIA transition functions, i.e. it counts multi-covering effects.

The resulting twistor space can be described by the covering \eqref{th-cover} where $\cU_{m,n}$
surround $\tzpcd{m,n}$ (see Fig. \ref{fig-D1inst}) and by the transition functions \eqref{transfun} with
\be
\begin{split}
G_{m,n}^{\rm D1}(\xi)& =  -\frac{\I}{(2\pi)^3}
\!\!\sum_{q_a\gamma^a\in H_2^+(\CYm)\cup\{0\}}\!\!\!\!\!\, n_{q_a}^{(0)}\,\frac{e^{-2\pi \I m q_a\xi^a}}{m^2(m\xi^0+n)},
\quad  m\ne 0,
\\
G_{0,n}^{\rm D1}(\xi) &= -\frac{\I (\xi^0)^2}{(2\pi)^3}
\!\!\sum_{q_a\gamma^a\in H_2^+(\CYm)\cup\{0\}}\!\!\!\!\!\, n_{q_a}^{(0)}\,\frac{e^{2\pi \I n q_a\xi^a/\xi^0}}{n^3}.
\end{split}
\label{trD1IIB}
\ee
It is easy to check that they do satisfy the transformation property
\be
G_{m,n}\mapsto\frac{G_{m',n'}}{c\xi^0+d}+\mbox{reg.}
\label{transGmn}
\ee
and the possibility to drop the regular terms is important.
Note also that $\sum_{n>0}G_{0,n}^{\rm D1}$ provides the sum of the one-loop and worldsheet instanton
$\alpha'$-corrections to the prepotential \eqref{lve}. Since $\cU_{0,n}=\cU_+$,
they are combined with its classical piece given by $\Hij{+0}$ to the full holomorphic
prepotential.

Although this twistor space looks completely different from the one we started with on the type IIA side,
it turns out that they are related by a certain gauge transformation.
This means that in different patches one should perform certain contact transformations
generated by regular functions. However, these generating functions can have singularities outside
their patch of definition. As a result, the transformed Darboux coordinates may have different
singularities than the initial ones and the covering should be adapted to the new analytic structure as well.
In this way a careful analysis allows to relate the melon-like type IIA picture to the type IIB formulation
given here.

\lfig{The covering and transition functions incorporating D(-1)-D1-instantons.}{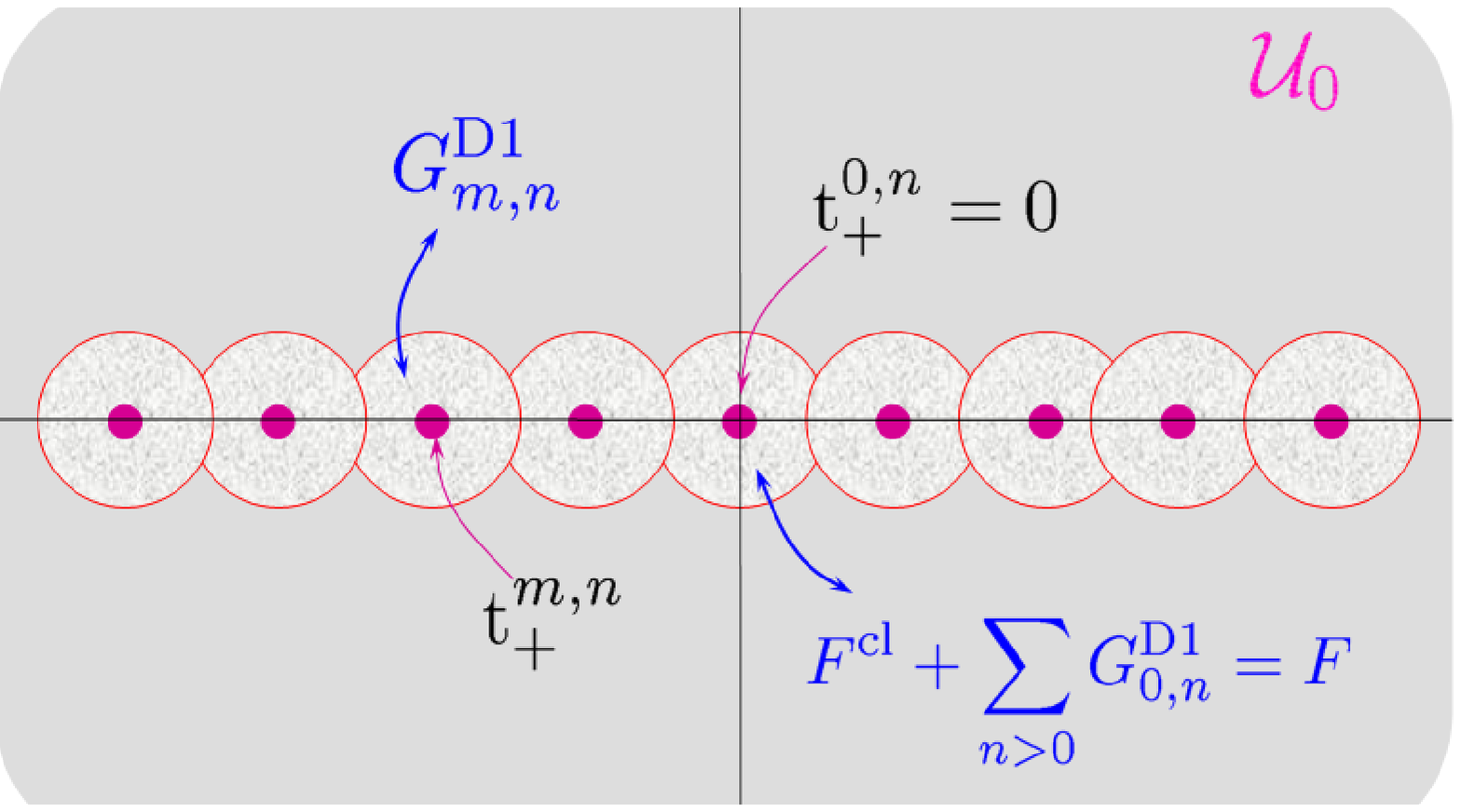}{12cm}{fig-D1inst}{-0.6cm}

Finally, note that the formula \eqref{contpotIIB} for the contact potential can be easily evaluated
and gives the following result
\be
e^{\phi} = \frac{\tau_2^2}{2}\,\cV(t^a)
+\frac{\sqrt{\tau_2}}{8(2\pi)^3}
\!\!\!\sum_{q_a\gamma^a\in H_2^+(\CYm)\cup\{0\}}\!\!\!\!\!\!\,
 n_{q_a}^{(0)}\,
{\sum\limits_{m,n}}'\frac{\tau_2^{3/2}}{|m\tau+n|^3}\(1+2\pi |m\tau+n|q_a t^a\)e^{-S_{m,n, q_a}} ,
\label{phiinvb}
\ee
which was first obtained in \cite{RoblesLlana:2006is} by starting from the one-loop expression and
by enforcing the $SL(2,\IZ)$ symmetry.
Here
\be
S_{m,n,q_a} =2\pi q_a | m \tau+n |\, t^a-2\pi \I q_a (m c^a +n  b^a)
\ee
is the classical action of a D1-D(-1)-instanton or, more precisely, an Euclidean
$(m,n)$-string wrapped around the effective curve $q_a\gamma^a$.

\subsection{D3-instantons}
\label{subsec-D3}

This sector is already not so well understood and at the moment only some preliminary results are
accessible in the linear instanton approximation. To present these results, we restrict ourselves to
the sector with fixed D3-brane charge $p^a$ because it should be preserved by S-duality.
Besides, we replace the integer-valued BPS indices $\gnkl{\gamma;v}$ by rational invariants
constructed as
\be
\label{defntilde}
\bgnkl{\gamma; v} = \sum_{d|\gamma}  \frac{1}{d^2}\,  \gnkl{\gamma/d; v} .
\ee
They encode multi-covering effects and allow to replace the dilogarithm in the transition functions \eqref{prepH}
by simple exponential. Thus, our starting point is
\be
\bH_\gamma=\frac{\bgnkl{\gamma; v}}{(2\pi)^2}\,
\sigma_{\text{D}}(\gamma) \expe{p^a\txi_a'-q_a'\xi^a-q_0'\xi^0},
\label{prepHpa}
\ee
with the charge vector restricted to $\gamma=(0,p^a,q_a,q_0)$.

The advantage of the rational invariants is the possibility to use a combination of
the spectral flow symmetry \eqref{chargesymp} and modular invariance.
In our approximation one can neglect the dependence of $\bgnkl{\gamma; v}$ on the K\"ahler moduli
and then their invariance with respect to the spectral flow symmetry allows to write that
\be
\bgnkl{0,p^a,q'_a,q'_0} \equiv \bOm{p^a,\mu_a}{\hat q_0} ,
\label{resOmeg}
\ee
where $\hat q_0$ is the invariant charge \eqref{invcharge} and
$\mu^a$ is the residue class of $q'_a\gamma^a$ modulo the sublattice
$\Lambda=\{ \gamma^a \kappa_{ab} \epsilon^b, \epsilon^a \gamma_a \in H_4(Y,\IZ) \}$.
These quantities can be used to construct the following function
\be
\label{defchimu}
h_{p^a,\mu_a}(\tau) = \sum_{\hat q_0 \leq \hat q_0^{\rm max}}
\bOm{p^a, \mu_a}{ \hat q_0}\,
\expe{-\hat q_0 \tau },
\ee
which is known to be a vector-valued modular form of weight $(-\frac{b_2}{2}-1,0)$
with multiplier system given by $\expe{c_{2,a} p^a \varepsilon(\gsl)}$
\cite{deBoer:2006vg,Manschot:2007ha,Manschot:2009ia}
where $\eps(\gsl)$ is the same multiplier system of the Dedekind eta function which appears in \eqref{SL2Z}.
The crucial property of the function $h_{p^a,\mu_a}(\tau)$ is that it is completely determined by its polar part
consisting of the terms in the sum which are singular in the limit $\tau_2\to\infty$.
As a result, all $\bOm{p^a, \mu_a}{ \hat q_0}$ for negative $\hat q_0$ can be expressed in terms of
just a finite number of rational BPS indices given by $\bOm{p^a, \mu_a}{ \hat q_0}$ for
$0\leq\hat q_0\leq \hat q_0^{\rm max}$.
We refer to \cite{Dijkgraaf:2000fq,Manschot:2007ha} for the explicit expression.

\lfig{The set of open contours and transition functions incorporating D3-instantons.}{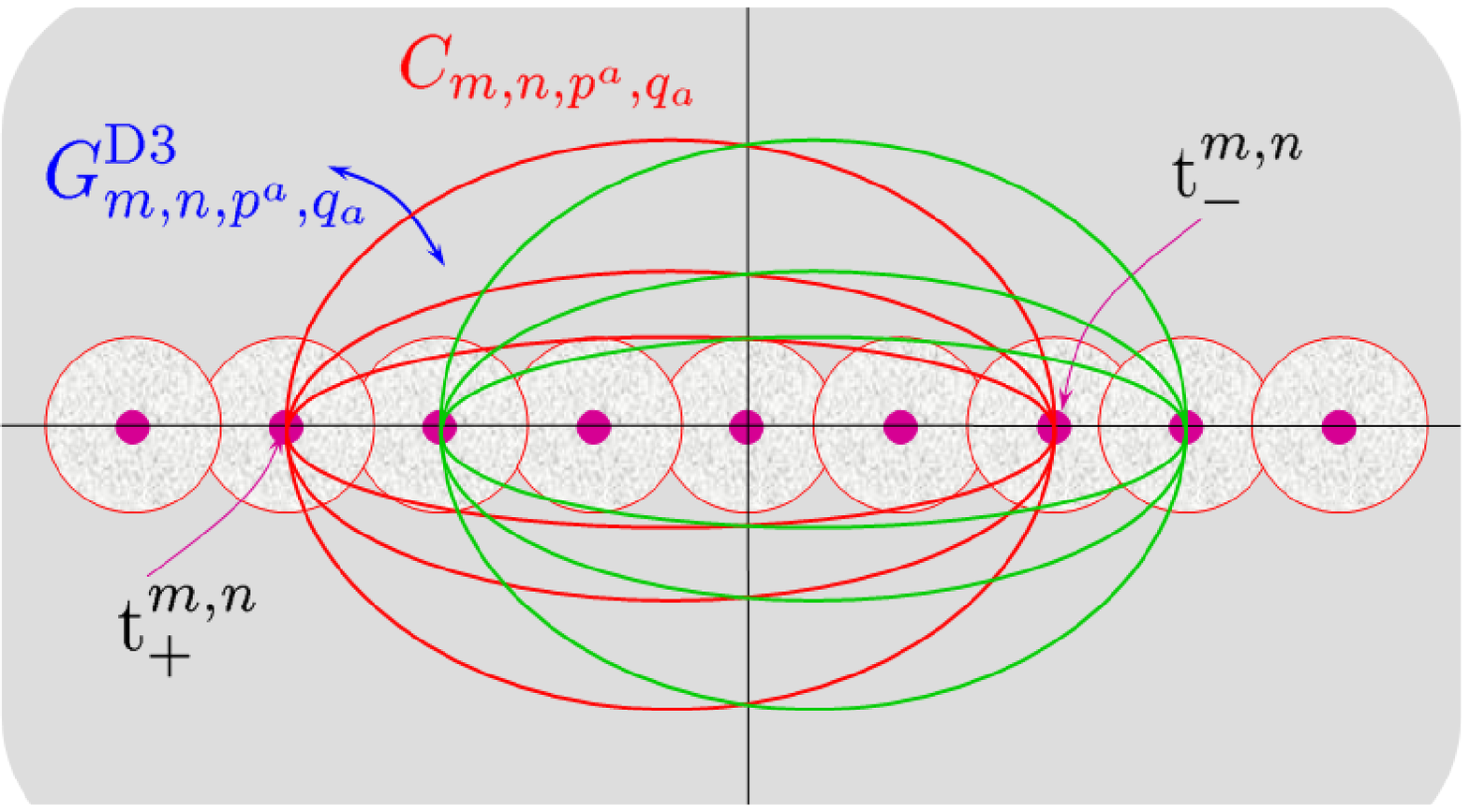}{12cm}{fig-D3inst}{-0.6cm}

After using this result for the rational invariants, one can show that the resummation over (negative) $\hat q_0$
leads to the following ansatz for the transition functions \cite{amp-in-progress}
\begin{eqnarray}
G_{m,n,p^a,q_a}^{\rm D3}\!\!\!\!&=&\!\!\!\!
\frac{1}{2\pi^2}\,(m\xi_0+n)\expe{-\frac{1}{2}A_{ab}p^ap^b}
\sum_{0\leq \hat q_0\leq \hat q_0^{\rm max}}
  \bOm{p^a,\mu_a}{\hat q_0}\,(-1)^{q'_a p^a}
\label{ans-trans} \\
&\times & \!\!\!\!\expe{ p^a
\gsl_{m,n}[\txi_a']
-\left( \hat q_0+\frac{1}{2} \kappa^{ab} q'_a q'_b \right) \gsl_{m,n}[\xi^0]
-q'_a\gsl_{m,n}[\xi^a]}
R_{-1-\frac{b_2}{2}}\(\frac{2\pi \I \hat q_0}{m(m\xi^0+n)}\),
\nn
\end{eqnarray}
where
\be
R_w(x)=1-\frac{1}{\Gamma(1-w)}\int_x^\infty e^{-z} z^{-w}\de z
\ee
and $\gsl_{m,n}$ is the $SL(2,\IZ)$ group element \eqref{Sdualde2} with $c=m$, $d=n$
acting on the Darboux coordinates by the transformation law \eqref{SL2Zxi}.
Up to the last factor, which was studied in detail in \cite{Manschot:2007ha},
it is immediate to see that under S-duality these functions
do transform as required by the theorem of section \ref{subsec-inclgener}, i.e.
according to \eqref{transGmn} and without the necessity to drop any regular contributions.
Thus, one gets the first sign that the D3-instanton sector admits a consistent implementation
of S-duality.

Moreover, note that the functions \eqref{ans-trans} have essential singularities at $\varpi=\tzpmcd{m,n}$.
This indicates that they should be associated with contours $C_{m,n,p^a,q_a}$
joining these two points on $\CP$ (see Fig. \ref{fig-D3inst}).
As a result, the complete set of contours is given by an infinite number of copies of
the melon-like picture, which we had in type IIA, rotated into each other by $SL(2,\IZ)$ transformations
so that it is consistent with the action of S-duality.
However, the details of this construction, as well as its extension to all orders
in the instanton expansion remain still unclear.

\section{Non-perturbative mirror map}

Once we understand how S-duality is realized at the level of the twistor space,
we can address another important issue: the non-perturbative mirror map.
The problem is that, to apply mirror symmetry, one needs a relation
between the type IIA and type IIB physical fields, which maps the hypermultiplet metrics
in the two mirror formulations into each other.
In the classical approximation such relation is furnished by the classical mirror map \eqref{symptob}.
But it is expected to receive all possible quantum corrections, both in $\alpha'$ and $g_s$.
The relation including these corrections will be called {\it non-perturbative mirror map} and
can be viewed as a realization of quantum mirror symmetry.

In general, to find quantum corrections to the mirror map
is an extremely complicated problem because the only condition which fixes the mirror map
is that after its application the metric must be invariant under the action of $SL(2,\IZ)$ \eqref{SL2Z}.
However, this condition becomes more explicit once we lift it to the twistor space because here
$SL(2,\IZ)$ should act not only isometrically, but also holomorphically.
As a result, this condition gives rise to a self-determining system where the consistency between mirror symmetry
and S-duality is enough to find both of them.
The consistency condition can be formulated as the following equation
\be
\Xi\({\bf m}[\gsl \cdot q_{\rm IIB}],\gsl\cdot\varpi\)=\gsl\cdot \Xi,
\label{consistcond}
\ee
where $\Xi$ denotes all Darboux coordinates, $q_{\rm IIB}$ are the type IIB physical fields, $\gsl\cdot$ is the action
of an $SL(2,\IZ)$ transformation and ${\bf m}$ is the non-perturbative mirror map.
This condition means that the $SL(2,\IZ)$ action on Darboux coordinates can be evaluated in two ways.
The first one, which is shown on the l.h.s., uses the explicit expression of the Darboux coordinates
following from our twistor construction in the type IIA formulation. Of course, it gives them as functions of the type IIA fields.
Therefore, to evaluate the action of $\gsl$,
one needs to use the (to be found) mirror map which expresses them in terms of the type IIB fields
whose transformations are known by definition. The second way, shown on the r.h.s., is a holomorphic
action of $SL(2,\IZ)$ directly on the Darboux coordinates.
In the classical limit it is given by \eqref{SL2Zxi} and in principle
one could expect that it gets modified by quantum corrections. However, all our results indicate that it actually
stays the same at the full non-perturbative level \cite{Alexandrov:2009qq}.
Thus, the equality \eqref{consistcond} of the two $SL(2,\IZ)$ actions provides a very non-trivial
condition on the mirror map ${\bf m}$.

In fact, the theorem from section \ref{subsec-inclgener} essentially gives a solution to the equation \eqref{consistcond}.
Indeed, the relations \eqref{mirror-map} substituted into the Darboux coordinates \eqref{twistlines}
show how the latter depend on the type IIB fields so that S-duality is correctly realized.
In particular, they ensure the holomorphic transformations \eqref{SL2Zxi}. Thus, the only thing which remains
to be found is a relation between $\zeta_{\rm cl}^a,\tzeta^{\rm cl}_\Lambda,\tsigma^{\rm cl},Y^a$
and the type IIA fields. This relation in turn follows from the change of the kernel
as described on page \pageref{modker}. For example, restricting to the D1-D(-1)-instanton sector
and in the presence of all $\alpha'$-corrections, one finds \cite{Alexandrov:2012bu}
\beq
\zeta^a_{\rm cl} & =&\zeta^a,
\qquad\qquad
Y^a=\frac{\tau_2}{2}\, z^a,
\nn
\\
\tzeta^{\rm cl}_\Lambda & =&\tzeta'_\Lambda -
\half\,  {\sum_{\gamma=(0,q_\Lambda)}} \int_{\ell_\gamma} \frac{\de \varpi}{2 \pi \I \varpi} \,
\frac{1/\varpi-\varpi}{1/\varpi+\varpi}
\, \p_{\xi^\Lambda}H_\gamma,
\label{qmirmapb}
\\
\tilde\sigma^{\rm cl} & =&\sigma
-\frac{\tau_2^2}{3}\, (z^\Lambda+\bz^\Lambda)(F'_\Lambda+\bF'_\Lambda)
+ {\sum_{\gamma=(0,q_\Lambda)}} \int_{\ell_\gamma} \frac{\de \varpi}{2 \pi \I \varpi} \,
\frac{1/\varpi-\varpi}{1/\varpi+\varpi}\,
\(1-\(\xi^\Lambda-\frc12\zeta^\Lambda\)\p_{\xi^\Lambda}\)H_\gamma.
\nn
\eeq
Resumming over $q_0$, one can express the same relation in terms of the transition functions
of the type IIB formulation \eqref{trD1IIB}. The instanton terms will be given by contour integrals
with the contours surrounding the points $\tzpcd{m,n}$.
Such integrals can be easily evaluated and therefore in this case it is possible to write
the mirror map in a very explicit form. We refer to \cite{Alexandrov:2009qq} for the explicit
expressions.

This result completely solves the problem of quantum mirror symmetry in the approximation where
five-brane and D3-instantons are neglected. The latter corrections are still described by the theorem,
so the relations \eqref{mirror-map} solve the most difficult part of the problem. However,
it still remains to relate $\zeta_{\rm cl}^a,\tzeta^{\rm cl}_\Lambda,\tsigma^{\rm cl}$
to the type IIA axionic fields. This should be done by a resummation of the Darboux coordinates
which should bring them from \eqref{exline} to the type IIB form \eqref{twistlines}.
Unfortunately, as was mentioned in section \ref{subsec-D3}, this has not been done yet in a satisfactory way.

For five-branes the situation is even more complicated because the corresponding corrections are expected
to affect even the $SL(2,\IZ)$ transformation of the fiber coordinate $\varpi$ \eqref{trans-t}.
Moreover, S-duality requires to include into consideration NS5-brane instantons, which remain unknown
on the type IIA side. Thus, even before addressing the issue of quantum mirror symmetry in the presence of five-branes,
we need to understand how to generate the NS5-brane instanton contributions.
This will be the subject of the next chapter.

\chapter{NS5-brane instantons}
\label{chap_NS5}

The instanton corrections coming from NS5-branes are
the last missing piece of the construction of the complete non-perturbative geometry
of the hypermultiplet moduli space (at least in type IIA picture).
In this chapter we present a progress in
understanding the general structure of these contributions and an attempt
to incorporate them into the twistor description of $\MH$.

There are several reasons why the NS5-brane contributions are difficult to find.
Let us point out some of them:
\begin{itemize}
\item
First of all, the NS5-brane is a complicated dynamical object whose dynamics is not well understood.
In particular, its world-volume theory, especially in the case of multiple fivebranes,
is not known to the full extent.

\item
In the type IIA formulation, it is expected that the world-volume theory involves
a self-dual 3-form flux $H$ which is an electric source for
the RR 3-form potential $\hA_3$ \cite{Callan:1991ky,Bandos:2000az}.
The self-duality constraint implies that fluxes on three-cycles $\gamma,\gamma'$, corresponding
to D2-NS5 bound states, cannot be measured (nor defined) simultaneously
if the intersection product $\langle \gamma,\gamma' \rangle$ is non-vanishing.
A mathematical consequence of this fact is that the NS5-brane partition function cannot
be a simple function, but only a section of a certain bundle.

\item
Another complication is that, upon inclusion of NS5-brane contributions, all continuous isometries
of the moduli space $\MH$ become broken. In particular, this implies that one does not have the QK/HK correspondence
at our disposal anymore and one has to apply the full machinery of the contact geometry.

\item
Finally, since the NS5-instantons introduce dependence on the NS-axion $\sigma$,
it becomes now important to take into account the full Heisenberg symmetry group \eqref{Heisenb},
including its non-commutative nature. Due to this, one can expect the same complications as
passing from a classical theory to its quantum version where the role of the quantization parameter
is played by the NS5-brane charge.

\end{itemize}

Besides these issues, there are many other subtleties all of which require
a special care. In particular, there is a subtle interplay between
the description of instanton contributions and the topology of the moduli space.
Therefore, before attacking the problem of NS5-brane instantons, we analyze
the topological structure of $\MH$ and its implications.

\section{Topological issues}
\label{sec_topissue}

The global structure of $\MH$ can be specified by providing a group of discrete identifications
of the coordinates parametrizing the moduli space which act isometrically on it.
Besides symplectic invariance and S-duality, one has two additional discrete symmetries
provided by Heisenberg and monodromy transformations (see section \ref{chap_moduli}.\ref{subsec_sym}).
What we need is to better understand their action and what it implies for NS5-brane contributions.

Let us concentrate for simplicity on the type IIA formulation. In type IIB the picture
is completely similar and can be obtained using mirror symmetry.
At fixed, vanishingly small string coupling $g_s$ and ignoring the NS-axion (i.e., modding out
by translations along $\sigma$, which are symmetries of the perturbative theory),
the ``reduced'' HM moduli space is known to be topologically the so called
intermediate Jacobian $\cJ_c(\CY)$. This is a torus bundle over the complex structure moduli space $\KC(\CY)$
with fiber $\cT\equiv H^3(\CY,\IR)/H^3(\CY,\IZ)$, which
parametrizes the space of harmonic RR three-form fields $C$ over $\CY$ modulo
large gauge transformations \cite{Morrison:1995yi}. This is precisely the topology consistent
with the structure of D-instanton corrections \eqref{d2quali} where the axionic couplings
arise as periods of $C$ over the cycle wrapped by the brane.

The next step is to include in this picture the NS-axion $\sigma$.
It parametrizes a certain circle bundle $\Cns$ over $\cJ_c(\CY)$
and our primary goal is to understand the topology of this bundle.
This can be done on the basis of two observations.
First, let us consider the general structure of NS5-brane corrections to the metric shown in \eqref{couplNS5}.
We can also write them in the following form
\be
\label{ns5quali}
\delta \de s^2\vert_{\text{NS5}}  \sim
e^{-4 \pi k e^{\phi}-\I\pi  k \sigma} \,
{\ZNS{k}(\cN, \zeta^\Lambda,\tzeta_\Lambda)}
\ee
where the tensorial nature of the metric was ignored. The second factor can be interpreted as
a partition function of $k$ NS5-branes. An important observation is that consistency
of the instanton correction \eqref{ns5quali} requires that $e^{\I\pi k\sigma}$ and $\ZNS{k}$
are valued in the same circle bundle.

Although it could seem to be a rather trivial statement, in fact, it leads to
certain restrictions on both the partition function and symmetry transformations of the NS-axion.
For example, let us consider the restriction of the circle bundle $\Cns$ to the torus $\cT$.
It is known \cite{Witten:1996hc} that the bundle where the fivebrane partition function $\ZNS{k}$ lives
has a restriction such that its first Chern class coincides with the K\"ahler class of $\cT$
\be
c_1(\Cns)\vert_{\cT} =  \omega_\cT \equiv \de\tzeta_\Lambda  \wedge \de\zeta^\Lambda.
\label{chernCbundle}
\ee
This means that the partition function should transform non-trivially under the large gauge transformations
of the RR-fields. The precise transformation, and correspondingly the fibration of $\Cns$ over the torus $\cT$,
is specified by a choice of quadratic refinement $\signs(\gamma)$, which satisfies
the same defining relation \eqref{qrifprop} as the one for D-instantons and can equally be expressed
in terms of characteristics $\Thns=(\thns^\Lambda,\phins_\Lambda)$ as in \eqref{quadraticrefinementpq}.
Although it is tempting to identify the characteristics $\Thns$ and $\ThD$ for two types of instantons,
we refrain from doing so as {\it a priori} there is no reason for them to coincide.
Then the partition function satisfies the following twisted periodicity condition
\be
\label{thperiod}
\ZNS{k} (\cN, C+H)
= (\sigma_\Thns(H))^k \, \expe{\frac{k}{2}(\eta^{\Lambda}\tilde\zeta_\Lambda-\tleta_\Lambda\zeta^{\Lambda})}
\ZNS{k}  (\cN,C),
\ee
where we abbreviated $C=(\zeta^\Lambda,\tzeta_\Lambda)$ and $H=(\eta^\Lambda,\tleta_\Lambda)\in H^3(\CY,\IZ)$.

Now we return to the instanton correction \eqref{ns5quali}.
Since it must be invariant under the large gauge transformations, the condition \eqref{thperiod}
immediately gives the shift of the NS-axion which must accompany the transformation of the RR-fields.
As a result, we have the following symmetry generators
\be
\begin{split}
T'_{H,\kappa}\ :\ \bigl(\zeta^\Lambda,\ \tzeta_\Lambda,\ \sigma\bigr)
\mapsto
\bigl(\zeta^\Lambda + \eta^\Lambda ,\
\tzeta_\Lambda+ \tleta_\Lambda ,\
\sigma + 2 \kappa
- \tleta_\Lambda \zeta^\Lambda+\eta^\Lambda \tzeta_\Lambda + 2 c(H)\bigr) ,
\end{split}
\label{heisext}
\ee
where $(H,\kappa) \in H^3(\CY,\IZ)\times \IZ$ and
\be
\label{propkappaH}
c(H)=- \frac12\, \eta^\Lambda \tleta_\Lambda + \tleta_\Lambda\theta^\Lambda
- \eta^\Lambda\phi_\Lambda \ \ \ {\rm mod}\   1,
\ee
so that $\sigma_\Theta(H)=(-1)^{2c(H)}$.
One observes that the generators $T'_{H,\kappa}$ are almost identical to the generators of
the Heisenberg algebra \eqref{Heisenb}. The precise relation is given by $T'_{H,\kappa}=T_{H,\kappa+c(H)}$.
Thus, our topological considerations have revealed an extra shift to be added to the naive
Heisenberg transformations \cite{Alexandrov:2010np}. This extra shift is important for consistency of
the group action and effectively abelianizes the non-commutative Heisenberg algebra when
one considers it on functions of the form $F_k(\zeta,\tzeta,\sigma)=e^{\I\pi k\sigma}F_k(\zeta,\tzeta)$.

So far we discussed how $\Cns$ fibers over the torus of the RR-fields, which is encoded
in the action of the Heisenberg symmetry \eqref{heisext}. It remains to understand the fibration of $\Cns$
over the complex structure moduli space $\KC(\CY)$, which is tied with the monodromy transformations.
To this aim, we use our second observation that the perturbative metric \eqref{hypmetone}
suggests that the connection defining $\Cns$ is determined by the one-form $D\sigma$ \eqref{Dsigone}
whose curvature is given by
\be
\label{c1L}
\de \left(\frac{D\sigma}{2}\right)= \omega_\cT + \frac{\chi_\CY}{24}\, \omega_\sk ,
\ee
where $\omega_\cT$ and $\omega_\sk=-\frac{1}{2\pi}\de\cA_K$ are the \kahler forms
on $\cT$ and $\KC(\CY)$, and the coefficient 1/2 takes into account that $\sigma$ has periodicity 2.
The first term is in nice agreement with \eqref{chernCbundle}.
On the other hand, the second term in \eqref{c1L}, originating from the one-loop correction
to the HM metric, has support on $\KC(\CY)$ and implies that the NS-axion picks up
anomalous variations under monodromies.
Indeed, it is easy to check that the following transformation
\be
\label{sigmon}
C\mapsto \rho(M)\cdot C,\qquad
\sigma \mapsto \sigma + \frac{\chi_\CY}{24\pi}  \Im f_M +2\kappa(M),
\ee
where $M$ is a monodromy in $\KC(\CY)$, $f_M$ is a local
holomorphic function determined by the rescaling
$\Omega\mapsto e^{f_M} \Omega$ of the holomorphic 3-form around the monodromy
and $\kappa(M)$ is an undetermined constant defined modulo 1, is a symmetry of
the metric \eqref{hypmetone}. It is very important to keep the constant
contribution given by the last term since without it one easily ends up with an inconsistency: a trivial rescaling
$\Omega\mapsto e^{2\pi \I} \Omega$ would lead to identifications
$\sigma\sim \sigma + \chi_\CY/12$ in conflict with the periodicity modulo 2 of $\sigma$.
Thus, consistency requires the introduction of a unitary character of the monodromy group,
$e^{2\pi\I\kappa(M)}$ \cite{Alexandrov:2010np,Alexandrov:2010ca}. Then the instanton correction \eqref{ns5quali}
implies that the fivebrane partition function should undergo a similar anomalous transformation.
As we will see in section \ref{sec_reltop}, this may have further implications for topological string
amplitudes. Unfortunately, the explicit form of the unitary character remains unknown so far.

\section{Fivebrane corrections to the contact structure}
\label{sec_NS5cor}

\subsection{S-duality on the job}
\label{subsec-Simages}

As follows from the duality chain presented in Fig. \ref{fig_scheme}, the most direct way
to get NS5-brane corrections to the HM moduli space metric is to use the $SL(2,\IZ)$ symmetry
of the type IIB formulation.
It exchanges D5 and NS5-branes and therefore allows to get the latter contributions once we know the former.
However, what we know exactly is not D5-brane corrections, but D2-instantons in type IIA.
D5-instantons follow by applying the mirror map, but as we discussed in the previous chapter,
the mirror map is not known at the full non-perturbative level. Without its non-perturbative completion,
we can get only a quasiclassical or one-instanton approximation to D5 and consequently to NS5-brane corrections.
Nevertheless, already such a result would represent a substantial progress.
Here we are going to present the corrections to the contact structure which result from such analysis and
which have been found in \cite{Alexandrov:2010ca}.

We proceed as follows: start from the known expression for D2-instanton contributions
with charge vector $\gamma=(p^0\neq 0,p^{a},q_a,q_0)$ neglecting all multi-instantons,
apply the classical mirror map \eqref{symptob} which maps D2-branes to
D5-D3-D1-D(-1) bound states, and sum over all images under $SL(2,\IZ)$.
The transformation matrix is chosen to be
\be
\label{Sdualde}
\gsl = \begin{pmatrix} a & b \\ - k/p^0 & p/p^0 \end{pmatrix} \in SL(2,\IZ) ,
\ee
where one takes $(c,d)=(-k/p^0,p/p^0)$ to run over coprime integers
and the integers $(a,b)$, ambiguous up to the addition of $(k/p^0,-p/p^0)$,
are chosen such that $a p + b k = p^0$.
Then $p^0$ is identified as $\gcd(p,k)$, whereas $k$ will be interpreted as NS5-brane charge.

One can perform this procedure directly at the level of the twistor space. This ensures that,
even working in the quasiclassical approximation, one satisfies all constraints of QK geometry.
In this approach an instanton correction is encoded in two pieces of data:
a contour on $\CP$ and a holomorphic function on the twistor space which describes
the transition function associated with the contour. For D-instantons these data
are given by the BPS rays $\ell_{\gamma}$ \eqref{rays} and the functions $H_{\gamma}$ \eqref{prepH}.
It will however be useful to pass to the rational BPS invariants $\bgnkl{\gamma}$ \eqref{defntilde},
as has been done for D3-instantons. Thus, our starting point is provided by
\be
\label{HDinstp}
\bH_{\gamma}=\frac{\bgnkl{\gamma}}{(2\pi)^2}\,\sigma_{\text{D}}(\gamma) \,
\expe{p^\Lambda\txi'_\Lambda-q'_\Lambda \xi^\Lambda} ,
\ee
where we passed also to the primed charges and Darboux coordinates, \eqref{chargeshift} and \eqref{txishift}.
After this change, no mirror map is already required, since the primed variables
are those which properly represent the action of $SL(2,\IZ)$ \eqref{SL2Zxi}.
As a result, the new data which are supposed to introduce NS5-brane corrections are given by
\be
\label{HtotD}
H_{k,p,\hgam} = \gsl\cdot \bH_{\gamma}  ,
\qquad\qquad
\ell_{k,p,\hgam}=\gsl\cdot \ell_{\gamma},
\ee
where $\hat\gamma=(p^a, \hat q_a,\hat q_0)$ denotes
the remaining D3-D1-D(-1)-charges and the hatted charges $\hat q_\Lambda$ denote the combinations \eqref{invcharge}
invariant under the spectral flow.

Using the S-duality action \eqref{SL2Zxi}, one easily finds \cite{Alexandrov:2010ca}
\be
\label{5pqZ2}
H_{k,p,\hgam}=-\frac{\bgnkl{\gamma}}{(2\pi)^2}\,\frac{k}{p^0}\(\xi^0 -n^0\)\sigma_{\text{D}}(\gamma)
\expe{k  S_\alpha+ \frac{p^0\( k \hat q_a (\xi^a-n^a) + p^0 \hat q_0\)}{k^2(\xi^0-n^0)}
+ a\,\frac{p^0 q'_0}{k}- c_{2,a} p^a\eps(\gsl)} ,
\ee
where
\be
\label{defSa}
S_\alpha= \alpha + n^\Lambda\txi_\Lambda+
\Fcl(\xi-n)-\hf\,A_{\Lambda\Sigma}n^\Lambda n^\Sigma ,
\ee
and we denoted $n^0=p/k$, $n^a=p^a/k$, valued in $\IZ/k$.
The factor $(p-k\xi^0)/p^0$ has been introduced to ensure that the functions \eqref{5pqZ2}
satisfy the transformation law \eqref{transGmn} with $G_{m,n}\sim H_{-m p^0,n p^0,\hgam}$
(without regular contributions).
As a result, they satisfy the extension of the theorem of section \ref{subsec-inclgener},
mentioned on page \pageref{ext-theor}, which allows the dependence of
transition functions on $\txi_0$ and $\alpha$.

On the other hand, the transformation \eqref{trans-t}, which is still valid in our approximation,
can be used to find the contour $\ell_{k,p,\hgam}$. It is easily seen to be a half-circle stretching between
the two zeros of $\xi^0-n^0$, which are nothing else but the points $t_\pm^{-k,p}$.
As a result, one obtains exactly the same picture as one had for D3-instantons in Fig. \ref{fig-D3inst}.
This shows that in type IIB fivebrane instantons are nicely combined with D3-instantons.

\subsection{Consistency checks}
\label{subsec-checks}

Altogether $H_{k,p,\hgam}$ and $\ell_{k,p,\hgam}$ provide the data which are expected to encode the fivebrane
corrections to the contact structure of the type IIB HM moduli space in the one-instanton approximation.
However, before we can actually claim that the deformation they induce
correctly introduces fivebrane instanton corrections, they should pass several consistency checks.
Here we present two very non-trivial tests which allow to consider them
as a serious candidate for the right description of fivebrane contributions.

\subsubsection{Heisenberg and monodromy transformations}

Besides S-duality, $\MH$ should carry an isometric action of
the Heisenberg symmetry and monodromy invariance.
In section \ref{sec_topissue} we found how these symmetries should be realized at the non-perturbative level:
they are given by \eqref{heisext} and \eqref{sigmon}, respectively.
As usual, they can be lifted to a holomorphic action on the twistor space
and should preserve the contact structure.
In our terms this implies that the transition functions and the associated contours are mapped to each other,
i.e. the image of $H_{k,p,\hgam}$ should be $H_{k,p',\hgam'}$ for a suitably chosen map $(p,\hgam)\mapsto (p',\hgam')$,
with the contours $\ell_{k,p,\hgam}$ following the same pattern.

In \cite{Alexandrov:2010ca} it was shown that our twistorial data, $H_{k,p,\hgam}$ and $\ell_{k,p,\hgam}$,
do satisfy this condition {\it up to some phase factors}. More precisely, one has
\be
T'_{H,\kappa}\cdot H_{k,p,\hgam}=\nu(\eta) H_{k,p-k\eta^0,\hgam[-k\eta^a/p^0]},
\qquad
M_{\epsilon^a}\cdot
H_{k,p,\hgam[p\epsilon/p^0]}
=\nu'(\epsilon) H_{k,p,\hgam[-p\epsilon/p^0]},
\label{symphase}
\ee
where $\hgam[\eps]$ denotes the image of the charge vector $\hgam$ under the spectral flow \eqref{chargesymp}
with parameter $\eps^a$,
and $\nu(\eta)$, $\nu'(\epsilon)$ are the constant phase factors which can be found in \cite{Alexandrov:2010ca}.
The cancelation of all dependence on the Darboux coordinates in the symmetry transformations \eqref{symphase}
is very promising, but the presence of the phases indicates some tension between S-duality,
Heisenberg and monodromy invariance and suggests that it may be necessary to relax some of
our assumptions about the way these symmetries are realized. In particular, it is possible
that S-duality is realized only as a finite index subgroup of the full $SL(2,\IZ)$.
In the absence of a satisfactory resolution of this problem, we simply ignore the anomalous phase factors
and proceed further. Moreover, it turns out that for $k=1$ the phase $\nu(\eta)$ can be canceled
by a suitable choice of characteristics
\be
\thD^{\Lambda}=0,
\qquad
\phins_0=A_{00}/2,
\qquad
\phins_a-\phiD_a=A_{aa}.
\ee
We will use this choice in the next section and the results obtained there seem to confirm
its correctness. Unfortunately, no choice was found which would lead to cancelations in other cases.

\subsubsection{Fivebrane instanton action}
\label{subsubsec-5inst}

The second consistency check is to show that the corrections to the metric
generated by the transition functions \eqref{5pqZ2} indeed have the form of the NS5-brane contributions \eqref{couplNS5}.
In other words, we need to compute the instanton action corresponding to \eqref{5pqZ2}.
The simplest way to do this is to evaluate the Penrose transform of $H_{k,p,\hgam}$
\be
\int_{\ell_{k,p,\hgam}}  \frac{\de\varpi}{\varpi}\,
H_{k,p,\hgam}(\xi(\varpi),\txi(\varpi),\alpha(\varpi)).
\label{Ptrans}
\ee
This is a typical integral which describes instanton corrections to various quantities. While the precise
form of the integration kernel can be different and vary from one quantity to another,
it does not affect the leading saddle point approximation we are interested in.

Substituting the tree level expressions for the Darboux coordinates \eqref{relABzeta}, it is straightforward
to find the saddle point and to evaluate the resulting instanton action $S_{k,p.\hgam}$.
The expression is a bit cumbersome and we refer to \cite{Alexandrov:2010ca}
for its explicit form. As a cross-check one can verify that the same expression is obtained
by applying the classical mirror map and S-duality to the classical D-instanton action \eqref{d2quali}
\be
S_{k,p.\hgam}=\gsl\cdot S_{\gamma},
\qquad
S_{\gamma}=2\pi\tau_2 |Z_\gamma|+ 2\pi\I (q_\Lambda \zeta^\Lambda-p^\Lambda\tzeta_\Lambda).
\label{actD}
\ee
The fivebrane action crucially simplifies in the small string coupling limit $\tau_2\to\infty$.
In this approximation one finds
\be
\label{SdV}
S_{k,p.\hgam}\approx \pi k\Bigl( 4 \, e^{\phi}
- \I (n^{\Lambda}-\zeta^\Lambda)\bar \cN_{\Lambda\Sigma} (n^{\Sigma}-\zeta^\Sigma)\Bigr)
+ \I\pi  k \(\sigma+(\zeta^\Lambda-2 n^\Lambda) \tzeta_\Lambda\) -2\pi\I m_\Lambda z^\Lambda,
\ee
where we used the following notations
\be
n^0=\frac{p}{k},
\qquad
n^a=\frac{p^a}{k},
\qquad
m_a=\frac{p^0}{k}\,\hat q_a,
\qquad
m_0=\frac{a p^0}{k}\, q'_0- c_{2,a}p^a \epsilon(\gsl).
\label{defcarge}
\ee
It is clear that the NS5-brane action \eqref{couplNS5} is a part of our result.
Moreover, the first term matches precisely the instanton action obtained in \cite{deVroome:2006xu},
based on a classical analysis of instanton solutions in $\cN=2$ supergravity,
while the other terms restore the appropriate axionic couplings.
Thus, the transition functions \eqref{5pqZ2} are in perfect agreement with
the supergravity description of NS5-branes.

\section{Fivebrane partition function and topological strings}
\label{sec_reltop}

\subsection{Relation to the topological string amplitude}
\label{subsec_ns5top}

During long time it was suspected that the NS5-brane partition function $\ZNS{k}$
is closely related to the topological string partition function $\Ztop$ \cite{Dijkgraaf:2002ac,Nekrasov:2004js}.
There are several reasons to expect such a relation. In particular, for $k=1$ they are both
governed by Donaldson-Thomas invariants and, in addition, they both behave as wave functions.
For topological strings this is a well known fact coming from
the holomorphic anomaly equations satisfied by $\Ztop$ and the possibility to define it
for different polarizations of the symplectic space $H^3(\CY,\IR)$ \cite{Witten:1993ed}.
For NS5-branes this is a consequence of the Heisenberg algebra or, equivalently,
the twisted periodicity condition \eqref{thperiod}. Indeed, the most general function
satisfying this condition can be written as a non-Gaussian theta series
\be
\label{thpl2f}
\cZ_{\Thns,\mu}^{(k)}(\cN, \zeta^\Lambda,\tzeta_\Lambda) =
\!\! \!\!\sum_{n\in \Gamma_m+\mu+\thns}\!\! \!
\Psi_\IR^{k,\mu}\left( \zeta^\Lambda - n^{\Lambda}\right)
\expe{  k (\tilde\zeta_{\Lambda}-\phins_\Lambda) n^{\Lambda}
+\frac{k}{2} (\thns^{\Lambda}\phins_{\Lambda}-\zeta^{\Lambda}\tzeta_{\Lambda})} ,
\ee
where $\Psi_\IR^{k,\mu}(\zeta^\Lambda)$ is a kernel of the theta series
which also depends on the complex structure moduli and the string coupling.
One observes that the unknown kernel depends only on a half of the variables
parametrizing the torus $\cT$. In the given case, these are $\zeta^\Lambda$.
On the other hand, one can give a similar representation where the kernel depends instead
on $\tzeta_\Lambda$ and is related to the original one by a Fourier transform.
This indicates that $\Psi_\IR^{k,\mu}(\zeta)$ should be viewed as a wave function in the $\zeta$-representation.
Moreover, it should transform in the metaplectic representation under changes of symplectic basis
similarly to the topological wave function in the real polarization.
Due to this, it is tempting to suggest that these two objects should be somehow identified.

It turns out that the results presented in the previous section allow to put these ideas on a firm ground.
To this aim, let us consider a formal sum of
the holomorphic functions \eqref{5pqZ2} for $k=1$ over the integer charges $p,p^a$ and $q_\Lambda$:
\be
H_{{\rm NS5}}^{(1)}(\xi,\txi,\alpha)=\sum_{p,p^a, q_\Lambda} H_{1,p,\hgam}(\xi,\txi,\alpha).
\label{fullH}
\ee
This sum is formal because each term is attached to a different contour on $\CP$.
Nevertheless, it can be meaningfully inserted into general formulas for Darboux
coordinates and the contact potential, provided the sum is performed {\it after} integration along $\CP$.
From the mathematical point of view, it just defines a section of $H^1(\cZ_{\MH^{\rm pert}},\cO(2))$.

As we know from the previous section, for $k=1$ there exists a choice of characteristics \eqref{Ptrans} such that
the section $H_{{\rm NS5}}^{(1)}$ is invariant under the Heisenberg symmetry. As a result, it can be recast
into the form very similar to the non-Gaussian theta series \eqref{thpl2f}
\be
\label{hthxin}
H_{{\rm NS5}}^{(1)}(\xi,\txi,\alpha)=\frac{1}{4\pi^2}
\sum_{n^\Lambda }
\cH_{{\rm NS5}}^{(1)}\( \xi^\Lambda - n^\Lambda\)\,
\expe{\alpha+n^\Lambda(\txi_\Lambda-\phi_\Lambda)} .
\ee
A remarkable fact is that the function $\cH_{{\rm NS5}}^{(1)}$ appearing in this representation
turns to be proportional to the wave function of the topological A-model
on $\CYm$ in the real polarization, which should be analytically continued to complex values of its argument
\cite{Alexandrov:2010ca},
\be
\cH_{{\rm NS5}}^{(1)}(\xi^\Lambda) \sim
 \Psi_\IR^{\rm top}(\xi^\Lambda) .
\label{NS5DTrelation}
\ee
The proportionality factor can be expressed in terms of Mac-Mahon function and depends only on $\xi^0$.

The relation \eqref{NS5DTrelation} provides a direct link between the NS5-brane instantons in type IIB
string theory on a Calabi-Yau $\CYm$ and A-model topological strings on the same Calabi-Yau.
By mirror symmetry, one expects that NS5-brane instantons in type IIA on $\CY$ should be related
to the topological B-model on $\CY$. On the other hand, despite the relation \eqref{NS5DTrelation}
is very concrete, the unclear status of the sum over charges
in the definition of $H_{{\rm NS5}}^{(1)}$ raises some questions about its physical interpretation
which certainly require a deeper understanding of the physics behind it.

\subsection{Fivebrane partition function from twistor space}
\label{subsec_nonlin}

Due to the complicated nature of NS5-branes, to define their partition function is a very non-trivial problem.
One of the standard approaches to this problem is a holomorphic factorization \cite{Witten:1997hc,Belov:2006jd}.
It does allow to find a partition function which has the form of the theta series \eqref{thpl2f} and reproduces the classical
fivebrane instanton action \eqref{SdV}. However, the validity of this approach is very restricted. First of all, it ignores
all non-linear effects on the fivebrane worldvolume so that it gives only a Gaussian approximation
of the full partition function. Then, in type IIA, it takes into account only
the degrees of freedom of the self-dual 3-form flux $H$,
whereas we expect also a non-trivial dependence on the complex structure moduli and the string coupling.
And finally, it involves some further simplifying assumptions which do not seem to hold in general \cite{Alexandrov:2010ca}.
Thus, while the holomorphic factorization likely captures the right topological properties of
the fivebrane partition function, it is not able to produce the complete result.

On the other hand, our twistor approach suggests a natural candidate for the full partition function:
it can be defined by summing the Penrose transform \eqref{Ptrans} over charges
\be
\hZNS{k,\mu}= \sum_{p,p^a, \hat q_a,\hat q_0} \int_{\ell_{k,p,\hgam}}  \frac{\de\varpi}{\varpi}\,
H_{k,p,\hgam}(\xi(\varpi),\txi(\varpi),\alpha(\varpi)),
\label{deffivepartf}
\ee
where the vector $\mu^a$ is the residue class defined below \eqref{resOmeg}.
Note that this definition absorbs the dependence on the NS-axion
so that it actually coincides with the r.h.s. of \eqref{ns5quali}.
In \cite{Alexandrov:2010ca} the sum over electric charges has been performed explicitly. Besides,
the saddle point evaluation of the Penrose transform in section \ref{subsubsec-5inst} has been generalized
to include the normalization factor which should correspond to the one-loop determinant.
As a result, for $k=1$ in the small coupling limit, this prefactor was found to be
\be
\cF \sim
\(\bar{z}^{\Lambda}\Im F_{\Lambda\Sigma}\bar{z}^{\Sigma}\)^{-1}
\varpi_s^{-2+\frac{\chi_\CY}{24}}
\, e^{f_1(z)},
\label{normalizationfactor}
\ee
where
\be
\varpi_s=4\, e^{\cK}\tau_2^{-1}\,(\zeta^{\Lambda}-n^{\Lambda})
\Im\cN_{\Lambda\Sigma} z^{\Sigma}
\ee
is the saddle point and $f_1(z)$ is the holomorphic part of the one-loop vacuum amplitude of the topological B-model.
The appearance of $f_1(z)$ further strengthens the connection with topological strings.
Moreover, it is consistent with the monodromy transformations of the NS-axion \eqref{sigmon}.
Indeed, denoting by $\cL$ the line bundle over $\KC(\CY)$ in which the holomorphic three-form $\Omega$ is valued,
$e^{f_1(z)}$ is a section of $\cL^{1-\frac{\chi_\CY}{24}+ \frac{b_3}{4}} $
which does not have a canonical definition unless $\chi_\CY$ is divisible by 24.
Due to this, one expects that its monodromy transformation also contains
a unitary character $\kappa(M)$ which appears in \eqref{sigmon}
and the condition that the total partition function $\hZNS{k}$ is well defined requires that
the two characters are the same.

Finally, we show how the fivebrane partition function \eqref{deffivepartf} can be encoded in
a certain state in the Hilbert space $\mathcal{H}$ of
square-integrable functions on $H^{3}(\CY, \mathbb{R})$.
To this aim, let us consider a specific state $| \Psi^{\Gamma_m,k,\mu} \rangle \in \cH$,
which has the following simple representation in terms of its wave function
$\Psi_\IR^{\Gamma_m,k,\mu}(\zeta^\Lambda)$ in the real polarization
\be
\Psi_\IR^{\Gamma_m,k,\mu}(\zeta^\Lambda)=
\sum_{n \in \Gamma_m+\mu}
\delta\left(\zeta^\Lambda - n^\Lambda\right)
\, e^{2\pi \I k \, \phi_\Lambda n^\Lambda }.
\label{PsiRNS5G}
\ee
Besides, we introduce two sets of operators,
$T^\Lambda$ and $\tilde T_\Lambda$, acting on $\cH$.
Their action on an arbitrary wave function in the real polarization reads
\be
\label{qalg}
T^\Lambda \cdot \Psi_\IR(\zeta^\Lambda) =2\pi k  \zeta^\Lambda  \Psi_\IR(\zeta^\Lambda)  ,
\qquad
\tilde T_\Lambda \cdot \Psi_\IR(\zeta^\Lambda) =  -\I  \p_{\zeta^\Lambda} \Psi_\IR(\zeta^\Lambda),
\ee
so that they satisfy the Heisenberg algebra
\be
[T^\Lambda,\tilde T_\Sigma]
=2\pi \I k \delta^\Lambda_\Sigma.
\label{Heisenbergalgebra}
\ee
with the NS5-brane charge playing the role of the Planck constant.
Then the fivebrane partition function \eqref{deffivepartf} appears as the Penrose transform
of a matrix element
\be
\label{matelp}
\hZNS{k,\mu}
=\int \frac{\de\varpi}{\varpi}\,
e^{2\pi\I k \alpha} \,
\langle \Psi^{\Gamma_m,k,\mu} |
\, e^{ \I(\xi^\Lambda \tilde T_\Lambda - \txi_\Lambda T^\Lambda)} |
\cH_{\rm NS5}^{(k,\mu)} \rangle ,
\ee
where $|\cH_{\rm NS5}^{(k,\mu)} \rangle$ is the state whose wave-function in the real polarization
is given by an extension of $\cH_{\rm NS5}^{(1)}$ from \eqref{hthxin}
to arbitrary NS5-brane charge $k$ and residue class $\mu^a$.
This is the state we were looking for.
As we know from \eqref{NS5DTrelation}, it essentially coincides with the state associated
to the topological string wave function.

\chapter{\ Example: universal hypermultiplet}
\label{chap-4d}

Until now we considered the HM moduli space which is obtained by compactifying type II string theory
on arbitrary Calabi-Yau. While our results uniquely determine the quantum corrected metric on $\MH$,
it is difficult to give its explicit form beyond the perturbative approximation.
However, there exists a special case where such an explicit representation is possible.
This is the case of type IIA string theory compactified on a {\it rigid} Calabi-Yau,
which has vanishing Hodge number $h^{2,1}$ and therefore does not have complex structure moduli.
In this case the effective theory contains only one hypermultiplet, which is known
by the name {\it universal hypermultiplet} (UHM) and was briefly mentioned on page \pageref{UHMremark}.
Since $n_H=1$, the HM moduli space is four-dimensional which is the simplest possible case of
a quaternion-K\"ahler manifold. This is the reason why the UHM was extensively studied in the literature,
see for instance \cite{Strominger:1997eb,Becker:1999pb,Gutperle:2000sb,Ketov:2002vr,Antoniadis:2003sw,Anguelova:2004sj,Davidse:2005ef},
and in fact the first insights to many of the results reported here have been obtained from these studies.

In this chapter we are going to specialize our general results to the particular case of the UHM.
Moreover, we use the fact that four-dimensional QK manifolds can be described in terms of solutions of
just one non-linear differential equation. We establish a connection between this description and our twistor approach,
which allows to immediately translate the results for the quantum corrections to the contact structure
on the twistor space to the corresponding corrections to the metric.

\section{Four-dimensional QK spaces}

\subsection{Przanowski description}
\label{subsec_Prz}

A four-dimensional QK manifold is an Einstein space with a non-vanishing cosmological constant
and a self-dual Weyl curvature. In the case of positive cosmological constant there are actually only
two possibilities: the four-sphere $S^4$ and the complex projective plane $\IC P^2$.
However, we are interested in the case of negative scalar curvature which is much more non-trivial
and allows for continuous deformations.

It is worth to mention that the limit of zero curvature corresponds to four-dimensional HK manifolds.
The metric on such manifolds, which are thus Ricci-flat and self-dual, is well known to be encoded
in solutions of the so called heavenly equations \cite{Plebanski:1975wn}. It is less known that
a similar result holds also for a non-vanishing cosmological constant. It was shown by Przanowski \cite{Przanowski:1984qq}
that locally it is always possible to find complex coordinates $z^\alpha$ such that
the metric takes the form
\be
\de s^2_\qkm= -\frac{6}{\Lambda}\(h_{\alpha \bar\beta}  \, \de z^{\alpha}\, \de z^{\bar\beta} +
2  e^h \, | \de z^2|^2  \)
\equiv 2 g_{\alpha\bar\beta}
\,\de z^{\alpha}\, \de z^{\bar\beta},
\label{metPrz}
\ee
where $h_\alpha= \partial h/\partial z^\alpha$, etc. It is completely determined in terms of a single
real function $h(z^\alpha,z^{\bar\alpha})$ and the constraints of quaternionic geometry require that
the function must satisfy the following non-linear
partial differential equation
\be
\Prz(h)\equiv h_{1\bar 1} h_{2\bar 2}
-h_{1\bar 2}h_{{\bar 1} 2}+\(2h_{1\bar 1}-h_1 h_{\bar 1}\)e^h=0.
\label{master}
\ee
To show that the metric \eqref{metPrz} is indeed quaternion-K\"ahler, it is sufficient to verify
the conditions \eqref{ompp}, where in four dimensions the coefficient $\nu$ is related to the cosmological constant
as $\nu=\Lambda/3$. They are satisfied provided the $SU(2)$-connection and
the quaternionic two-forms are expressed in terms of $h$ as follows \cite{Looyestijn:2008pg,Alexandrov:2009vj}
\begin{equation}
\begin{split}
p^+=e^{h/2}{\rm d}z^2, &
\qquad
p^3=\frac{\I}{2}(\partial -{\bar \partial})h,
\qquad
p^-=(p^+)^*,
\\
\omega_\qkm^+=\frac{6}{\Lambda}\,e^{h/2}h_1\,\de z^1\wedge \de z^2,
& \qquad
\omega_\qkm^3 =2\I \,g_{\alpha\bar\beta}\,\de z^\alpha \wedge \de z^{\bar \beta},
\qquad
\omega_\qkm^-= (\omega_\qkm^+)^*.
\end{split}
\label{su2-conn}
\end{equation}

It is important to emphasize that solutions of the Przanowski equation \eqref{master} are not in a one-to-one
correspondence with the self-dual Einstein metrics. In general, there are infinitely many ways of expressing
a given metric in the form \eqref{metPrz}. One can distinguish between two types of ambiguities.
The first comes from the following {\it holomorphic} change of coordinates
\begin{equation}
\begin{split}
\label{diff-equiv}
& \qquad
z^1\rightarrow {z'}^{1} = f(z^1,z^2),
\qquad
z^2 \rightarrow {z'}^{2} = g(z^2),
\\
&
h(z^1,z^2) \rightarrow {h'}(z^1,z^2) =
h\(f(z^1,z^2),g(z^2)\) - \log \left|g_2(z^2)\right|^2
\end{split}
\end{equation}
which, supplemented by the above change of the Przanowski function, provides a symmetry of the metric \cite{Przanowski:1990gf}.
The second ambiguity roots in the choice of complex structure defined by the coordinates $z^\alpha$.
A QK space has infinitely many {\it local} integrable
complex structures and any of them can be used to achieve the Przanowski ansatz \eqref{metPrz}.
In general, different complex structures lead to different solutions of the Przanowski equation which are related
to each other through a non-holomorphic change of variables.
As we will see, this phenomenon has a natural interpretation in the framework of the twistor approach.

\subsection{QK spaces with isometries}
\label{subsec_4disom}

If a four-dimensional QK manifold has an isometry, there exists even a simpler description.
It is provided by the Tod ansatz \cite{MR1423177} which can be written in local coordinates $(\rho,z,\bz, \psi)$
in the form
\be
\label{dstoda}
\de s_\cM^2 = -\frac{3}{\Lambda}\[
\frac{P}{\rho^2} \left( \de \rho^2 + 4 e^ \todaQ \de z \de\bar z \right)
+ \frac{1}{P\rho^2}\,(\de \sigp + \Theta )^2\] ,
\ee
where the isometry acts as a shift in the coordinate $\sigp$.
Here, $ \todaQ$ is a function of $( \rho,z,\bz)$,  $P\equiv 1- \hf\,\rho \p_\rho  \todaQ $,
and $\Theta$ is a one-form such that
\be
\label{dth}
\de \Theta = \I (\p_z P \de z - \p_{\bz} P \de \bar z)\wedge \de  \rho
- 2\I\, \p_\rho(P {\rm e}^ \todaQ)\de z\wedge \de\bar z.
\ee
The Einstein self-duality condition of the metric imposes on the function $\todaQ$
the three-dimensional continuous Toda equation
\be
\p_{z} \p_{\bz}  \todaQ +\p_ \rho^2 \, {\rm e}^ \todaQ = 0,
\label{Toda}
\ee
which also plays the role of the integrability condition for \eqref{dth}.

This description can be related to the more general one due to Przanowski
if one chooses the coordinates $z^\alpha$ such that
the Killing vector $\p_\sigp$ acts by shifting the imaginary part of $z^1$.
Then the condition that this action is isometric is equivalent to $h_1=h_{\bar 1}$
and the relation between  $h,z^1,z^2$ and $ \todaQ,\rho,z$ is given by
the Lie-B\"acklund transformation \cite{Przanowski:1991ru,Tod:2006wj}
\be
z=z^2,
\qquad
\rho=1/(2h_1),
\qquad
 \todaQ=h+ 2 \log \rho.
\label{Backlund}
\ee

Similarly, in the case of two commuting isometries, there is an ansatz due to Calderbank and Pedersen \cite{MR1950174}
which encodes the metric in terms of solutions of the Laplace equation on the hyperbolic plane.
We refer to \cite{Alexandrov:2009vj} for the relation of this ansatz to the Toda and Przanowski equations.

\subsection{Relation to the twistor description}

The above results provide an explicit parametrization of the four-dimensional QK metrics
in terms of solutions of differential equations. However, this leaves us with the problem
of solving these non-linear equations and an additional problem that different solutions
can in fact describe the same manifold. This should be contrasted with the twistor approach
where any consistent set of holomorphic transition functions defines a quaternionic manifold,
but to get the metric, in general, requires solving some (integral) equations.

A relation between these two descriptions has been established in \cite{Alexandrov:2009vj}.
It comes from the observation that the metric on $\qkm$ can be expressed through
the K\"ahler potential and the contact one-form on its twistor space \cite{deWit:2001dj}
\be
\label{dszx}
\de s^2_{\qkm}\equiv \frac{12}{\Lambda} \left(\de s^2_{\qkt}-e^{-2K_{\qkt}}|\hCX|^2\right),
\ee
where the r.h.s. should be restricted to any real-codimension 2 (local) complex submanifold $\cC$ transverse
to the contact distribution, i.e. such that $\hCX\vert_\cC\neq 0$.
This form of the metric essentially coincides with the Przanowski ansatz \eqref{metPrz}
and allows to identify the Przanowski function with the restriction of the K\"ahler potential on $\qkt$
to the submanifold $\cC$ up to a K\"ahler transformation
\be
h=-2K_{\cC}+F_{\cC}+\bar F_{\cC}.
\label{identh}
\ee
Moreover, one can show that a set of constraints to be satisfied by the K\"ahler potential,
which include in particular the Monge-Amp\`ere equation representing the K\"ahler-Einstein
property of the metric on $\qkt$, reduces to the single Przanowski equation \eqref{master}.

This derivation also demonstrates that choosing different complex submanifolds, one generates
different solutions of the Przanowski equation. A particular choice of $\cC$ specifies a local
integrable complex structure on $\qkm$. As a result, this choice appears to be the origin of the non-holomorphic
ambiguity in the Przanowski description mentioned in the end of section \ref{subsec_Prz}.
In practice, the complex submanifold $\cC$ can be specified by the vanishing of
some holomorphic function $\cC(\xi,\txi,\alpha)=0$ on $\qkt$.
Then it is enough to solve this condition with respect to one of the holomorphic coordinates and
substitute the solution into the K\"ahler potential.

A particular convenient choice of the complex submanifold is provided by $\cC=1/\xi$ because
this is the only choice which can be equivalently described as a hypersurface of constant fiber coordinate
$\varpi$, namely $\varpi=0$.
In this case, one can explicitly identify the Przanowski coordinates $z^\alpha$ and function $h$
with coefficients of the Darboux coordinates \eqref{txiqlineQK} in the small $\varpi$-expansion.
More precisely, one has \cite{Alexandrov:2009vj}
\be
\begin{split}
& z^1=\frac{\I}{2}\,\ai{+}_0+c_\alpha \log Y-\ctxi A,
\qquad
z^2=\frac{\I}{2}\,\txii{+}_0+\ctxi\log Y,
\\
& \qquad\qquad\qquad\quad
h=-2\phi+2\log\frac{Y}{2}.
\end{split}
\label{relhPhi}
\ee
Conversely, some of the first coefficients and the constant part of the contact potential can be
expressed through the Przanowski function
\be
Y=\frac{{\rm e}^{h/2}}{|h_1|},
\qquad
\xi^{[+]}_{0}=Y\phi^{[+]}_1+\frac{h_2}{h_1},
\qquad
\phi^{[+]}_0=-\log(2h_1).
\label{resolgen}
\ee
In the presence of an isometry these relations can be further simplified. For example, the contact potential
becomes globally defined and $\varpi$-independent so that $\Phi^{[i]}=\phi$
and one can drop the first term in the expression for $\xi^{[+]}_{0}$.
Using \eqref{Backlund}, one also finds an explicit relation between the twistor data and the Toda potential
\be
\rho=e^\phi,
\qquad
\todaQ=2\log(Y/2) .
\label{Backlund_tw}
\ee
In particular, the radial coordinate is identified with the contact potential and therefore in the context of string theory
coincides with the four-dimensional string coupling.

\section{Geometry of the universal hypermultiplet}

\subsection{Perturbative universal hypermultiplet}
\label{subsec_4dpert}

The perturbative metric on the moduli space of the universal hypermultiplet can be obtained from
the one-loop metric \eqref{hypmetone} where there are no complex structure moduli $z^a$ and
there is only one pair of the RR fields $\zeta,\tzeta$. In addition, since a degree 2 homogeneous function
of one argument is necessarily quadratic, the holomorphic prepotential must be of the form
$F=\frac{\tau}{2}\, X^2$. The complex parameter $\tau$ is the modulus of the intermediate Jacobian $\cJ_c(\CY)$
defined in section \ref{chap_NS5}.\ref{sec_topissue}. Although it does affect the global structure
of the moduli space, locally it can be absorbed into a redefinition of the RR scalars.
For simplicity, we restrict it to be $\tau=-\I/2$ so that $K=N=1$ and $\cN=-\I/2$.
As a result, the perturbative metric takes the form \cite{Antoniadis:2003sw,Anguelova:2004sj}
\be
\de s_{{\rm UHM}}^2=\frac{r+2c}{r^2(r+c)}\,\de r^2+
\frac{r+2c}{r^2} \left|\de{\tilde \zeta} +\frac{\I}{2}\,\de \zeta \right|^2
+\frac{r+c}{16r^2(r+2c)}\( \de\sigma+{\tilde\zeta}\de\zeta -\zeta \de{\tilde\zeta}\)^2 .
\label{UHM}
\ee

It is easy to recognize that the metric \eqref{UHM} falls in the Tod ansatz \eqref{dstoda}
where one should take \cite{Davidse:2005ef}
\be
\todaQ=\log(\rho+c),
\label{UHM-T}
\ee
identify the coordinates as
\be
\rho=r,
\qquad
z=-\frac{1}{4}\,(\zeta-2\I\tzeta),
\qquad
\sigp=-\frac{1}{8}\,\sigma,
\ee
and put $\Lambda=-3/2$. It is trivial to check that \eqref{UHM-T} indeed satisfies the Toda equation \eqref{Toda}
being its one of the most trivial solutions. The one-loop deformation appears as
a simple shift of the radial coordinate, which is a symmetry of the Toda equation.

Similarly, one can put the metric into the Przanowski form provided one takes
\cite{Alexandrov:2006hx}
\be
\begin{split}
z^1=-\bigl( r+c\log (r+c)-\tfrac{1}{8}\,\zeta^2 \bigr)& - \tfrac{\I}{4}(\sigma+\zeta\tzeta) ,
\qquad
z^2=-\tfrac{1}{4}(\zeta-2\I\tzeta),
\\
&
h=-\log\frac{r^2}{r+c}.
\end{split}
\label{h-one-loop}
\ee
These equations perfectly agree with the identifications \eqref{relhPhi} where the Darboux coordinates
are taken from chapter \ref{chap_pert} with $\tau_2$ excluded in favor of the contact potential
$r=e^\phi$ \eqref{oneloopres}. Thus, the Darboux coordinates we are working with, in the patch $\cU_0$, are given by
\be
\label{gentwiUHM}
\begin{split}
\xii{0} &= \zeta + 2\sqrt{r+c}\left( \varpi^{-1}  -\varpi \right) ,
\\
\txii{0} &= \tzeta -\I\sqrt{r+c}\left( \varpi^{-1}+\varpi \right),
\\
\tai{0}&= \sigma -\I\sqrt{r+c}\left(\varpi^{-1} (\zeta-2\I\tzeta)+ \varpi (\zeta+2\I\tzeta) \right)
-8\I c\log\varpi.
\end{split}
\ee
To check that the function $h$ \eqref{h-one-loop} does solve the Przanowski equation \eqref{master},
one should express $r$ as a function of $z^\alpha$. In presence of the one-loop correction this can be done
only implicitly, which is however sufficient to compute the derivatives of $h$ and to verify the equation.

Finally, we mention that there is a surprising duality between
the one-loop corrected UHM described here and the $c=1$ non-critical string theory compactified
on the self-dual radius \cite{Alexandrov:2012bn}
(for a review of $c=1$ strings, see \cite{Klebanov:1991qa,Alexandrov:2003ut}).
Both systems are represented by the same solution of the Toda equation and the twistor description
of the UHM turns out to be equivalent to the Lax formulation of the integrable structure
of the $c=1$ string theory.

\subsection{Instanton corrections as perturbations}
\label{subsec_4dinst}

To write the instanton corrected UHM metric, one should find a solution
of the Przanowski equation \eqref{master} which would incorporate these contributions.
Of course, it is almost impossible task given the complicated nature of this equation.
If one ignores the NS5-brane instantons, so that the isometry along the NS-axion remains preserved,
it is possible to use the Toda description of section \ref{subsec_4disom}.
However, it does not much simplify the problem as the equation to be solved
is still a non-linear equation in partial derivatives.

Nevertheless, both these equations become very powerful once one is interested only
in linear deformations of the one-loop corrected metric \eqref{UHM}.
This restriction is equivalent to the one-instanton approximation which was discussed repeatedly
in the previous chapters. In this approximation, one can expand the ``master" equation, \eqref{master} or \eqref{Toda},
around a solution representing the perturbative metric. As a result, one obtains a {\it linear} differential
equation which is much easier to solve. Of course, not all solutions of these linearized equations are physically relevant
and some conditions which select the admissible solutions should be imposed.

This procedure has been carried out in \cite{Davidse:2005ef} for the Toda equation and in
\cite{Alexandrov:2006hx} for the Przanowski equation. The following three types of solutions have been found:
\beq
\delta h^{(2)}_{p,q}& =& \frac{C}{r}\,e^{-2\pi \I(q\zeta-p\tzeta)} K_0\left( 4\pi \sqrt{(r+c)(4q^2+p^2)} \right) ,
\label{d2}
\\
\delta h^{(2')}_{p,k,\pm} &=& \frac{C}{r}\,
\frac{ \(\sqrt{1+\frac{\zeta^2}{16(r+c)}}-\frac{\zeta}{ 4\sqrt{r+c}}\)^{16\pi c k}}{\sqrt{16(r+c)+\zeta^2}}
\, {\rm e}^{\pm 2\pi\I\( p\tzeta- k(\sigma+\zeta\tzeta)\)}
e^{-\pi \(p-k\zeta\)\sqrt{16(r+c)+\zeta^2}} ,
\label{solmemex}
\\
\delta h^{(5)}_{k,\pm}&=& \frac{C}{r}\, e^{\pm \pi \I k \sigma} \,
\frac{e^{\pi k\( 4 r-(\frac14\,\zeta^2+\tzeta^2)\)}}{ (r+c)^{4c\pi k}}\,
\int_1^{\infty}e^{-8\pi k(r+c)t}\frac{{\rm d}t}{t^{1+8\pi ck}},
\label{solns5}
\eeq
where the charges are restricted to satisfy $p>0,\ k\zeta<0$
for the second solution and $k>0$ for the last one.
These solutions are not unique. In particular, any function obtained by applying a Heisenberg transformation
to \eqref{d2}-\eqref{solns5} or a linear combination thereof is also a solution of the Przanowski
equation linearized around \eqref{h-one-loop}.

Furthermore, the first solution has a transparent physical meaning: this is just an instanton correction due to
a D2-brane with charge $(p,q)$. Its leading large $r$ asymptotics, corresponding to the small $g_s$ limit,
perfectly agrees with \eqref{d2quali}, whereas the Bessel function provides all subleading contributions.
Similarly, comparing the third solution with \eqref{couplNS5}, one recognizes an NS5-brane instanton.
Applying the Heisenberg shift of $\tzeta$ and integrating over the shift parameter, one can obtain another solution
which differs only by the exponential factor. Its leading asymptotics reads
$e^{\pm \pi \I k(\sigma+\zeta\tzeta)-\pi k( 4r+\hf\,\zeta^2)}$ and reproduces the NS5-brane action \eqref{SdV}
from the previous chapter. The possibility to represent NS5-brane instantons in different forms can be
considered as a realization of the wave function property discussed in section \ref{chap_NS5}.\ref{subsec_ns5top}.

Finally, the second solution does not have an obvious interpretation. Its axionic couplings indicate
that it should represent a bound state of NS5 and D2-branes, but the real part of the instanton action
has a somewhat unusual form.
Although there is a BPS solution in classical supergravity which reproduces such instanton action
\cite{Davidse:2003ww}, the microscopic interpretation of \eqref{solmemex} remains unclear.

More generally, it turns out that the linearized Przanowski equation around a solution $h$
can be written in the following form \cite{Alexandrov:2009vj}
\be
\dPrz_h \, (\delta h)=  {\rm e}^h \, |h_1|^4 \,
\[  -\frac{3}{2\Lambda} \Delta + 1 \]\frac{\delta h}{|h_1|^2}=0.
\label{geneq}
\ee
In the case with an isometry when $h_1=h_{\bar 1}$,
the second order differential operator $\Delta$ coincides with the Laplace-Beltrami operator
defined by the metric \eqref{metPrz} one expands around. Otherwise, $\Delta$ is given
by its generalization involving a certain connection term, which unfortunately has not been
understood from a geometric point of view.
In any case, the deformations of $h$ appear as eigenmodes of this operator.
The specific eigenvalue $2\Lambda/3=R/6$ corresponds to a conformally coupled massless scalar field.
In this case, the eigenmodes of the Laplace-Beltrami operator are known to be generated by Penrose-type
contour integrals \cite{Neitzke:2007ke}. This establishes a link with our twistor approach.
Indeed, using identifications \eqref{relhPhi}, it was shown in \cite{Alexandrov:2009vj}
that if the transition functions can be represented as in \eqref{genf}
where $\Hpij{ij}$ is treated as a perturbation, the corresponding variation
of the Przanowski function is given by
\be
{\delta h}=-h_1\sum_j \oint_{C_j} \frac{\d \varpi}{2\pi \I \varpi}\, \Hpij{ij}.
\label{delh}
\ee
Moreover, for the UHM metric \eqref{UHM} one can explicitly verify that \eqref{delh} satisfies the
linearized equation \eqref{geneq} for any set of holomorphic functions and associated closed contours.
One may also consider open contours provided one makes sure that
all boundary contributions coming from the integration by parts and the action of the Laplace
operator on the limits of integration cancel each other.
This turns out to be the case when the end points of the contour lie
on any complex submanifold of the twistor space.

The simple formula \eqref{delh} allows to translate our results on deformations of the contact structure
on the twistor space induced by instanton corrections directly to the variation of the metric, without
necessity to solve any equations. It is sufficient to take the transition functions associated with a given
type of instantons, evaluate them on the Darboux coordinates \eqref{gentwiUHM} and integrate along
the corresponding contour. Substituting the result into \eqref{metPrz} then produces
an instanton corrected metric on the UHM moduli space in the one-instanton approximation.
Note that all open contours appeared so far do satisfy the condition that their ends belong to a complex
submanifold: for the contours $\ell_\gamma$ associated to D2-branes the submanifold is defined by $1/\xi=0$
and for the contours relevant for D3 and fivebranes in type IIB these are $m\xi+n=0$.

In this way, in particular, one can reproduce the solutions of the linearized Przanowski equation given above.
For example, the simple exponential function $\Hp\sim e^{-2\pi \I(q\xi-p\txi)}$, which is a single term from
the dilogarithm sum \eqref{prepH}, reproduces $\delta h^{(2)}_{p,q}$.
On the other hand, to get $\delta h^{(5)}_{k,\pm}$, one should take
\be
\Hp\sim (\xi \pm 2\I\txi )^{8\pi c k} \,
e^{\pm \pi \I k\tilde\alpha -\pi k\(\frac{1}{4}\xi^2+\txi^2\)} ,
\label{NS5fun}
\ee
and choose a contour $C$ connecting $\varpi=\infty\, (\varpi=0)$
to the point $\varpi_{\pm}$ corresponding to the complex submanifold $\xi\pm 2\I\txi=0$, namely
\be
\varpi_{\pm}=-\[\frac{4\sqrt{r+c}}{\pm\zeta+ 2\I\tzeta}\]^{\pm 1}.
\ee
Although such holomorphic functions did not appear in our studies of fivebrane corrections
based on S-duality, it is tempting to suggest that they may be relevant for the type IIA description
of NS5-brane instantons. These functions correspond to the so called symmetric gauge
where none of the Heisenberg shifts \eqref{Heisenb}, except the shift of the NS-axion, is an explicit symmetry.
Therefore, the complete answer should involve a double sum over both shifts of $\xi$ and $\txi$
generalizing the theta function representation \eqref{hthxin} \cite{Bao:2009fg}.
Although it is possible to relate it to the usual $\xi$ and $\txi$-representations,
the symmetric gauge is plausible due to the simplicity of the transition function \eqref{NS5fun}.
Moreover, it appears to be symplectic invariant since the quadratic term in the exponent is nothing else but
the Hesse potential evaluated on the symplectic vector $(\xi,\txi)$. This fact also opens
a possibility for a generalization of \eqref{NS5fun} to higher dimensions.
However, this part of the full picture has not been understood yet
and our story about the twistor approach to the non-perturbative HM moduli space stops here.

\begin{chapternon}{Conclusions}

In this review we summarized a progress in understanding quantum corrections
to the hypermultiplet moduli space.
Just five years ago only the tree level metric and
some fragmentary facts about its quantum corrections were known,
whereas now we are already very close to the complete non-perturbative picture!
We have shown how such a coherent picture arises step by step: starting from
the one-loop correction to the tree level metric, incorporating D-instantons,
and finishing with NS5-brane contributions, all included consistently with the
constraints of quaternion-K\"ahler geometry.

To achieve these results, it was crucial to apply the twistor approach to quaternionic geometries,
which has been developed precisely for this purpose. It originated as a generalization of the projective
superconformal approach, which marked many important findings in theories with $N=2$ supersymmetry,
but is restricted to quaternionic spaces with sufficient number of commuting isometries.
In contrast, the twistor approach does not have any restrictions
and in its framework many non-trivial physical effects obtain a simple and nice geometric description.

On the way, we have also gained new insights into the action of S-duality,
realization of quantum mirror symmetry, and connection to integrable models.
The last point is especially encouraging since it gives a hope that the whole problem
possesses a hidden integrable structure. Its discovery would certainly have a great impact
on the subject and could provide new hints to the fundamental structure of string theory.

Furthermore, the study of NS5-brane corrections revealed their close connection
to topological strings. Although such a connection was suspected to exist already long ago,
we were able to give it a precise form. However, its implications have not been investigated yet and
are still to be understood.

Of course, there still remains a lot of work to be done before the main problem
addressed here can be claimed to be completely solved.
Moreover, in the course of achieving this goal, one opens new research directions
which raise new issues and new questions.
Therefore, we would like to finish this review with a list of problems
which, from our viewpoint, are either very interesting or indispensable for further progress:
\begin{itemize}
\item
The most urgent problem is to extend the results reported here on instanton corrections
due to D3, D5 and NS5-branes to the multi-instanton level.
Such extension should also provide the complete non-perturbative mirror map
and make a conclusion about the fate of S-duality: Is it realized as the full $SL(2,\IZ)$ group or only
as some of its subgroups?

\item
In chapter \ref{chap_NS5} we found instanton corrections due to NS5-branes in the linear approximation.
The resulting picture is adapted to the type IIB formulation as it explicitly respects S-duality.
On the other hand, very often the results, which exist in the two mirror formulations, appear in type IIA
in a simpler form than in type IIB, as is the case, for example, for D-instantons.
Therefore, it is natural to ask how the NS5-brane contributions look like
in the type IIA picture? Note that this formulation is expected to be explicitly invariant under
symplectic transformations, but it is not clear how to reconcile this requirement
with the theta function representation \eqref{thpl2f}, which is a consequence of the Heisenberg symmetry.
Some hints to a possible resolution of this puzzle can be found in the study of instanton corrections
to the universal hypermultiplet presented in the last chapter.

\item
An important problem is to show that the full non-perturbative moduli space metric is free of any singularities.
In particular, this implies that the instanton corrections must resolve the one-loop singularity at $r=-2c$.
This requirement gives a very non-trivial consistency check on our construction.

At present, it is not clear how such a resolution can be demonstrated.
Probably, it should involve some resummation of the sum over charges and taking into account multi-instanton
contributions. This in turn raises the problem that, in contrast to the gauge theory case, the BPS indices
$\gnkl{\gamma}$ in string theory are exponentially growing and therefore the sum over charges is divergent.
In \cite{Pioline:2009ia} it was argued that this sum can be considered as an asymptotic series and be resumed
using a generalized Borel technique. As a result, the ambiguity of the asymptotic series turns out to be of the order
of NS5-brane instantons so that the latter are expected to convert this series into a convergent one.
To show that this is indeed the case is another important open problem.

\item
A series of questions arises from the observed relation of D-instanton corrections to integrable models
described by Thermodynamic Bethe Ansatz.
First of all, it would be interesting to understand what is the precise integrable structure
behind this TBA and what is its physical interpretation from the string point of view.
In particular, how to extract the two dimensions, where the integrable model is supposed to live, in string theory?

It should be noted that for $N=2$ supersymmetric gauge theories there is a special chamber in the moduli space
where the spectrum takes a particularly simple form \cite{Shapere:1999xr}.
In this canonical chamber the BPS states can be associated with the nodes of a Dynkin diagram
and the scalar product between their charges are provided by the corresponding Cartan matrix.
The phases of the central charges turn out to be ordered also in a special way and, as a result,
the TBA equations for the metric on the instanton corrected moduli space
can be mapped to the Y-system of the standard ADE integrable models
\cite{Cecotti:2010fi}.
Whether similar results hold in the string theory context remains unclear.

\item
A very interesting question is whether the integrability of D-instantons is extended to include NS5-brane contributions?
An affirmative answer to this question could have important implications.
Whereas the relation between D-instantons and integrability is fascinating, in a sense,
it is not completely unexpected. As we know, in this approximation the HM moduli space
is dual by the QK/HK correspondence to a HK manifold and
a relation between the hyperk\"ahler geometry and integrability has been observed long ago \cite{Donagi:1995cf}.
On the other hand, we are not aware of any examples where the former can be replaced by the contact geometry.
Thus, the HM moduli space corrected by NS5-brane instantons has a chance to provide the first
example of such type.

Moreover, in the presence of NS5-branes, the Heisenberg symmetry becomes isomorphic
to the quantum torus algebra, where the fivebrane charge plays the role of the quantum deformation parameter.
Therefore, it is natural to suspect that the inclusion of NS5-branes
corresponds to some quantum deformation of a classical integrable system.
In particular, a natural hypothesis is that the D-instanton transition functions on the type IIA side
get modified to be given by the quantum dilogarithm. However, much more work is required to establish whether
this is indeed the case.

\item
A closely related question is whether the refined BPS invariants, which appear naturally in
$N=2$ gauge theories in presence of line defects \cite{Gaiotto:2010be} and
are subject to the motivic wall-crossing formula \cite{ks,Dimofte:2009bv}, play some role also in our story?
S-duality used to get fivebrane contributions seems to imply that the usual Donaldson-Thomas invariants
are enough to describe all instanton corrections.
However, if the above idea about the quantum dilogarithm is not completely false,
the refined invariants do have a chance to appear.

\item
Finally, one may expect that the results presented in this review will have some important implications for other domains
of string theory which include, in particular, topological strings and the physics of BPS black holes.
Hopefully, at least some of these results can also be extended to vacua with $N=1$ supersymmetry and be useful for
phenomenological model building.

\end{itemize}

\vskip 1cm

\centerline{\bf \large \itshape Acknowledgements }

\vskip 0.5cm

It is a great pleasure to thank all my collaborators who shared with me
during last years their deep knowledge and exciting ideas.
My big gratitude goes to Jan Manschot, Daniel Persson, Philippe Roche, Frank Saueressig, Stefan Vandoren,
and especially to Boris Pioline whose limitless energy and devotion to physics
were indispensable for many developments presented here.
I would like to thank also Vladimir Fateev, Nikolay Gromov, Vladimir Kazakov, Ivan Kostov, Andrew Neitzke,
Vasily Pestun, Sylvain Ribault, Pedro Vieira and Konstantin Zarembo
for many useful discussions.

\end{chapternon}

\renewcommand{\baselinestretch}{1} \normalsize
\renewcommand{\bibname}{References}

\clearemptydoublepage

\addcontentsline{toc}{chapter}{References}

\providecommand{\href}[2]{#2}\begingroup\raggedright\endgroup

\end{document}